\documentclass[traditabstract]{aa}
\usepackage{graphicx}
\usepackage{txfonts}
\usepackage{natbib}

\usepackage{bm}
\usepackage{url}
\def\farcs{\hbox{$.\!\!^{\prime\prime}$}}
\def\farcm{\hbox{$.\mkern-4mu^\prime$}}
\def\mt{}
\def\ro{}

\begin{document}
\titlerunning{Evidence for the accelerated expansion 
of the Universe from 3D weak lensing with COSMOS}
\title{Evidence for the accelerated expansion of the Universe from weak lensing tomography with COSMOS\thanks{Based on observations made with the NASA/ESA \textit{Hubble Space Telescope}, obtained from the data archives at the Space Telescope European Coordinating Facility and the Space Telescope Science Institute, which is operated by the Association of Universities for Research in Astronomy, Inc., under NASA contract NAS 5-26555. 
}}
   \author{Tim~Schrabback
          \inst{1,2}\and
  Jan~Hartlap\inst{2}\and
  Benjamin~Joachimi\inst{2}\and
  Martin~Kilbinger\inst{3,4}\and
  Patrick~Simon\inst{5}\and
  Karim Benabed\inst{3}\and
  Maru\v{s}a~Brada\v{c}\inst{6,7} \and 
   Tim~Eifler\inst{2,8} \and
   Thomas~Erben\inst{2} \and
Christopher~D.~Fassnacht\inst{6} \and
   F.~William~High\inst{9}\and
   Stefan~Hilbert\inst{10,2}\and
   Hendrik~Hildebrandt\inst{1}\and
   Henk~Hoekstra\inst{1} \and
  Konrad~Kuijken\inst{1} \and
  Phil~Marshall\inst{7,11}  \and
  Yannick~Mellier\inst{3} \and
  Eric~Morganson\inst{11} \and
  Peter~Schneider\inst{2} \and
  Elisabetta~Semboloni\inst{2,1} \and
   Ludovic~Van~Waerbeke\inst{12} \and
   Malin~Velander\inst{1}
      }

   \offprints{T. Schrabback}
    \institute{
     Leiden Observatory, Leiden University, Niels Bohrweg 2, NL-2333 CA Leiden, The Netherlands, \email{schrabback@strw.leidenuniv.nl}
             \and
       Argelander-Institut f\"ur Astronomie, Universit\"at Bonn,
           Auf dem H\"ugel 71, D-53121 Bonn, Germany
              \and
            Institut d'Astrophysique de Paris, CNRS UMR 7095 \& UPMC, 98 bis, boulevard Arago, 75014 Paris, France\and
              Shanghai Key Lab for Astrophysics, Shanghai Normal University,  Shanghai 200234, P.~R.~China\and
                The Scottish Universities Physics Alliance
              (SUPA), Institute for Astronomy, School of Physics, 
 University of Edinburgh, Royal Observatory, Blackford Hill, Edinburgh EH9 3HJ, UK \and
Physics Dept., University of California, Davis, 1  
Shields Ave., Davis, CA 95616\and
          Physics department, University of California, Santa Barbara,
              CA 93601 \and
              Center for Cosmology and AstroParticle Physics, The Ohio State University, 
Columbus, OH 43210, USA \and
              Department of Physics, Harvard University, Cambridge, MA 02138\and
              Max Planck Institute for Astrophysics, Karl-Schwarzschild-Str. 1, 85741 Garching, Germany
              \and
              KIPAC, P.O. Box 20450, MS29, Stanford, CA 94309
\and
              University of British Columbia, Department of Physics and Astronomy, 6224 Agricultural Road, Vancouver, B.C. V6T 1Z1, Canada
      }

   \date{Received 31 October 2009 / Accepted 8 March 2010}
         
   \abstract
{
We present a comprehensive {\ro analysis of weak gravitational lensing by large-scale structure} in the
\textit{Hubble Space Telescope} Cosmic Evolution Survey (COSMOS), in which 
we combine
space-based galaxy shape measurements 
with ground-based photometric redshifts to study the redshift dependence of 
the lensing signal
and constrain cosmological
parameters.
After applying our weak lensing-optimized data reduction, principal component interpolation for the spatially and
temporally  varying ACS point-spread function, and improved modelling of charge-transfer inefficiency,
we measure a lensing signal which is consistent with pure gravitational modes and no {\ro significant shape} systematics.
We carefully estimate 
the statistical uncertainty 
from simulated COSMOS-like fields
 obtained from ray-tracing through the Millennium Simulation, 
including the  full non-Gaussian sampling variance.
We test our lensing pipeline on simulated space-based data,
recalibrate  non-linear power spectrum corrections using the ray-tracing analysis,
employ photometric redshift information to reduce potential contamination by
intrinsic galaxy alignments, and  marginalize over systematic uncertainties.
We find that the weak lensing signal scales with redshift as expected
from General Relativity for a concordance
$\Lambda$CDM cosmology,
including the full cross-correlations between different redshift bins.
Assuming a flat $\Lambda$CDM cosmology, we measure
\mbox{$\sigma_8\left(\Omega_\mathrm{m}/0.3\right)^{0.51}=0.75\pm0.08$}
 from lensing,
in perfect agreement with WMAP-5, yielding joint constraints
\mbox{$\Omega_\mathrm{m}= 0.266^{+0.025}_{-0.023}  $}, 
\mbox{$\sigma_8=0.802^{+0.028}_{-0.029}$} 
(all 68.3\% conf.). 
Dropping the assumption of flatness
and using priors from the HST Key Project and Big-Bang nucleosynthesis only,
we find a negative deceleration parameter $q_0$ at 
94.3\%
confidence from the
tomographic lensing analysis, providing independent evidence for the accelerated expansion of the Universe.
For a flat $w$CDM cosmology 
and prior \mbox{$w\in[-2,0]$}, we obtain
\mbox{$w<-0.41$} (90\% conf.).
Our dark energy constraints are still relatively weak solely due to the limited area of COSMOS.
{\ro However, they provide an important demonstration for the usefulness of tomographic weak
lensing measurements from space.}
}
   \maketitle
\section{Introduction}
During the last decade 
strong evidence for an accelerated expansion of the Universe has been
found with
several independent cosmological probes 
including type Ia supernovae  \citep[][]{rfc98,pag99,rsc07,kra08,hwb09}, cosmic microwave background \citep{bab00,svp03,kdn09}, galaxy clusters \citep{ars08,mae08,mar09,vkb09},
baryon acoustic oscillations \citep{ezh05,pce07,pre09}, integrated Sachs-Wolfe
effect \citep{gsc08,gns08,hhp08},
and strong gravitational lensing \citep{sma09}.
Within the standard cosmological framework this can be described with the ubiquitous presence of a
 new constituent named dark energy, which counteracts the attractive force of gravity on the largest scales and contributes  \mbox{$\sim 70\%$} to the total energy budget today.
There are various attempts to explain dark energy, ranging from Einstein's cosmological constant, via a dynamic fluid named
quintessence, to a possible breakdown of General Relativity \citep[e.g.][]{hul07,aab09}, 
all of which lead to profound implications for fundamental physics.
In order to make substantial progress and to be able to distinguish between the different scenarios,
several large dedicated surveys are currently being designed.

One  technique holding particularly high promise to constrain dark energy \citep{abc06,pse06,aab09} is weak gravitational lensing, which utilizes the subtle image distortions imposed onto the observed shapes of distant galaxies while their light bundles pass through the gravitational potential of foreground structures \citep[e.g.][]{bas01}.
The strength of the lensing effect depends on the total foreground mass distribution, independent of the relative contributions of luminous and dark matter.
Hence, it provides a unique tool to study the statistical properties of large-scale structure directly \citep[for reviews see ][]{sch06,hoj08,mvw08}.

Since its first detections by \citet{bre00,kwl00,wme00} and \citet{wtk00}, substantial progress has been made with the measurement of this cosmological weak lensing effect, which is also called cosmic shear.
Larger surveys have significantly reduced statistical uncertainties
\citep[e.g.][]{hyg02,btb03,jbf03,mrb05,wmh05,hmw06,smw06,hss07,fsh08},
while tests on simulated data have led to a better understanding of PSF systematics
\citep[][and references therein]{hwb06,mhb07,bbb09}. 
Finally, being a geometric effect, gravitational lensing depends 
on the source redshift distribution, where most earlier measurements had to
rely on external redshift calibrations from the small \textit{Hubble} Deep Fields.
Here, the impact of sampling variance was demonstrated
by 
\citet{bhs07}, who recalibrated earlier measurements using photometric redshifts from the much larger CFHTLS-Deep, significantly 
improving derived cosmological constraints.

Dark energy affects the distance-redshift relation and suppresses
 the time-dependent growth of structures.
Being 
sensitive to both effects,  weak lensing is 
 a powerful probe of dark energy properties, also 
providing important tests for theories of
modified gravity \citep[e.g.][]{beb01,bew04,stu07,dmm07,jaz08,sch08}. 
Yet, in order to significantly constrain these 
redshift-dependent 
effects, the shear signal must be measured as a function of source redshift,
an analysis often 
called
 weak lensing tomography or 3D weak lensing 
\citep[e.g.][]{hu99,hu02,hut02,jat03,hea03,huj04,bej04,sks04,taj04,hkt06,tkb07}.
Redshift information is additionally required to eliminate potential
contamination of the lensing signal from intrinsic galaxy alignments \citep[e.g.][]{kis02,his04,hwh06,jos08}.
In general, weak lensing studies have to rely on photometric redshifts \citep[e.g.][]{ben00,iam06,hwb08} given that most of the studied galaxies are too faint for spectroscopic measurements.

{\mt So far, tomographic cosmological weak lensing techniques 
were
 applied to  
real data by \citet{btb05,smw06,kht07,mrl07}}.
Dark energy constraints from previous 
weak lensing surveys 
were
limited by the lack of the required
individual photometric redshifts \citep{jjb06,hmw06,smw06,kbg09}
or small survey area \citep{kht07}.
The currently best data-set for 3D weak lensing is given by the COSMOS
Survey \citep{saa07}, which is 
the largest continuous area ever imaged with the \textit{Hubble Space Telescope} (HST), comprising 1.64 deg$^2$ of deep imaging with the Advanced Camera for Surveys (ACS).
Compared to ground-based measurements, 
the HST point-spread function (PSF) yields substantially increased number densities of sufficiently resolved galaxies and better control for systematics due to smaller PSF corrections.
Although HST has  been used for earlier cosmological weak lensing analyses 
\citep[e.g.][]{rrg02,rrc04,meh05,hbb05,ses07}, 
these studies lacked the area and deep photometric redshifts
which are available for COSMOS \citep{ics09}.
This combination of superb space-based imaging and ground-based photometric redshifts makes COSMOS the perfect test case for 3D weak lensing studies.
\citet{mrl07} conducted an earlier 3D weak lensing analysis of COSMOS, in
which they 
{\mt correlated the shear signal between three redshift bins 
} and constrained the matter density $\Omega_\mathrm{m}$ and power spectrum normalization $\sigma_8$.
In this paper we present a new analysis of the data, with 
several differences compared to the earlier study: we employ a new,
exposure-based model for the spatially and temporally varying ACS PSF, which
has been derived from dense stellar fields using a principal component analysis (PCA). 
Our new parametric correction for the impact of charge transfer inefficiency (CTI) on stellar images eliminates earlier PSF modelling uncertainties caused by confusion of CTI- and PSF-induced stellar ellipticity. 
Using the latest photometric redshift catalogue
of the field \citep{ics09},
we split our galaxy sample into five individual redshift bins and
additionally estimate the redshift distribution for very faint galaxies
forming a sixth bin without individual photometric redshifts, doubling the number of galaxies used in our cosmological analysis. 
We study the redshift scaling of the shear signal between these six  bins in
detail, employ an accurate covariance matrix obtained from ray-tracing
through the Millennium Simulation, which we also use to recalibrate
non-linear power spectrum corrections, and marginalize over parameter uncertainties.
In addition to $\Omega_\mathrm{m}$ and  $\sigma_8$, we also constrain the
dark energy equation of state parameter $w$ for a flat $w$CDM cosmology, and
the vacuum energy density $\Omega_\Lambda$ for a general (non-flat)
$\Lambda$CDM cosmology, yielding constraints for the deceleration parameter $q_0$.

This paper is organized as follows.
We summarize the most important information on the data and photometric redshift catalogue
in Sect.\thinspace\ref{se:data}, while further details on the ACS data reduction are given in
App.\thinspace\ref{app:data}.
Section \ref{se:wl} summarizes the weak lensing measurements
including our new correction schemes for PSF and CTI, for which we provide details in App.\thinspace\ref{app:psf_cti}.
We conduct various tests for shear-related systematics in Sect.\thinspace\ref{se:wl:tests}. 
We then present the weak lensing tomography analysis in
Sect.\thinspace\ref{se:tomo}, and cosmological parameter estimation in
Sect.\thinspace\ref{se:cosmo}.
We discuss our findings and conclude in Sect.\thinspace\ref{se:dis}.

Throughout this paper all magnitudes are given in the AB system, where  $i_{814}$ denotes the \texttt{SExtractor} \citep{bea96} \mbox{$\mathtt{MAG\_AUTO}$} magnitude measured from the ACS data (Sect.\thinspace\ref{se:data:acs}), while $i^+$ is the {\ro \texttt{MAG\_AUTO}}
magnitude determined by \citet{ics09} from the  Subaru data (Sect.\thinspace\ref{se:tomo:zcatsbright}).
In several tests we employ a reference {\ro WMAP-5-like \citep{dkn09}} flat $\Lambda$CDM cosmology characterized by \mbox{$\Omega_\mathrm{m}=0.25$}, \mbox{$\sigma_8=0.8$}, \mbox{$h=0.72$},  \mbox{$\Omega_\mathrm{b}=0.044$}, \mbox{$n_\mathrm{s}=0.96$}, where we use the transfer function by  \citet{eih98} and non-linear power spectrum corrections according to \citet{spj03}.

\section{Data}
\label{se:data}
\subsection{HST/ACS data}
\label{se:data:acs}

The COSMOS Survey \citep{saa07} is the largest contiguous field observed with the \textit{Hubble Space Telescope}, spanning
 a total area of \mbox{$\sim 77^\prime\times 77^\prime$} \mbox{(1.64\,$\mathrm{deg}^2$)}.
It comprises 579 ACS
 tiles, each observed  in F814W for 2028s using four dithered 
exposures.
The survey is centred at
\mbox{$\alpha=10^\mathrm{h}00^\mathrm{m}28.6^\mathrm{s}$}, 
\mbox{$\delta=+02^\circ 12^\prime 21\farcs0$} (J2000.0), and data were taken between 
October 2003 and November 2005.

We 
have 
reduced the ACS/WFC data starting from the flat-fielded 
images.
We
apply updated bad pixel masks,
 subtract the sky background, 
and compute optimal weights as detailed in
App.\thinspace\ref{app:data}.
For the image registration, distortion correction, cosmic ray rejection, and stacking we use
\texttt{MultiDrizzle}\footnote{MultiDrizzle version 3.1.0} \citep{kfh02},
applying the latest time-dependent distortion solution from \citet{and07}.
We iteratively align exposures within each tile by cross-correlating the
positions of compact sources and applying 
 residual shifts and rotations.

In tests with dense stellar fields we found that the default cosmic ray
rejection parameters 
of \texttt{MultiDrizzle} can lead to false flagging of central stellar
pixels as cosmic rays, especially if telescope breathing 
introduces 
significant
PSF variations (see Sect.\thinspace\ref{se:wl}) between combined
exposures.
Hence, stars  will
be partially rejected in exposures with deviating PSF properties.
On the contrary, galaxies will not be
flagged due to their shallower light
profiles, leading to different effective stacked PSFs for stars and galaxies.
To avoid any influence on
the lensing analysis, we create separate stacks for the shape measurement of
galaxies and stars, where we use close to default cosmic ray rejection
parameters for the former (\texttt{driz\_cr\_snr}=''4.0 3.0'',
\texttt{driz\_cr\_scale}=''1.2 0.7'', see \citealt{kfh02,kac07}), but less aggressive masking for the
  latter (\texttt{driz\_cr\_snr}=''5.0 3.0'',
\texttt{driz\_cr\_scale}=''3.0 0.7'').
As a result, the false masking of stars is substantially reduced. 
On the downside some actual cosmic rays lead to imperfectly corrected artifacts in the
``stellar'' stacks. 
This is not problematic given the very low fraction of affected stars, for which
 the artifacts only introduce additional noise in the shape measurement.

For the final image stacking we employ the \texttt{LANCZOS3}
interpolation kernel and a pixel scale of 0\farcs05, which minimizes 
noise correlations and aliasing without unnecessarily broadening the
PSF \citep[for a detailed comparison to other kernels see][]{jbs07}.
Based on our
input noise models (see App.\thinspace\ref{app:data})
we
compute a correctly scaled RMS image for the stack.
We match the stacked image WCS to the ground-based catalogue by \citet{ics09}.

We employ our RMS noise model for object detection with  \texttt{SExtractor} \citep{bea96}, where we require 
a minimum of 8 adjacent pixels being at least $1.4\sigma$
above the background, employ deblending parameters \mbox{$\mathtt{DEBLEND\_NTHRESH}=32$}, \mbox{$\mathtt{DEBLEND\_MINCONT}=0.01$}, and measure \mbox{$\mathtt{MAG\_AUTO}$} magnitudes
$i_{814}$, which we correct for a mean galactic extinction offset of 0.035 \citep{sfd98}.
Objects near the field boundaries or containing noisy pixels, for which fewer than two good input exposures contribute, 
are automatically excluded. We also create magnitude-scaled polygonal masks for saturated stars and their diffraction spikes.
Furthermore, we reject scattered light and large, potentially incorrectly deblended galaxies by running \texttt{SExtractor} with a low $0.5\sigma$ detection threshold for 3960 adjacent pixels,
where we further expand each object mask
by six pixels.
The combined masks {\ro for the stacks} were visually inspected and adapted if necessary.

Our  fully filtered   mosaic shear catalogue contains a total of
$446\,934$ galaxies with \mbox{$i_{814}<26.7$}, corresponding to \mbox{76
  galaxies$/\mathrm{arcmin}^2$}, 
where we {\ro  exclude double detections
in overlapping tiles  and reject the fainter component in the case of close galaxy pairs with separations \mbox{$<0\farcs5$}}.
For details on the weak lensing galaxy selection criteria see App.\thinspace\ref{se:wl:galcor}.

In addition to the stacked images, our fully time-dependent PSF analysis (see
Sect.\thinspace\ref{se:wl}, App.\thinspace\ref{su:starpca})
 makes use of individual exposures, for which we use the cosmic ray-cleansed
\textit{COR} images before resampling,
provided by \texttt{MultiDrizzle} during the run with less aggressive cosmic ray masking.
{\ro These
are only used for the analysis 
of 
high signal-to-noise stars, which can be identified automatically in the half-light radius versus signal-to-noise space\footnote{$\Delta r_\mathrm{h}=0.25$ pixel wide kernel; \mbox{$S/N>20$}, defined as in \citet{ewb01}; peak flux \mbox{$<25\,000 \mathrm{e}^-$}}. 
Here we employ simplified field masks 
only excluding the outer regions of a tile with poor cosmic ray masking.}

\subsection{Photometric redshifts}

\label{se:tomo:zcats}

\subsubsection{Individual photometric redshifts for \mbox{$i^+<25$} galaxies}
\label{se:tomo:zcatsbright}

We  use 
the public COSMOS-30 photometric redshift catalogue from \citet{ics09}, 
which covers the full ACS mosaic and is magnitude limited to \mbox{$i^+<25$} (Subaru 
{\ro \texttt{SExtractor MAG\_AUTO}}
magnitude).
It is based on the  30 band photometric catalogue, which includes
imaging  in 20 optical bands, as well as  near-infrared and deep IRAC data (Capak et al. 2009, in preparation).
\citet{ics09} computed photometric redshift using the \textit{Le Phare} code
\citep[S.\thinspace Arnouts \& O. Ilbert; also][]{iam06}, 
 reaching an excellent 
accuracy  of \mbox{$\sigma_{\Delta z/(1+z)}=0.012$} for \mbox{$i^+ < 24$}
and \mbox{$z < 1.25$}. The near-infrared (NIR) and infrared coverage extends the capability for reliable photo-$z$ estimation to higher redshifts, where the Balmer break moves out of the optical bands. Extended to \mbox{$z\sim 2$}, 
\citet{ics09}
find an accuracy of \mbox{$\sigma_{\Delta z/(1+z_\mathrm{s})}=0.06$} at \mbox{$i^+ \sim 24$}.
The comparison to spectroscopic redshifts from the zCOSMOS-deep sample \citep{llr07} with \mbox{$i^+_\mathrm{median}=23.8$} indicates a 20\% catastrophic outlier rate (defined as \mbox{$|z_\mathrm{phot}-z_\mathrm{spec}|/(1+z_\mathrm{spec})>0.15$}) for galaxies at \mbox{$1.5<z_\mathrm{spec}<3$}. In particular, for 7\% of the high-redshift (\mbox{$z_\mathrm{spec}>1.5$}) galaxies a low-redshift photo-$z$ (\mbox{$z_\mathrm{phot}<0.5$}) was assigned.
{ \mt
This degeneracy is expected for faint (\mbox{$i^+\gtrsim24$}) high-redshift galaxies,
for which the Balmer break cannot be identified
if they are undetected in the  NIR data
(limiting depth \mbox{$J\sim 23.7$}, \mbox{$K\sim
  23.7$} at $5\sigma$).
}
Due to the employed magnitude prior the contamination is expected to be
mostly uni-directional from high to low redshifts.

{\ro We tested this by comparing the  COSMOS-30 catalogue
to photometric redshifts estimated by \citet{hpe09} in the overlapping CFHTLS-D2 field using only optical \mbox{$u^*griz$} bands and the \textit{BPZ} photometric redshift code \citep{ben00}. Here we indeed find that 56\% 
of 
the matched \mbox{$i^+<25$} galaxies with  COSMOS-30 photo-$z$s  in the range \mbox{$2\le z_\mathrm{C30} \le 4$}
are identified at \mbox{$z_\mathrm{D2} \le 0.6$} in the D2 catalogue, if only a weak cut to reject
galaxies with double-peaked D2 photo-$z$ PDFs (\mbox{$\mathrm{ODDS}>0.7$}) is applied\footnote{\ro A more stringent cut \mbox{$\mathrm{ODDS}>0.95$} reduces this fraction to 14\%. Yet, it also reduces the absolute number of galaxies by a factor $4.7$. Note that, in contrast, 26\% (22\%) of the matched galaxies with a D2 photo-$z$ \mbox{$2\le z_\mathrm{D2} \le 4$} are placed at \mbox{$z_\mathrm{C30} \le 0.6$} for \mbox{$\mathrm{ODDS}>0.7$} (\mbox{$\mathrm{ODDS}>0.95$}). These could be explained by Lyman-break galaxies, which are better constrained by the deeper $u^*$ observations in the CFHTLS-D2. In any case we expect negligible influence on our results given our treatment for faint \mbox{$z_\mathrm{C30} \le 0.6$} galaxies.}. 
}

If not accounted for, such a contamination of a  low-photo-$z$ sample with high-redshift galaxies would be particularly severe for weak lensing tomography, given the strong dependence of the lensing signal on redshift.
In Sect.\thinspace\ref{se:tomo} we will therefore split galaxies with assigned \mbox{$z_\mathrm{phot}<0.6$} into sub-samples with expected low (\mbox{$i^+<24$}) and high (\mbox{$i^+>24$}) contamination,
where we only include the former in the cosmological analysis. 
Matching our shear catalogue to the fully masked COSMOS-30 photo-$z$ catalogue
yields a total of 194\,976 unique matches.

\subsubsection{Estimating the redshift distribution for $i^+>25$ galaxies}
\label{se:tomo:extrapolate}

\begin{figure}
   \centering
   \includegraphics[width=8cm]{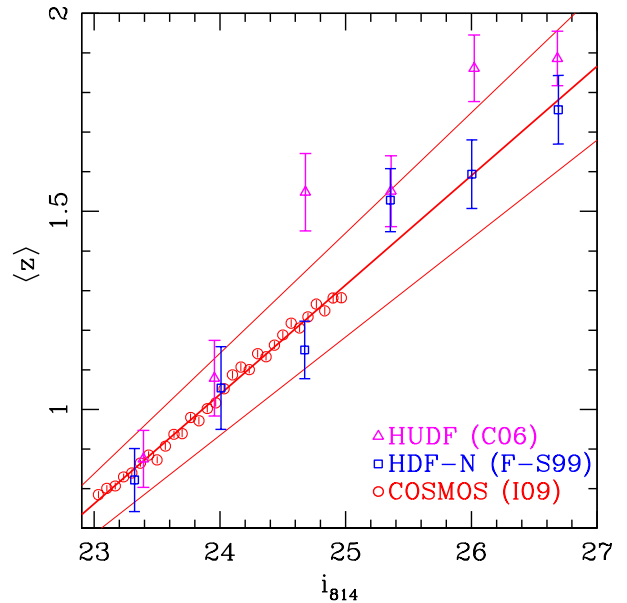}
   \caption{Relation between the mean photometric redshift and $i_{814}$ magnitude for COSMOS, HUDF, and HDF-N, where the error-bars indicate the error of the mean assuming Gaussian scatter and neglecting sampling variance.
The best fit (\ref{eq:zmag}) to the COSMOS data from \mbox{$i_{814}<25$} is
shown as the bold line, whereas the thin lines indicate the 
conservative $10\%$
uncertainty considered for the extrapolation in the cosmological analysis. The HDF-N data agree with the relation
very well, whereas the mean redshifts are higher in the HUDF both for
\mbox{$i_{814}<25$} and \mbox{$i_{814}>25$}, demonstrating the influence of
sampling variance in such small fields. 
   }
   \label{fi:zmag}
    \end{figure}

In order to include galaxies 
without individual photo-$z$s in our analysis, we need to estimate their redshift distribution.
Fig.\thinspace\ref{fi:zmag} shows the mean photometric COSMOS-30 redshift for galaxies in our shear catalogue
as a function of $i_{814}$.
In the whole magnitude range \mbox{$23<i_{814}<25$}
the data are very well described by the relation
\begin{equation}
\label{eq:zmag}
\langle z \rangle = (0.276\pm0.003) (i_{814}-23) + 0.762\pm0.003
\end{equation}
For comparison we also plot points from the \textit{Hubble} Deep Field-North
 \citep[HDF-N,][]{fly99} and \textit{Hubble} Ultra Deep Field \citep[HUDF,][]{cbs06}\footnote{For the HUDF we interpolate $i_{814}$ from the $i_{775}$ and $z_{850}$ magnitudes provided in the \citet{cbs06} catalogue.}  
for the extended magnitude range \mbox{$23<i_{814}<27$}, where both catalogues are redshift complete. The \mbox{HDF-N} data agree very well with the COSMOS fit over the whole extended range, on average to 2\%.
In contrast, the mean photometric redshifts in the HUDF are on average
higher than 
(\ref{eq:zmag}) by 16\% for \mbox{$23<i_{814}<25$} and 10\% for
\mbox{$25<i_{814}<27$}. 
The difference between the \mbox{HDF-N} and HUDF can be regarded as a rough estimate for the impact of sampling variance in such small fields.
The fact that the HUDF galaxies
systematically deviate from  
(\ref{eq:zmag})
 not only for  \mbox{$i_{814}>25$} but \emph{also} for \mbox{$i_{814}<25$} where \mbox{COSMOS-30} photo-$z$s are available, indicates that it is most likely affected by sampling variance containing a relative galaxy over-density at higher redshift.
Given the excellent fit for the COSMOS galaxies and very good agreement for the \mbox{HDF-N} data we are thus confident to
use 
(\ref{eq:zmag})
 for a limited extrapolation to \mbox{$i_{814}<26.7$} for our shear galaxies.
This is also motivated by the fact that \mbox{$i_{814}<25$} and \mbox{$i_{814}>25$} galaxies are not completely independent, but partially probe the same large-scale structure at different luminosities.

Due to the non-linear dependence of the shear signal on redshift it is not only necessary to estimate the correct mean redshift 
of the galaxies, but also their actual redshift distribution.
In weak lensing studies  the redshift distribution is often parametrized as
$
p(z)\propto (z/z_0)^\alpha \exp{\left[-(z/z_0)^\beta\right]} 
$
\citep[e.g. ][]{bbs96}, which \citet{ses07} extended by fitting $\alpha,\beta$ in combination with a linear dependence of the median redshift on magnitude, leading to a magnitude-dependent $z_0$.
Yet, it was noted that this fit was not fully capable to reproduce the  shape of the redshift distribution of the fitted galaxies. Given the higher accuracy needed for the analysis of the larger COSMOS data, we use a modified parametrization
\begin{equation}
\label{eq:zdist_tim}
p(z|i_{814})\propto \left(\frac{z}{z_0}\right)^\alpha \left( \exp{\left[-\left(\frac{z}{z_0}\right)^\beta\right]} 
+ cu^d \exp{\left[-\left(\frac{z}{z_0}\right)^\gamma\right]}  \right) \,,
\end{equation}
where \mbox{$z_0=z_0(i_{814})$}, and 
\mbox{$u=\mathrm{max}[0,(i_{814}-23)]$}.
{\ro Using a maximum likelihood fit\footnote{\ro We employ the CERN Program Library \texttt{MINUIT} (\url{http://wwwasdoc.web.cern.ch/wwwasdoc/minuit/}).} 
 we determine best-fitting parameters
\mbox{$(\alpha,\beta,c,d,\gamma)=$} \mbox{$(0.678,5.606,0.581,1.851,1.464)$}
{\ro from the individual magnitudes, photo-$z$s, and (symmetric) 68\%   photo-$z$ errors of 
all galaxies with \mbox{$23<i_{814}<25$}}.
From 
 Eqs. (\ref{eq:zmag}) and (\ref{eq:zdist_tim}) 
we then numerically compute the}
 non-linear relation between
 $z_0$ and $i_{814}$, for which we provide the fitting formulae
   \begin{eqnarray}
      z_0&=& 0.446(i_{814}-23)+1.235 \quad \mathrm{for}\quad 22<i_{814}\le23\\
     z_0&=& \sum_{j=0}^{j=7} a_j [(i_{814}-23)/4]^j \quad \quad\,  \mathrm{for}\quad 23<i_{814}<27
   \end{eqnarray} 
with 
$(a_0,...,a_7) = (1.237,
1.691,
-12.167,
43.591,
-76.076,$
$72.567,
-35.959,
7.289)$.
The total redshift distribution of the survey is then simply given by the mean distribution \mbox{$\phi(z)=\sum_{k=1}^{N}p(z|i_{814,k})/N$}.

\begin{figure}
   \centering
   \includegraphics[width=9cm]{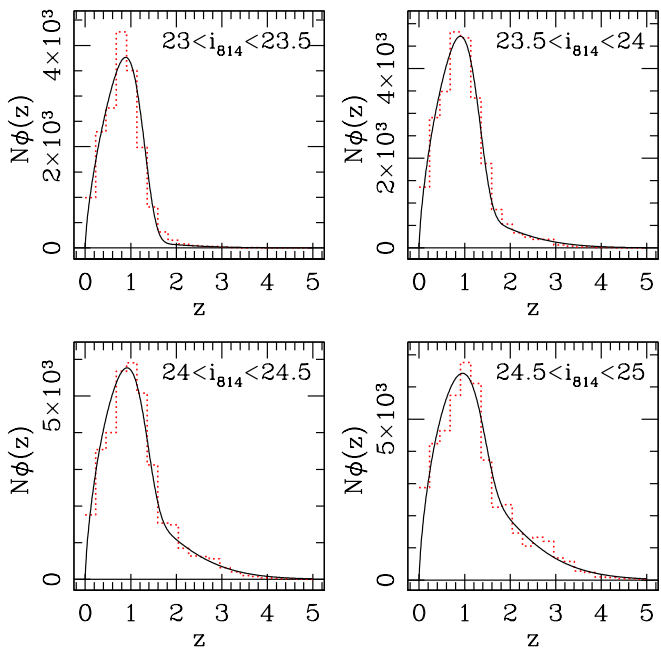}
   \caption{Redshift histogram for galaxies in our shear catalogue with
     COSMOS-30 photo-$z$s (dotted), split into four magnitude bins. The
     solid curves show the fit according to (\ref{eq:zmag}) and (\ref{eq:zdist_tim}), which is capable to describe both the peak and high redshift tail.
   }
   \label{fi:brightNz}
    \end{figure}

\begin{figure}
   \centering
   \includegraphics[width=6.5cm]{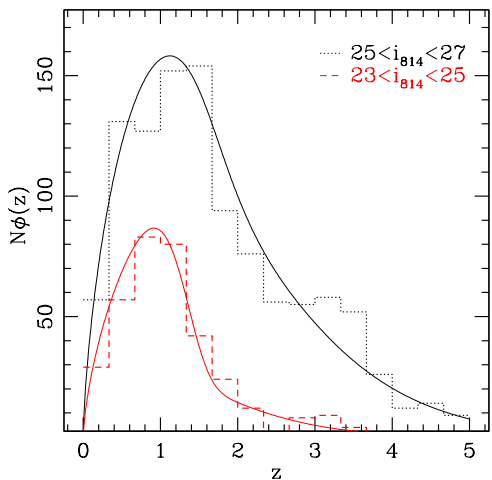}
   \caption{Combined redshift histogram for the HDF-N and HUDF photo-$z$s, split into two magnitude bins. The solid curves show the prediction according to (\ref{eq:zmag}), (\ref{eq:zdist_tim}) and the galaxy magnitude distribution. The good agreement for \mbox{$25<i_{814}<27$} galaxies confirms the applicability of the model in this magnitude regime.
   }
   \label{fi:HDFUDFNz}
    \end{figure}

{\ro We  chose the functional form of (\ref{eq:zdist_tim})
because its first addend allows  for a good description of the peak of the redshift distribution, while the second addend fits the magnitude-dependent tail at higher redshifts;
%
}
see Fig.\thinspace\ref{fi:brightNz} for a comparison of the data and model in four magnitude bins.
In Fig.\thinspace\ref{fi:HDFUDFNz} we  compare the actual redshift
distribution for the combined HDF-N and HUDF data to the one we \emph{predict} from their magnitude distribution and the fit to the COSMOS data, finding very good agreement also for \mbox{$25<i_{814}<27$}.
The only major deviation is given by 
a galaxy over-density in the HUDF photo-$z$ data near \mbox{$z\sim 3.2$},
which is also partially responsible for the higher mean redshift in
Fig.\thinspace\ref{fi:zmag} and which may be attributed to large-scale structure.

Our fitting scheme assumes that the COSMOS-30 \mbox{photo-$z$s} provide unbiased
estimates for the true galaxy redshifts. However, in
Sect.\thinspace\ref{se:tomo:zcats} we suspected that \mbox{$i^+ \gtrsim 24$}
galaxies with assigned 
\mbox{$z< 0.6$} 
might contain a 
significant contamination with high-redshift galaxies.
To assess the impact of this uncertainty, we derive the fits for
(\ref{eq:zmag}) and (\ref{eq:zdist_tim})
 using only galaxies with
\mbox{$23<i^+ < 24$}, reducing the estimated mean redshift of 
shear galaxies without COSMOS-30 photo-$z$ by $4\%$.
As an alternative test, we assume that 20\% of the
\mbox{$z< 0.6$} 
galaxies with \mbox{$24<i^+ < 25$} are truly at \mbox{$z=2$}, increasing the
estimated mean redshift by $8\%$.
Compared to the fit uncertainty in 
(\ref{eq:zmag})
(\mbox{$\sim
  1\%$}) this constitutes the main source of error for our redshift
extrapolation. 
In the cosmological parameter estimation (Sect.\thinspace\ref{se:cosmo}), we constrain this uncertainty and marginalize over it using a nuisance parameter,
which rescales the redshift distribution 
within a conservatively chosen $\pm 10\%$ interval.
Note that the $+4\%$ difference between the measured and
predicted mean redshift of the combined HDF-N and HUDF data in
Fig.\thinspace\ref{fi:HDFUDFNz} actually suggests a smaller uncertainty.

\section{Weak lensing shape measurements}
\label{se:wl}

To measure an accurate lensing signal, we have to carefully correct for
instrumental signatures.
Even with the high-resolution space-based data at hand, we have to accurately account for both PSF blurring and ellipticity, which introduce spurious shape distortions. 
To do so, one  requires both a good model for the PSF, and a method which accurately employs it to measure unbiased
estimates for the (reduced) gravitational shear $g$ from noisy galaxy images.

For the latter,
we use the KSB+
formalism \citep{ksb95,luk97,hfk98}, see  \citet{ewb01,ses07} and App.\thinspace\ref{app:ksbimp} for details on our implementation.
As found with simulations of ground-based weak lensing data, KSB+ 
can significantly
underestimate gravitational shear 
\citep[][]{ewb01,brc01,hwb06,mhb07}, where
the  calibration bias $m$ and possible PSF anisotropy residuals $c$, defined via
\begin{equation}
\label{eq:calibbias}
g^\mathrm{obs}-g^\mathrm{true}=m g^\mathrm{true} + c\,,
\end{equation}
 depend on the details of the implementation.
\citet[STEP2]{mhb07} detected a shear measurement
degradation  
for faint objects for our pipeline,
which is not surprising given the fact that the KSB+ formalism does not account for noise.
While \citet{ses07} simply corrected for the resulting mean calibration bias, the 3D
weak lensing analysis performed here requires unbiased shape measurements
not only on average, but also as function of redshift, and hence galaxy
magnitude and size \citep[see e.g.][]{kth08,kaa09,stw09}.
We therefore empirically account for this degradation
with a power-law fit to the signal-to-noise dependence of the calibration bias 
\begin{equation}
\label{eq:sndep}
m=-0.078 \left( \frac{S/N}{2}\right)^{-0.38}\,,
\end{equation}
where $S/N$ is computed with the {\ro galaxy size-dependent} KSB weight function \citep{ewb01}, and
corrected for
noise correlations
 as done in \citet{hss09}.
{\ro As \mbox{$S/N$} relates to the significance of the galaxy shape measurement, it provides a more direct correction for noise-related bias than fits as a function of magnitude or size.}
We have determined this correction using the STEP2 simulations of ground-based weak lensing data \citep{mhb07}.
In order to test if it performs reliably for the ACS data,
we have 
analysed a set of
simulated ACS-like  data (see App.\thinspace\ref{app:sims}).
In summary, we find that the remaining calibration bias is \mbox{$m=+0.008\pm0.002$} on average, and
\mbox{$|m|<0.02$} over
the entire magnitude range used, which 
is negligible compared to the statistical uncertainty for COSMOS.
Likewise, PSF anisotropy residuals, which are characterized in (\ref{eq:calibbias}) by $c$, are found to be negligible in the simulation (dispersion \mbox{$\sigma_c=0.0006$}), assuming accurate PSF interpolation.

Weak lensing analyses  usually create PSF models from the observed images of
stars, which have to be interpolated for the position of each galaxy. 
Typically, a high galactic latitude ACS field
contains only $\sim 10-20$ stars with sufficient $S/N$, 
which are too few for the spatial polynomial interpolation  
commonly used in ground-based weak lensing studies.
In addition, a stable PSF model cannot be used,
given that substantial temporal 
PSF variations have been detected,
mostly 
caused by
focus changes resulting from  orbital temperature variations (telescope breathing), mid-term seasonal effects, and long-term shrinkage of the optical telescope
assembly (OTA) \citep[e.g.][]{kri03,lmc06,ank06,ses07,rma07}.
To circumvent this problem, we have implemented a PSF correction scheme based on principal component analysis (PCA), as 
first suggested by \citet{jaj04}.
We have analysed 700 $i_{814}$ exposures of dense stellar fields, interpolated the PSF variation in each exposure with polynomials, and performed a PCA analysis of the polynomial coefficient variation.
We find that $\sim97\%$ of the total PSF ellipticity variation in random pointings can be described with a single
parameter related to the change in telescope focus, confirming earlier results \citep[e.g.][]{rma07}.
However, we find that additional  variations 
are still significant. In particular, we detect a dependence 
on the relative angle 
between the pointing and the orbital telescope movement\footnote{Technically speaking, we show a dependence on the velocity aberration plate scale factor in Fig.\thinspace\ref{fi:pca:pc1_pc2}.},
suggesting that heating in the sunlight does not only change the telescope focus, but also creates slight additional aberrations dependent on  the relative sun angle. 
These deviations may be coherent between COSMOS tiles observed under similar orbital conditions.
To account for this effect, we split the COSMOS data into 24 epochs of observations taken 
closely in time,
and determine a low-order, focus-dependent residual model from all stars within one epoch.
We provide further details on our PSF correction scheme in
App.\thinspace\ref{su:starpca}.

As an additional observational challenge, the COSMOS data suffer from 
defects in the ACS CCDs, which are caused by the continuous cosmic ray bombardment in space.
These defects act as charge traps reducing the charge-transfer-efficiency (CTE), an
effect referred to as charge-transfer-inefficiency (CTI).
When the image of an object is transferred across such a defect during
parallel read-out, a fraction of its charge is trapped and
 statistically released, effectively creating charge-trails following objects in
the read-out $y$-direction \citep[e.g.][]{rma07,clk09,msl10}.
For weak lensing measurements the dominant effect of CTI is the introduction of a spurious 
ellipticity component
in the read-out direction.
In contrast to PSF effects, CTI affects objects non-linearly due to the limited depth of charge traps. 
Hence, the two effects 
must be corrected  separately.
As done by \citet{rma07}, we employ an empirical correction for galaxy shapes,
but also take the dependence on sky background into account.
Making use of the CTI flux-dependence, we additionally determine and apply a parametric CTI model for stars, which is important  as PSF and CTI-induced ellipticity get mixed otherwise. 
We present details on our CTI correction schemes for stars in
App.\thinspace\ref{sec:wl:starcte} and for galaxies in
App.\thinspace\ref{se:wl:galcor}.
Note that \citet{msl10} recently presented a method to correct for CTI directly on the image level. 
We find that the methods employed here are sufficient for our science analysis, as also confirmed by the tests presented in Sect.\thinspace\ref{se:wl:tests}.
However, for weak lensing data with much stronger CTE degradation, such as ACS data taken after Servicing Mission 4, their pixel-based correction should be superior.

\section{2D shear-shear correlations and tests for systematics}
\label{se:wl:tests}

To measure the cosmological signal and conduct tests for systematics we compute the second-order shear-shear correlations
\begin{equation}
\label{eq:xipm2d}
\xi_\pm(\theta)= \frac{\sum_{i,j} (\gamma_{\mathrm{t},i}
\gamma_{\mathrm{t},j}\pm \gamma_{\times,i} \gamma_{\times,j}) \Delta_{ij}
}{\sum_{i,j}\Delta_{ij}} 
\end{equation}
from  galaxy pairs separated by
\mbox{$\vartheta=|\boldsymbol{\vartheta}_i-\boldsymbol{\vartheta}_j|$}.
Here, \mbox{$\Delta_{ij}=1$} if the  galaxy separation $\vartheta$ falls
within the considered angular bin around $\theta$, and \mbox{$\Delta_{ij}=0$} otherwise.
In (\ref{eq:xipm2d}) we approximate our reduced shear estimates
\mbox{$g=\gamma/(1-\kappa)\simeq \gamma$} with the shear $\gamma$ as
commonly done in cosmological weak lensing 
(typically \mbox{$|\kappa|\sim 1\%-3\%$}; correction employed in Sect.\thinspace\ref{se:cosmo:constraints:mr:wmap}),  
decompose it into the
tangential
component $\gamma_\mathrm{t}$ and the 45 degree rotated cross-component
$\gamma_\times$ relatively to the separation vector,
and employ uniform weights.

  \begin{figure*}
   \centering
   \includegraphics[width=5.9cm]{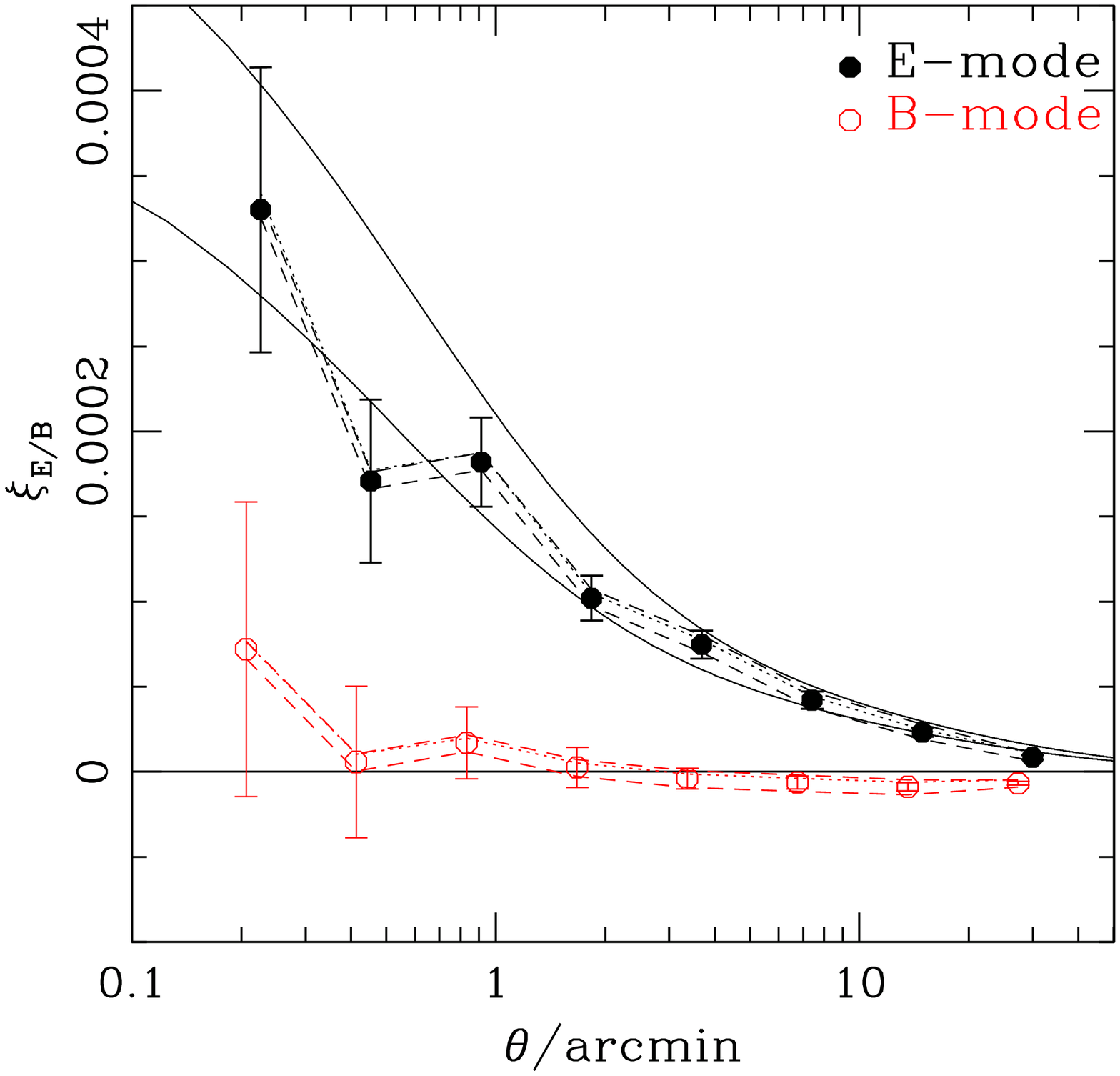}
   \includegraphics[width=5.9cm]{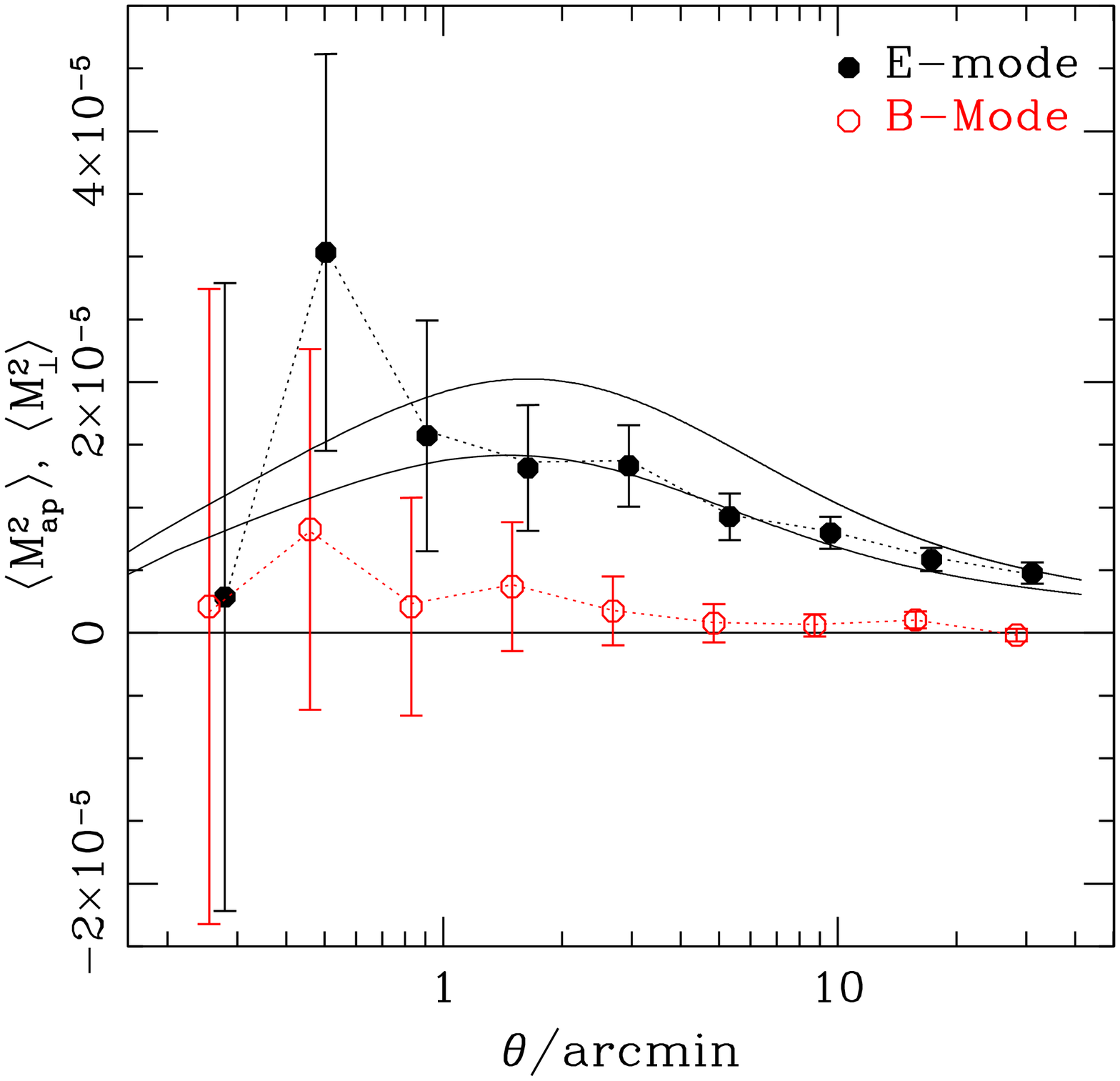}
   \includegraphics[width=5.9cm]{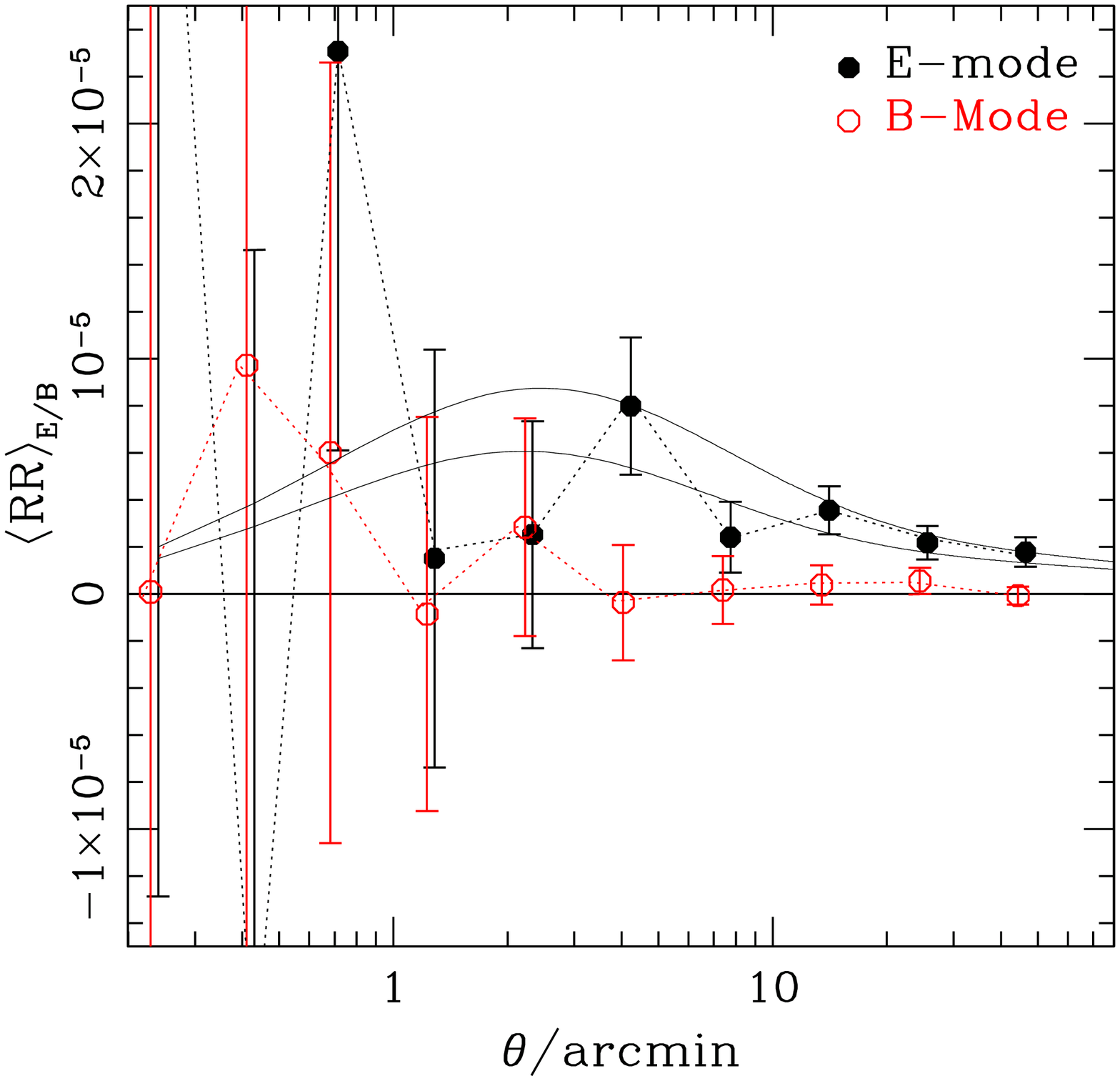}
   \caption{Decomposition of the shear field into E- and B-modes using the shear correlation function $\xi_{E/B}$ (\textit{left}), aperture mass dispersion $\langle M_\mathrm{ap/\perp}^2 \rangle$ (\textit{middle}), and ring statistics $\langle\mathcal R \mathcal R \rangle_{E/B}$ (\textit{right}). 
Error-bars have been computed from 300 bootstrap resamples of the shear catalogue, accounting for shape and shot noise, but not for sampling variance.
The solid curves indicate model predictions for \mbox{$\sigma_8=(0.7,0.8)$}.
In all cases the B-mode is consistent with zero, confirming the success of our correction for instrumental effects.
For  $\xi_{E/B}$ the E/B-mode decomposition is model-dependent, where we have assumed \mbox{$\sigma_8=0.8$} for the points, while the dashed curves have been computed for \mbox{$\sigma_8=(0.7,0.9)$}.
The dotted curves indicate the signal if the residual ellipticity correction discussed in App.\thinspace\ref{se:wl:galcor} is not applied, yielding nearly unchanged results.
Note that the correlation between points is strongest for $\xi_{E/B}$ and
weakest for $\langle\mathcal R \mathcal R \rangle_{E/B}$.
   }
   \label{fi:tests:eb}
    \end{figure*}

As an important consistency check in weak gravitational lensing, the signal can be 
decomposed into a curl-free component (E-mode) and a curl component (B-mode).
Given that lensing creates only E-modes, the detection of a significant B-mode indicates the presence of uncorrected residual systematics in the data.
\citet{cnp02} show that $\xi_\pm$ can be decomposed into E- and B-modes as
\begin{equation}
\label{eq:xi_eb}
  \xi_{E/B}(\theta)=\frac{\xi_+(\theta)\pm\xi^\prime(\theta)}{2} \,,
\end{equation}
with
\begin{equation}
  \label{eq:xi_prime}
  \xi^\prime(\theta)=\xi_-(\theta)+4\int_\theta^\infty \frac{\mathrm{d}\vartheta}{\vartheta}\xi_-(\vartheta) -12 \theta^2 \int_\theta^\infty \frac{\mathrm{d}\vartheta}{\vartheta^3}\xi_-(\vartheta)  \,.
\end{equation}
We plot this decomposition for our COSMOS catalogue in  the left panel of Fig.\thinspace\ref{fi:tests:eb}.
Given that the integration in (\ref{eq:xi_prime}) extends to infinity,
we employ $\Lambda$CDM predictions for \mbox{$\theta>40^\prime$},
leading to a slight model-dependence, which
is indicated
by the dashed curves corresponding to \mbox{$\sigma_8=(0.7,0.9)$}, whereas the points have been computed for \mbox{$\sigma_8=0.8$}.
Within this section, error-bars and covariances are estimated from 300
bootstrap resamples of our galaxy shear catalogue, 
which accounts for both shot noise and shape noise.
As seen in Fig.\thinspace\ref{fi:tests:eb}, we detect no significant B-mode
$\xi_B$.
However, note that different angular scales are highly correlated for $\xi_{E/B}$, which mixes power on a broad range of scales
and potentially smears out the signatures of systematics.

An
E/B-mode decomposition, 
for which the correlation between different scales is weaker, 
is provided by the dispersion of the aperture mass \citep{sch96m}
\begin{equation}
\label{eq:map}
\langle M_\mathrm{ap/\perp}^2 \rangle (\theta)= \frac{1}{2}\int_0^{2\theta} \frac{\mathrm{d} \vartheta \, \vartheta}{\theta^2} \left [\xi_+ (\vartheta) T_+ \left( \frac{\vartheta}{\theta} \right)
\pm \xi_- (\vartheta) T_- \left( \frac{\vartheta}{\theta} \right) \right]\,,
\end{equation}
with   $T_\pm$ given in \citet{swm02},  where we employ the aperture mass weight function proposed by \citet{swj98}. 
The computation of (\ref{eq:map}) requires integration from zero, which is not practical for real data.
We therefore truncate $\xi_\pm$ for $\theta<0\farcm05$, where the introduced bias is small compared to
our statistical errors \citep{kse06}.
\citet{mrl07} measure a significant B-mode component $\langle
M_\mathrm{\perp}^2 \rangle$ at scales \mbox{$1^\prime\lesssim \theta
  \lesssim 3^\prime$}, whereas this signal is negligible in the present analysis.
We 
quantify the on average slightly positive $\langle M_\mathrm{\perp}^2
\rangle$
by fitting
a mean offset taking the bootstrap covariance into account {\ro (correlation between neighbouring points \mbox{$\simeq 0.5$})}, yielding 
an average
\mbox{$\overline{\langle M_\mathrm{\perp}^2 \rangle}=(1\pm4)\times 10^{-7}$} 
if all points are considered, 
and \mbox{$\overline{\langle M_\mathrm{\perp}^2 \rangle(\theta < 6^\prime)}=(1.0\pm1.4)\times 10^{-6}$} {\ro or \mbox{$\overline{\langle M_\mathrm{\perp}^2 \rangle (\theta < 2^\prime)}=(4.0\pm4.7)\times 10^{-6}$}
 if only small scales 
are included, 
}
consistent with no B-modes.

The cleanest  E/B-mode decomposition is given by the  
ring statistics (\citealt{sck07,esk09}; see also \citealt{fuk10}), which can be computed from the correlation function using a finite interval with  non-zero lower integration limit
\begin{equation}
\label{eq:ring}
\langle \mathcal R \mathcal R \rangle_{E/B} (\Psi)= \frac{1}{2}\int_{\eta \Psi}^{\Psi} \frac{\mathrm{d} \vartheta }{ \vartheta} \left [\xi_+ (\vartheta) Z_+ (\vartheta, \eta )
\pm \xi_- (\vartheta) Z_- ( \vartheta, \eta  ) \right]\,,
\end{equation}
with functions $Z_\pm$ given in \citet{sck07}.
We compute $\langle \mathcal R \mathcal R \rangle_{E/B}$  using a
scale-dependent integration limit $\eta$ as outlined in \citet{esk09}.
As can be seen from the right panel of Fig.\thinspace\ref{fi:tests:eb}, also  $\langle \mathcal R \mathcal R \rangle_{B}$ is consistent with no B-mode signal.

The non-detection of significant B-modes in our shear catalogue
is an important confirmation for our
correction schemes for
instrumental effects
and 
suggests that the measured signal is truly of cosmological origin.

\begin{figure}
   \centering
   \includegraphics[width=8cm]{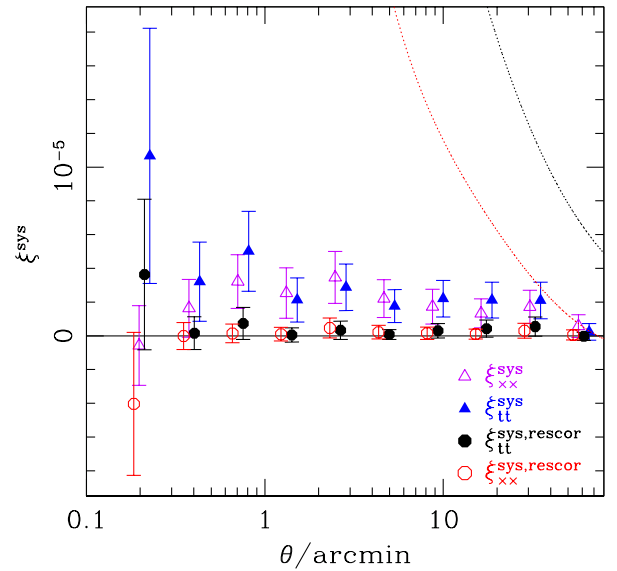}
   \caption{Cross-correlation between galaxy shear estimates and uncorrected stellar ellipticities as defined in (\ref{eq:xisys}).
The signal is consistent with zero if the residual ellipticity correction discussed in App.\thinspace\ref{se:wl:galcor} is applied (circles). 
Even without this correction (triangles) it is at a level negligible compared to the expected cosmological
     signal (dotted curves), except for the largest scales, where the error-budget is anyway dominated by sampling variance. 
 }
   \label{fi:xisys}
    \end{figure}

As a final test for shear-related systematics we compute the correlation between corrected galaxy shear estimates $\gamma$ and uncorrected stellar ellipticities $e^*$
\begin{equation}
\label{eq:xisys}
\xi^\mathrm{sys}_{tt/\times\times}(\theta) =\frac{\langle \gamma_{t/\times} e^*_{t/\times} \rangle | \langle \gamma_{t/\times} e^*_{t/\times} \rangle|}{ \langle e^*_{t/\times} e^*_{t/\times} \rangle}\,,
\end{equation}
which we normalize using the stellar auto-correlation as suggested by \citet{bmr03}.
As detailed in App.\thinspace\ref{se:wl:galcor}, we employ a somewhat ad hoc residual correction for a very weak remaining instrumental  signal.
We find that $\xi^\mathrm{sys}$  is indeed only consistent with zero if this correction is applied  (Fig.\thinspace\ref{fi:xisys}), yet even without correction, $\xi^\mathrm{sys}$ is  negligible compared to the expected cosmological signal.
The negligible impact can also be seen from the two-point statistics in
Fig.\thinspace\ref{fi:tests:eb}, where the points are computed including
residual correction, while the dotted lines indicate the measurement without
it.
We suspect that this residual instrumental signature could either be
caused by the limited capability of KSB+ to fully correct for a complex 
space-based PSF, or a residual PSF modelling uncertainty due to the low
number of stars per ACS field.
In any case we have verified that this residual correction has a negligible impact on the
cosmological parameter estimation in Sect.\thinspace\ref{se:cosmo}, changing
our constraints on $\sigma_8$ at the $2\%$ level, 
well within the statistical uncertainty.

\section{Weak lensing tomography}
\label{se:tomo}

In this section we present our analysis of the redshift dependence of 
the lensing signal
in COSMOS. We start with the 
definition 
of redshift bins in Sect.\thinspace\ref{se:tomo:zbins}, summarize the 
theoretical framework in  Sect.\thinspace\ref{se:tomo:formalism}, describe
our angular binning and treatment of intrinsic galaxy alignments in
Sect.\thinspace\ref{se:angularbinning}, elaborate on the covariance estimation in Sect.\thinspace\ref{se:tomo:covariance}, present 
the measured redshift scaling in  Sect.\thinspace\ref{se:tomo:zscale}, and discuss indications for a contamination of faint $z_\mathrm{phot}<0.6$ galaxies with high redshift galaxies in  Sect.\thinspace\ref{se:tomo:conta}.

\subsection{Redshift binning}
\label{se:tomo:zbins}

\begin{figure}
   \centering
   \includegraphics[width=8cm]{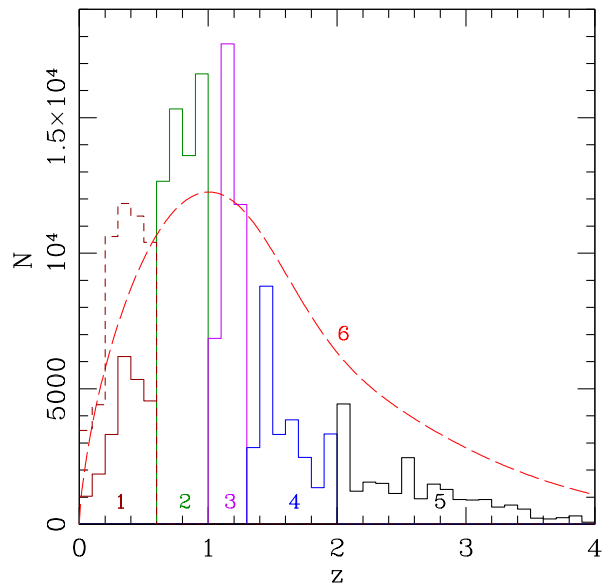}
   \caption{
Redshift distributions for our tomography analysis. 
The solid-line histogram shows the individual COSMOS-30 redshifts used for bins 1 to 5, while
the  difference between the dashed and solid histograms indicates the \mbox{$24<i^+<25$} galaxies with  \mbox{$z_\mathrm{phot}< 0.6$}, which are excluded in our analysis due to potential contamination with high-redshift galaxies.
The long-dashed curve corresponds to the estimated redshift distribution for \mbox{$i_{814}<26.7$} shear galaxies without individual COSMOS-30 photo-$z$, which we use as bin 6.
   }
   \label{fi:zbins}
    \end{figure}

We split the galaxies with individual COSMOS-30 photo-$z$s into
five 
 redshift bins, as summarized in Table \ref{tab:zbins} and illustrated in Fig.\thinspace\ref{fi:zbins}.
{\ro We chose the intermediate limits \mbox{$z=(0.6,1.0,1.3)$} 
such that 
the Balmer/4000\AA } break is
approximately located at the centre of one of the broadband $r^+i^+z^+$
filters.
This minimizes the impact of possible artifical clustering in \mbox{photo-$z$} space 
{\ro and hence scatter between redshift bins}
for 
galaxies too faint to be detected in the Subaru medium bands.
{\mt 
Given {\ro our} chosen limits,
most catastrophic redshift
errors are faint bin 5 galaxies identified as bin 1 (Sect.\thinspace\ref{se:tomo:zcatsbright}).}
Thus, we do not include 
\mbox{$z<0.6$} 
galaxies with \mbox{$i^+>24$} in our analysis due to their potential contamination with high redshift galaxies, but study their lensing signal separately in Sect.\thinspace\ref{se:tomo:conta}.
We use all  galaxies without individual photo-$z$ estimates with
\mbox{$22<i_{814}<26.7$}\footnote{Including galaxies with \mbox{$i^+<25$}
  which are located in masked regions for the ground-based photo-$z$ catalogue, but not
 for the space-based lensing catalogue.} as a broad bin 6, for which we
estimated the redshift distribution in Sect.\thinspace\ref{se:tomo:extrapolate}.

\begin{table}[tb]
\begin{center}
\caption{Definition of redshift bins, number of contributing galaxies, and mean redshifts.
\label{tab:zbins}
}
\vspace{0.2cm}
\begin{tabular} {cccccc}
\hline
Bin & $z_\mathrm{min}$ & $z_\mathrm{max}$ & $N$&  $\langle z \rangle$ \\
\hline
1 & 0.0 & 0.6 & \mbox{$i^+<24:$} 22\,294$^*$
&    0.37\\
 &  &  & \mbox{$i^+>24:$} 29\,817 &   \\
2 & 0.6 & 1.0 & 58\,194 &  0.80\\
3 & 1.0 & 1.3 & 36\,382 &  1.16\\
4 & 1.3 & 2.0 & 25\,928 &  1.60\\
5 & 2.0 & 4.0 & 21\,718 &  2.61\\
6 & 0.0 & 5.0 &  251\,958  & $1.54\pm0.15$  \\
\hline
\end{tabular}
\end{center}

$^*$: Here  we also exclude 259 galaxies with
 \mbox{$i^+<24$}, which have a significant secondary peak in their
redshift probability distribution at \mbox{$z_{\mathrm{phot,}2}>0.6$}. 
\end{table}

\subsection{Theoretical description}
\label{se:tomo:formalism}

Extending the formalism from Sect.\thinspace\ref{se:wl:tests}, we 
split the galaxy sample into redshift bins and cross-correlate shear estimates between 
bins $k$ and $l$
\begin{equation}
\label{eq:xipm_from_pkappa}
\hat{\xi}_\pm^{kl}(\theta)= \frac{\sum_{i,j} (\gamma_{\mathrm{t},i}^k
\gamma_{\mathrm{t},j}^l\pm \gamma_{\times,i}^k \gamma_{\times,j}^l)\Delta_{ij}
}{\sum_{i,j}\Delta_{ij}} \,,
\end{equation}
where the summation extends over all galaxies $i$ in bin $k$, and all galaxies $j$ in bin $l$.
These 
are estimates for the shear cross-correlation functions $\xi_\pm^{kl}$, 
which are filtered versions of the convergence cross-power spectra
\begin{equation}
\label{eq:xipm_pkappa}
    \xi_{+/-}^{kl} (\theta) = \frac{1}{2 \pi} \int_0^\infty \mathrm{d} \ell \, \ell \, \mathrm{J}_{0/4} (\ell \theta) P_\kappa^{kl} (\ell) \, ,
\end{equation}
where  $\mathrm{J}_n$ denotes the $n^\mathrm{th}$-order Bessel function of the first kind and $\ell$ is the modulus of the two-dimensional wave vector.
These can be calculated from line-of-sight integrals over the three-dimensional {\ro (non-linear)} power spectrum $P_\delta$ {\ro(see Sect.\thinspace\ref{se:constraints:modelcalc})} as
\begin{equation}
\label{eq:pkappa}
     P_\kappa^{kl}(\ell) = \frac{9 H_0^4 \Omega_\mathrm{m}^2}{4 c^4} \int_0^{\chi_\mathrm{h}} \mathrm{d} \chi \frac{g_k(\chi)g_l(\chi)}{a^2(\chi)} P_\delta \left( \frac{\ell}{f_K(\chi)}, \chi \right) \, ,
\end{equation}
with the Hubble parameter $H_0$, matter density $\Omega_\mathrm{m}$, scale factor $a$, comoving radial distance $\chi$, comoving distance to the horizon $\chi_\mathrm{h}$, 
and comoving angular diameter distance $f_K(\chi)$.
The geometric lens-efficiency factors 
\begin{equation}
\label{eq:lenseff}
g_{k}(\chi) \equiv \int_\chi^{\chi_\mathrm{h}} \mathrm{d} \chi' \, p_{k}(\chi') \frac{f_K(\chi'-\chi)}{f_K(\chi')} 
\end{equation}
are weighted according to the redshift distributions $p_{k}$ of the two considered redshift bins
\citep[see e.g.][]{kai92,bas01,sks04}.

\subsection{Angular binning and treatment of intrinsic galaxy alignments}
\label{se:angularbinning}
{\mt
Our six redshift bins define a total of 21 combinations of redshift bin
pairs (including auto-correlations).
For each redshift bin
pair ($k,l$), we compute the shear cross-correlations $\xi_+^{kl}$ and $\xi_-^{kl}$ in six logarithmic angular bins
between 0\farcm2 and $30^\prime$. 
We include all of these angular and redshift bin combinations in the analysis of the weak
lensing redshift scaling presented in this section, to keep it as general as possible.
Yet, for the cosmological parameter estimation in
Sect.\thinspace\ref{se:cosmo}, we 
carefully select the included bins
to minimize potential bias by intrinsic galaxy alignments and uncertainties
in theoretical model predictions.

In order to minimize potential contamination by intrinsic alignments 
of physically associated galaxies, we  exclude the auto-correlations
of the relatively narrow redshift bins 1 to 5. These contain
the highest fraction of galaxy pairs at similar redshift, and hence
carry the strongest potential contamination. 

An additional contamination may originate from 
alignments between intrinsic galaxy shapes and their surrounding density field causing the gravitational shear \citep[e.g.][]{his04,hmi07}.
A complete removal of this effect requires more advanced analysis schemes \citep[e.g.][]{jos08},
which we postpone to a future study.
Yet, following the suggestion by \citet{mhi06}, we  exclude luminous red
galaxies (LRGs) in the computation of the shear-shear correlations used for the
parameter estimation.
This  reduces potential contamination, given that LRGs were found to carry the strongest alignment signal 
{\ro \citep[][]{mhi06,mbb09,hmi07}}.
We select these galaxies from the \citet{ics09} 
photo-$z$ catalogue with cuts in the photometric type
\mbox{$\mathrm{mod}_\mathrm{gal}\le 8$} (``ellipticals'')
and absolute magnitude \mbox{$M_\mathrm{V}<-19$}, excluding a total of
5\,751 galaxies\footnote{{\mt In the cross-correlation between two redshift bins, it would be sufficient to exclude LRGs in the lower redshift bin only. However, for convenience we generally exclude them.}}.
We accordingly adapt the redshift distribution for the parameter estimation.

In the cosmological parameter estimation,
we additionally exclude the smallest angular
bin (\mbox{$\theta < 0\farcm5$}), for which
the theoretical model predictions have the largest uncertainty due to
required  non-linear corrections
(Sect.\thinspace\ref{se:constraints:modelcalc}) and  the influence of
baryons \citep[e.g.][]{rzk08}.

While we do not exclude LRGs and the smallest angular bin 
for the redshift scaling analysis presented in the current section, we have
verified that their exclusion leads to only very small changes,
which are well within the statistical errors and do not affect our  
conclusions.
}

\subsection{Covariance estimation}
\label{se:tomo:covariance}

In order to interpret our measurement and constrain cosmological parameters, we need to reliably estimate the data 
covariance matrix and its inverse.
\citet{mrl07}  estimate a
covariance for their analysis from the variation between the four COSMOS quadrants.
This approach 
yields too few independent realisations and may
substantially underestimate the true errors \citep{hss07b}.
We also do not employ a covariance for Gaussian statistics \citep[e.g.][]{jse08}
due to the neglected influence of non-Gaussian sampling variance.
This is particularly important for the small-scale signal probed with COSMOS \citep{kis05,swh07}.
Instead, we estimate the 
covariance matrix from 288 realisations of COSMOS-like fields obtained from
ray-tracing through the Millennium Simulation \citep{swj05}, which combines a large simulated volume 
yielding many quasi-independent lines-of-sight with
a relatively high spatial and mass resolution. The latter is needed to fully utilize the
small-scale signal measureable in a deep space-based survey.

The details of the ray-tracing analysis are given in \citet{hhw09}.
In brief, we use  tilted lines-of-sight through the simulation to avoid repetition of structures along the backwards lightcone, providing us with $32$
quasi-independent $4\deg\times 4\deg$ fields, which we further subdivide into nine COSMOS-like subfields, {\ro yielding a total of 288 realisations}.
We randomly populate the fields with galaxies,
employing the same galaxy number density, field masks, shape noise,
and redshift distribution as in the COSMOS data.
We incorporate photometric redshift errors for bins 1 to 5 by randomly
misplacing galaxy redshifts assuming a (symmetric) Gaussian scatter according to the $1\sigma$ errors in the photo-$z$ catalogue.
In contrast, the redshift  calibration uncertainty for bin 6 is not a stochastic but a systematic error, which we account for in the cosmological model fitting in Sect.\thinspace\ref{se:cosmo}.

The value of \mbox{$\sigma_8=0.9$} used for the Millennium Simulation is slightly high compared
to current estimates.
This will lead to an overestimation of
the errors, hence our analysis can be considered slightly conservative.
We have to neglect the cosmology dependence of the covariance
\citep{esh09} in the parameter estimation, given that 
we have currently only one simulation with high resolution and large volume at hand.

We need to invert the covariance matrix for the  cosmological parameter estimation 
in Sect.\thinspace\ref{se:cosmo}.
While the covariance estimate $\boldsymbol{\hat{C_*}}$
from the  ray-tracing  realizations is unbiased, a bias is introduced by correlated noise  in the matrix inversion.
To obtain an unbiased estimate for the inverse covariance $\boldsymbol{C}^{-1}$, 
we apply the correction 
\begin{equation}
\label{eq:inv_cov}
  \boldsymbol{\hat{C}}^{-1} = c\thinspace\boldsymbol{\hat{C_*}}^{-1} = \frac{n-p-2}{n-1}\boldsymbol{\hat{C_*}}^{-1} \quad \mathrm{for}\thinspace p<n-2
\end{equation}
discussed in \citet{hss07b}, where  \mbox{$n=288$} is the number of independent realisations and
$p$ is
 the dimension of the data vector.
As discussed in Sect.\thinspace\ref{se:angularbinning},
we exclude the smallest angular bin and
auto-correlations of redshift bins 1 to 5,
yielding  \mbox{$p=160$} and a moderate correction factor \mbox{$c\simeq
  0.4390$}.
In contrast, for the full data vector including all bins and correlations
(\mbox{$p=252$}), a very substantial correction factor 
\mbox{$c\simeq  0.1185$} would be required. 
Hence, our optimized data vector also leads to a more robust covariance 
inversion.

In order to limit the required correction for the covariance inversion,
we do not include more angular
bins in our analysis. 
We have 
therefore 
optimized the bin limits using Gaussian
covariances \citep{jse08} and a Fisher-matrix analysis aiming at maximal
sensitivity to cosmological parameters.

\subsection{Redshift scaling of shear-shear cross-correlations}
\label{se:tomo:zscale}

  \begin{figure}
   \centering
   \includegraphics[width=8cm]{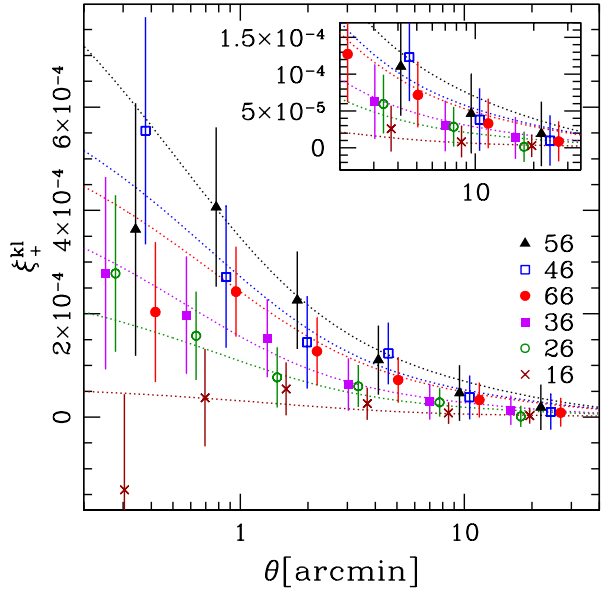}
   \caption{Shear-shear cross-correlations $\xi_+^{k6}$ between bins 1 to 6
     and bin 6, where points are plotted at their \emph{effective} $\theta$, weighted within one bin according to the $\theta$-dependent number of contributing galaxy pairs.
 The curves indicate $\Lambda$CDM predictions for our reference cosmology
 with $\sigma_8=0.8$. Corresponding points and curves have been equally
 offset along the $x$-axis for clarity. The error-bars correspond to the
 square root of the diagonal elements of the full ray-tracing covariance. Note that the points
 are substantially correlated both between angular and redshift bins,
 leading to the smaller scatter than naively expected from the
 error-bars.
   }
   \label{fi:xicross}
    \end{figure}

We plot the shear-shear cross-correlations $\xi_+^{k6}$  between all redshift bins and the broad
bin 6  in Fig.\thinspace\ref{fi:xicross}.
{\mt These cross-correlations carry the lowest 
shot noise and
shape noise due to the large number of galaxies in bin 6. }
The good agreement between the data and $\Lambda$CDM model already indicates
that the weak lensing signal roughly scales with redshift as expected. 
The errors correspond to the square root of the diagonal elements of the full ray-tracing
covariance. 
Points are correlated not only within a redshift
bin pair, but also between  different redshift combinations,  as their lensing signal is partially
caused by the same foreground structures. In addition,  galaxies in bin 6 contribute to different cross-correlations.
Note that our relatively broad angular bins lead to a significant variation of the theoretical models \emph{within} a bin.
When computing an average model prediction for a bin, we therefore weight according to the $\theta$-dependent number of galaxy pairs within this bin.
Likewise, we plot points at their \emph{effective} $\theta$, which has been weighted accordingly.

\begin{figure*}[]
   \centering
   \includegraphics[width=8.8cm]{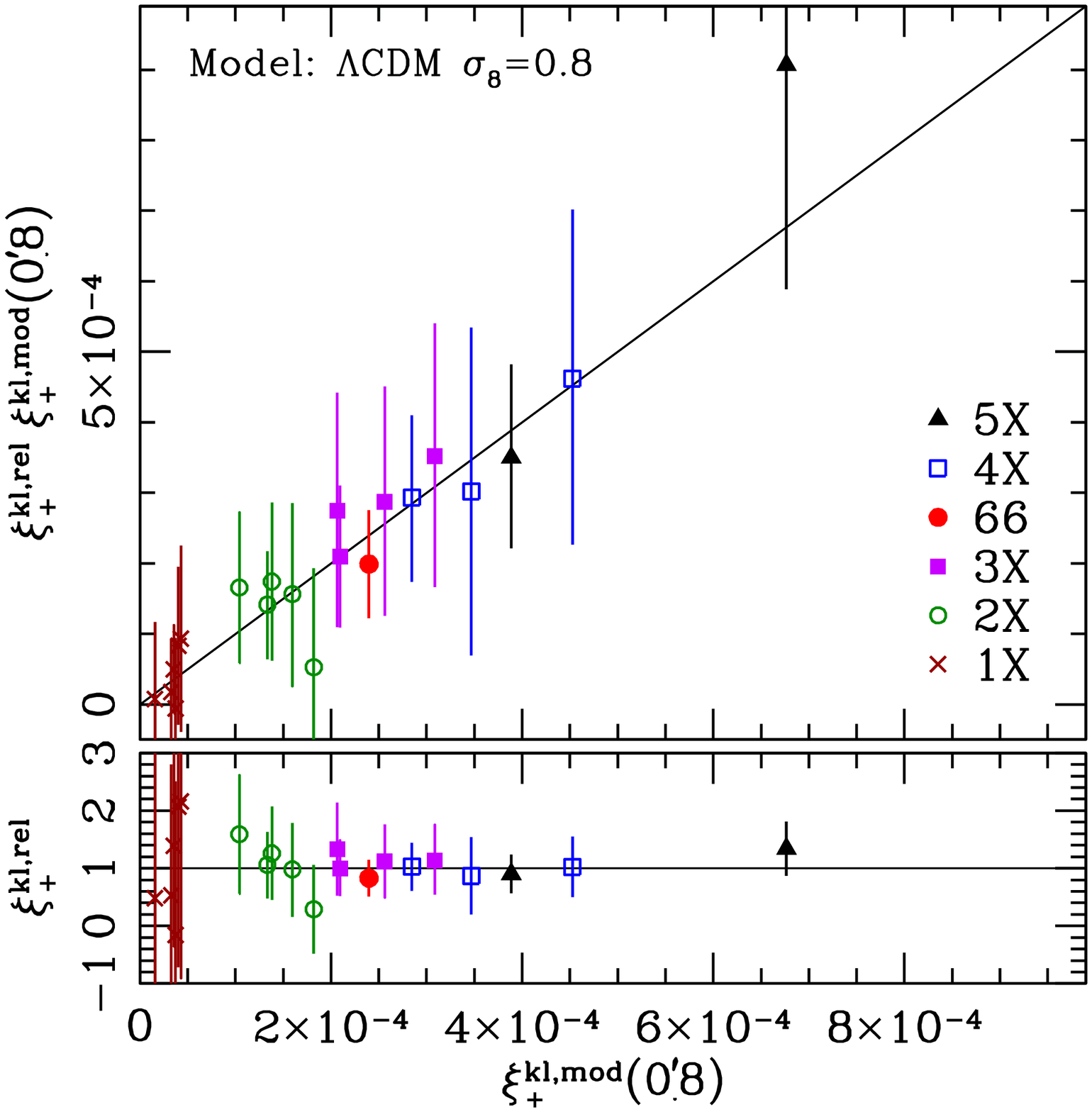}
    \includegraphics[width=8.8cm]{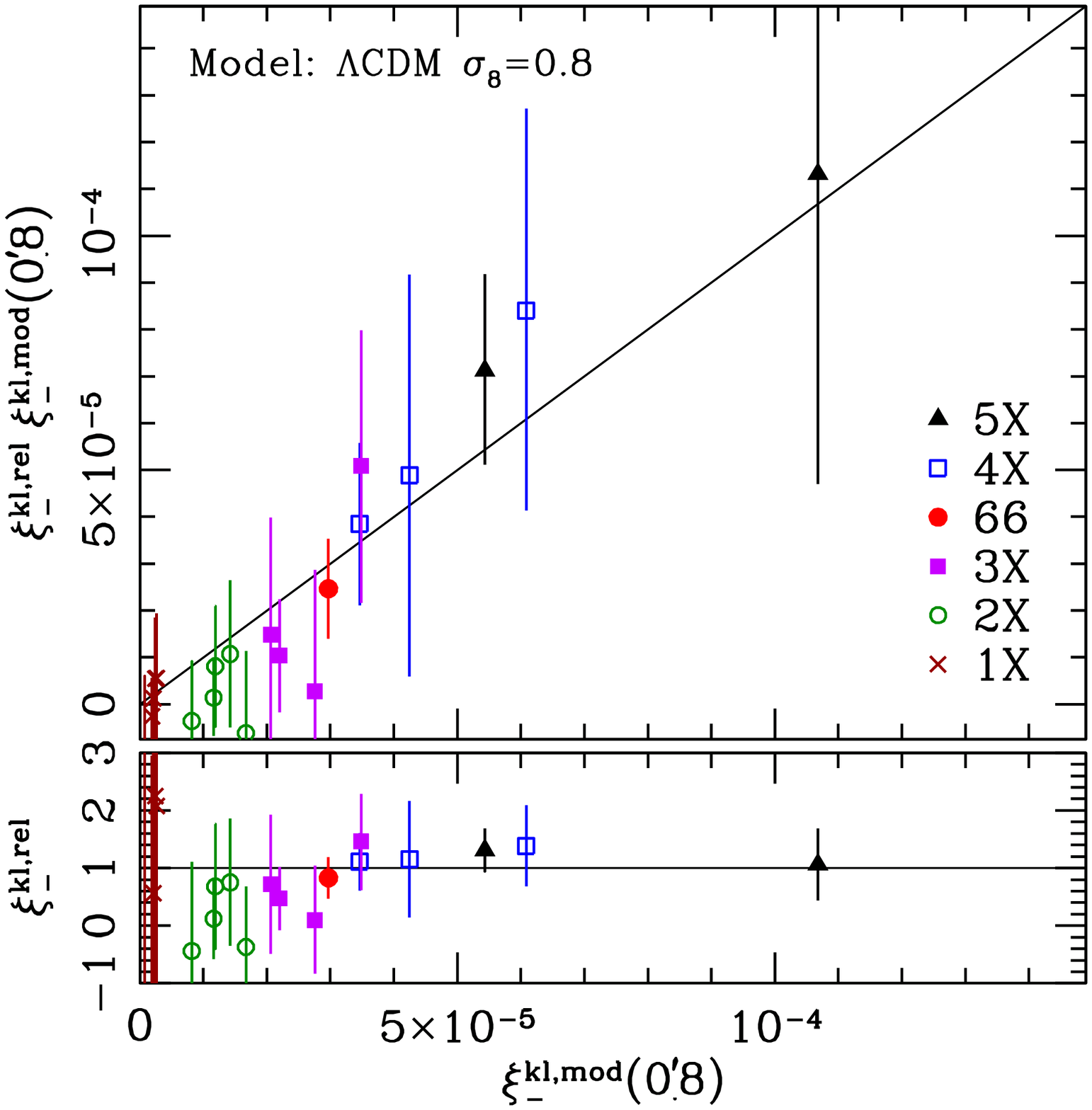}
  \caption{Shear-shear redshift scaling for $\xi_+$ (\textit{left}) and $\xi_-$ (\textit{right}). Each point corresponds to one redshift bin combination, where we 
have 
combined different angular scales by fitting the signal amplitude  \mbox{$\xi_\pm^{kl,\mathrm{rel}}$}
relative to the model prediction $\xi_\pm^{kl,\mathrm{mod}}(\theta)$
for our reference $\Lambda$CDM cosmology with \mbox{$\sigma_8=0.8$}.
The \textit{lower} plots show the relative amplitude as a function of the model prediction $\xi_\pm^{kl,\mathrm{mod}}(0\farcm8)$ for a reference angular bin centred at \mbox{$\theta=0\farcm8$}, whereas 
the amplitude has been 
scaled with  $\xi_\pm^{kl,\mathrm{mod}}(0\farcm8)$ for the \textit{upper} plots.
Symbols of one kind correspond to cross-correlations of one bin with all higher-numbered bins. Within one symbol the 
partner redshift bins sort according to the mean lensing efficiency, from left to right as 1, 2, 3, 6, 4, 5.
Note that points are correlated as each redshift bin is used for   six
bin combinations, and given that foreground structures contribute to the signal of all bin combinations at higher redshift.
The error-bars are computed from the full ray-tracing covariance, accounting
for this influence of large-scale structure.
   }
   \label{fi:zscale}
    \end{figure*}

Instead of plotting 21 separation-dependent, noisy  cross-correlations,
we condense the information into a single plot showing the redshift dependence of the signal.
Here we assume that the predictions for our reference cosmology describe the relative \emph{angular} dependence of
the signal sufficiently well, 
and fit the data points as
\begin{equation}
\label{eq:xiratio}
\xi_\pm^{kl,\mathrm{fit}}(\theta)=\xi_\pm^{kl,\mathrm{rel}}\xi_\pm^{kl,\mathrm{mod}}(\theta)\,,
\end{equation}
where \mbox{$\xi_\pm^{kl,\mathrm{mod}}(\theta)$} is the model for the reference cosmology with \mbox{$\sigma_8=0.8$},
and \mbox{$\xi_\pm^{kl,\mathrm{rel}}$} is the fitted relative amplitude.
In this fit, we take the full ray-tracing covariance between the angular scales into account.
We plot the resulting 21 ``collapsed'' cross-correlations 
for both $\xi_+$ and $\xi_-$
in Fig.\thinspace\ref{fi:zscale}, as a function of their model prediction at
a reference angular scale of 0\farcm8, where points are again correlated.
For both cases the redshift scaling of the signal is fully consistent with  $\Lambda$CDM expectations, showing a strong increase with redshift.
This demonstrates that 3D weak lensing does indeed perform as expected.
We note that 
for $\xi_-$ the signal is somewhat low for lower redshift combinations (smaller $\xi_\pm^{kl,\mathrm{mod}}$), whereas it is slightly increased compared to predictions at higher redshifts.
This behaviour is not surprising as most massive structures in COSMOS are located at \mbox{$0.7\lesssim z \lesssim 0.9$} \citep{sab07}, which create a lensing signal only for the higher redshift source bins.
Slight differences between  $\xi_+$ and $\xi_-$ are also expected, given 
that they probe the power spectrum with different filter functions, see Eq.\thinspace(\ref{eq:xipm_pkappa}). 

\subsection{Contamination of the excluded faint \mbox{$z<0.6$} sample with high-$z$ galaxies}
\label{se:tomo:conta}

\begin{figure}
   \centering
   \includegraphics[width=8cm]{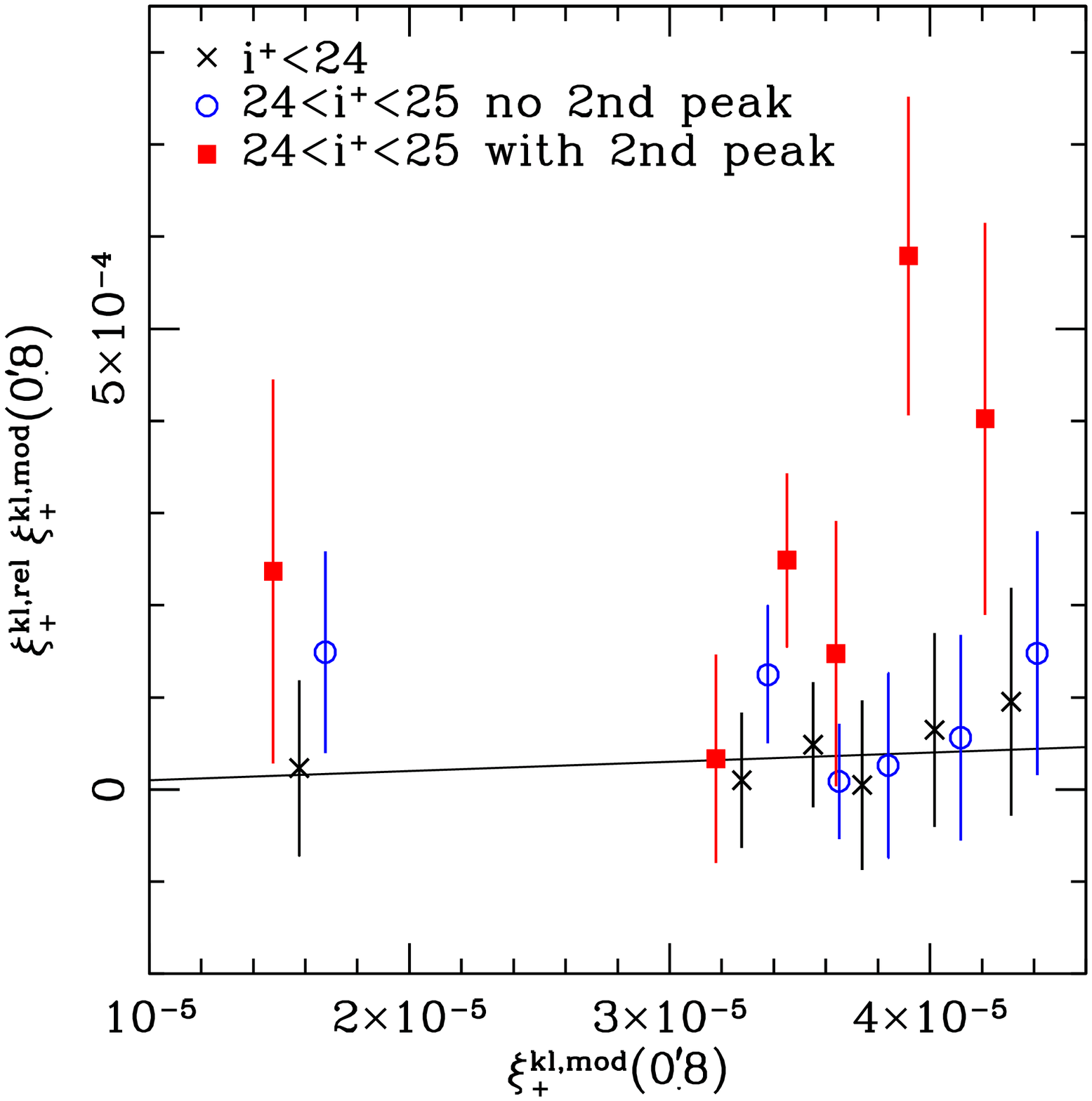}
  \caption{Shear-shear redshift scaling for $\xi_+^{1l}$ as in
    Fig.\thinspace\ref{fi:zscale}, but now only cross-correlations with bin
    1 (\mbox{$z<0.6$}) are shown, hence the different axis scale. The signal from the
    \mbox{$i^+<24$}  galaxies used in our cosmological analysis (crosses),
    is well consistent with the $\Lambda$CDM prediction (curve). Galaxies
    with \mbox{$24<i^+<25$}  and a single-peaked photo-$z$ probability distribution
    (circles) show a mildly increased but still consistent signal. In
    contrast,  \mbox{$24<i^+<25$}  galaxies with a significant secondary
    peak at \mbox{$z_{\mathrm{phot},2}>0.6$} in their individual photo-$z$ probability distribution,
    show a strong signal excess (squares), suggesting strong contamination with high-redshift galaxies.
   }
   \label{fi:zscale_lowz}
    \end{figure}

As discussed in Sect.\thinspace\ref{se:tomo:zcats}, we expect a significant
fraction of faint \mbox{$i^+\gtrsim 24$} galaxies with assigned
photometric redshift
\mbox{$z_\mathrm{phot}<0.6$}
 to be truly located at high redshifts
\mbox{$z_\mathrm{true}\gtrsim 2$}. To test this hypothesis, we plot the collapsed 
shear cross-correlations 
 for different samples of galaxies with assigned
 \mbox{$z_\mathrm{phot}<0.6$}    in Fig.\thinspace\ref{fi:zscale_lowz}.
For the \mbox{$i^+< 24$} galaxies used in the cosmological analysis the
signal is well consistent with expectations, suggesting negligible
contamination. For a \mbox{$24<i^+<25$} sample with single-peaked photo-$z$
probability distribution 
  a mild increase is detected.
This is still consistent with
  expectations, suggesting at most low contamination.
We also study a sample of galaxies each of which has a
significant secondary peak in their photometric redshift probability
distribution at
\mbox{$z_{\mathrm{phot},2}>0.6$}, amounting to 36\% of all \mbox{$24<i^+<25$} galaxies with \mbox{$z_\mathrm{phot}<0.6$}. This sample shows a strong boost in the lensing signal,
suggesting strong contamination with high-redshift galaxies.

We can obtain a rough estimate for this contamination if we assume that the
shear signal does actually scale as in our reference $\Lambda$CDM cosmology.
For simplicity we assume that the cross-contamination can be described as a
uni-directional scatter from bin 5 to bin 1, and that the true redshifts of
the misplaced galaxies follow the distribution within bin 5.
The expected contaminated signal is then given as a linear superposition of
the cross-correlation predictions with bin 1 and bin 5 respectively, according 
to the relative number of contributing galaxy pairs
\begin{eqnarray}
\label{eq:contamination}
\xi_+^{11,\mathrm{cont}}&=&(1-r)^2\xi_+^{11,\mathrm{mod}}+r^2
\xi_+^{55,\mathrm{mod}}+2r(1-r)\xi_+^{15,\mathrm{mod}}\\
\xi_+^{1l,\mathrm{cont}}&=&(1-r)\xi_+^{1l,\mathrm{mod}}+r
\xi_+^{l5,\mathrm{mod}}\,, \quad\mathrm{for}\,\,l>1\,,\nonumber
\end{eqnarray}
where $r$ is the contamination fraction, 
{\mt 
i.e. the fraction of the 
bin 1 galaxies with
\mbox{$24<i^+<25$}
and a significant secondary peak in their photo-$z$ PDF, which should have been placed into bin 5}.
We fit the measured shear-shear cross-correlations $\xi_+^{1l}$ with (\ref{eq:contamination}) as a
function of $r$, where we fix the reference $\Lambda$CDM cosmology and employ
a special ray-tracing covariance (generated for \mbox{$r=0.5$}), yielding an estimate for the
contamination 
\mbox{$r=0.7\pm0.2\,(\mathrm{stat.})\pm0.1\,(\mathrm{sys.})$}, where the systematic error indicates the response to a change in $\sigma_8$ by $0.1$.
This translates to a total contamination of $(25\pm7\pm4)\%$ for the
\mbox{$24<i^+<25$} galaxies with \mbox{$z_\mathrm{phot}<0.6$}, which is
consistent with our estimate for the redshift calibration uncertainty for
bin 6 (Sect.\thinspace\ref{se:tomo:extrapolate}).
Note that we also measure an increased signal in $\xi_-^{1l}$ for the sample with secondary photometric redshift peak, but do not include it in the fit (\ref{eq:contamination})  due to the stronger deviations for $\xi_-^{kl,\mathrm{rel}}$ in Fig.\thinspace\ref{fi:zscale}. An adequate inclusion would then require a more complex analysis scheme, with a comparison not to the model predictions, but to all measured cross-correlations.

Our analysis provides an interesting confirmation for 
 the photometric redshift analysis by \citet{ics09}, which apparently
 succeeds in identifying sub-samples of (mostly) uncontaminated and potentially
 contaminated galaxies quite efficiently. 

\section{Constraints on cosmological parameters}
\label{se:cosmo}
\subsection{Parameter estimation and considered cosmological models}
\label{se:cosmo:paraestimation}

The statistical analysis of the shear tomography correlation
functions, assembled as data vector $\vec{d}$,
is
based on a standard Bayesian approach
\cite[e.g.][]{2003Book...MACKAY}. Therein, prior knowledge of model
parameters $\vec{p}$ is combined with the information on those
parameters inferred from the new observation and expressed as
posterior probability distribution function (PDF) of $\vec{p}$:
\begin{equation}
  P(\vec{p}|\vec{d})= \frac{P(\vec{d}|\vec{p})P(\vec{p})}{P(\vec{d})}\,.
\end{equation}
Here,  $P(\vec{p})$ is the prior based on theoretical
constraints 
and previous observations, and $P(\vec{d})$ denotes the evidence.
The  likelihood function $P(\vec{d}|\vec{p})$ is the statistical model of the measurement noise, for which we choose a
Gaussian model 
\begin{equation}
\ln
P(\vec{d}|\vec{p})=-\frac{1}{2}\left[\vec{d}-\vec{m}(\vec{p})\right]^\mathrm{t}\boldsymbol{C}^{-1}\left[\vec{d}-\vec{m}(\vec{p})\right]+ \mathrm{const}\,,
\end{equation}
where $\vec{m}(\vec{p})$ is the parameter-dependent model, and $\boldsymbol{C}^{-1}$ the inverse covariance, which we estimated from the ray-tracing realizations in Sect.\thinspace\ref{se:tomo:covariance}. 

In our analysis we consider different cosmological models,
which are characterized by the parameters
\mbox{$\vec{p}=(\Omega_{\rm DE},\Omega_{\rm m},\sigma_8,h,w,f_{\rm z})$}, with the 
dark energy density $\Omega_{\rm DE}$,
matter
  density $\Omega_{\rm m}$,  power spectrum normalization $\sigma_8$, Hubble parameter $h$, and (constant) dark energy equation of state parameter $w$. Here, $f_{z}$ denotes a nuisance parameter 
encapsulating the uncertainty in the redshift calibration for bin 6 as
\mbox{$p_6(z,f_{z})\equiv p_6(f_{z}z)$}, 
which was discussed in Sect.\thinspace\ref{se:tomo:extrapolate}.
We consider 
\begin{itemize}
\item a \textbf{flat $\boldsymbol{\Lambda}$CDM} cosmology with fixed \mbox{$w=-1$}, \mbox{$\Omega_{\rm  m}\in[0,1]$}, and \mbox{$\Omega_{\rm  DE}=\Omega_\Lambda=1-\Omega_{\rm  m}$},
\item a \textbf{general (non-flat) $\boldsymbol{\Lambda}$CDM} cosmology with fixed \mbox{$w=-1$} and \mbox{$\Omega_{\rm  DE}=\Omega_\Lambda\in[0,2]$}, \mbox{$\Omega_{\rm  m}\in[0,1.6]$}, and
\item a \textbf{flat $\boldsymbol{w}$CDM} cosmology with \mbox{$w\in[-2,0]$}, \mbox{$\Omega_{\rm  m}\in[0,1]$}, and \mbox{$\Omega_{\rm  DE}=1-\Omega_{\rm  m}$}.
\end{itemize}
In all cases,
we employ priors with flat PDFs for 
\mbox{$\sigma_8\in[0.2,1.5]$} and
\mbox{$f_{z}\in[0.9,1.1]$}.
{\mt
In our default analysis scheme we also apply a  
Gaussian prior
for \mbox{$h=0.72\pm0.025$},
and
assume a fixed baryon density \mbox{$\Omega_\mathrm{b}=0.044$} and spectral
index \mbox{$n_\mathrm{s}=0.96$} as consistent with \citet{dkn09}, where the
small 
uncertainties on $\Omega_\mathrm{b}$ and $n_\mathrm{s}$ are negligible for our analysis.
Note that
we relax these priors for parts of the analysis in
Sect.\thinspace\ref{sec:constraints:nonflatlcdm} and Sect.\thinspace\ref{se:cosmo:constraints:mr:wmap}.
}

The practical challenge of the parameter estimation is to evaluate the posterior within a
reasonable time, as the computation of one model vector
for shear tomography correlations is time-intensive.
For an efficient sampling of the parameter space, we 
employ the Population Monte Carlo (PMC)
method as described in \cite{WK09}.
 This algorithm is an adaptive
importance-sampling technique \citep{CDGMR07}: instead of creating a
sample under the posterior as done in traditional Monte-Carlo Markov chain (MCMC) techniques \citep[e.g.][]{2001CQGra..18.2677C}, points are sampled from a
simple distribution, the so-called proposal, in our case a mixture of
eight Gaussians. Each point is then weighted by the ratio of the
proposal to the posterior at that point. In a number of iterative
steps, the proposal function is adapted to give better and better
approximations to the posterior.
We run the PMC
algorithm for up to eight iterations, using 5\,000 sample points in each iteration.
To reduce the Monte-Carlo variance, we
use 
larger samples with 10\,000 to 20\,000 points for the final iteration.
These are used to create density histograms, mean parameter values, and
confidence regions.
Depending on the experiment, the effective sample size of the final
 importance sample was between 7\,500 and 17\,700.
 We also cross-checked parts of the analysis with an
independently developed code which is based on the traditional but less
efficient MCMC
approach, finding fully consistent results.

\subsection{Non-linear power spectrum corrections}
\label{se:constraints:modelcalc}

  \begin{figure}
   \centering
     \includegraphics[width=7.5cm]{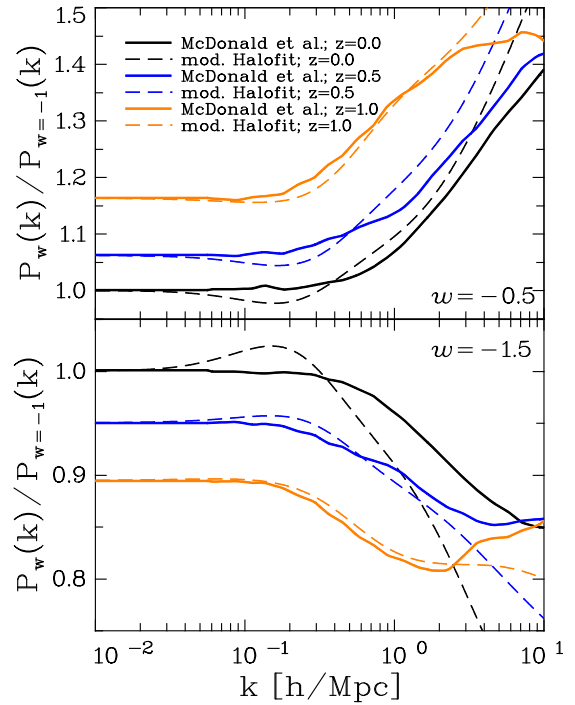}
  \caption{Comparison of the fit formulae for the non-linear growth of structure in $w$CDM cosmologies. Shown is the three-dimensional matter power spectrum, normalized by the corresponding $\Lambda$CDM power spectrum, as a function of the wave vector $k$.
 In the upper panel we consider a $w$CDM cosmology with \mbox{$w=-0.5$}, in the lower panel one with \mbox{$w=-1.5$}. 
Solid curves show the fit to the simulations by \citet{mtc06}, while the dashed lines have been 
obtained by interpolating the \citet{spj03} fitting formulae between the cases of an OCDM and a $\Lambda$CDM cosmology as outlined in Sect.\thinspace\ref{se:constraints:modelcalc}.
Each fit formula has been computed at redshifts \mbox{$z=0$} (black), \mbox{$z=0.5$} (blue), and \mbox{$z=1$} (orange). 
While deviations are substantial at \mbox{$z=0$}, the lensing analysis of the deep COSMOS data is mostly sensitive to structures at  \mbox{$z\gtrsim 0.4$}, where deviations are  reasonably small.
Note that the remaining cosmological parameters have been set to their default 
WMAP5-like values, except for \mbox{$\sigma_8=0.9$}.
   }
   \label{fi:wcorr}
    \end{figure}

To calculate model
predictions for the correlation functions according to (\ref{eq:xipm_pkappa}),
(\ref{eq:pkappa}), and (\ref{eq:lenseff}), we need to 
evaluate the involved distance ratios and compute the non-linear power spectrum \mbox{$P_\delta(k,z)$}.
Given a set of parameter values, the computation of the distances and the linearly extrapolated power spectrum
is straightforward.
We employ the transfer function by \citet{eih98} for the latter, taking baryon damping but no oscillations  into account (`shape fit').
 
For $\Lambda$CDM models we estimate the full
non-linear power spectrum according to \citet{spj03}.
\citet{mtc06} also provide non-linear power spectrum corrections for \mbox{$w\ne -1$}, but these were
tested for a  narrow range in  \mbox{$\sigma_8=0.897 \pm 0.097$} only.
We want to keep our analysis as general as possible, not having to assume such a strong prior on $\sigma_8$.
Following the \texttt{icosmo} code 
\citep{rak08} 
we instead
interpolate the non-linear corrections from \citet{spj03} between the cases
of a $\Lambda$CDM cosmology \mbox{($w=-1$)} and an  OCDM cosmology,
acting as a dark energy with \mbox{$w=-1/3$}.
This is achieved by replacing the parameter \mbox{$f=\Omega_\Lambda/(1-\Omega_{\rm m})$} in the halo model fitting function \citep{spj03}. 
This parameter is used to interpolate between spatially flat models with
dark energy (\mbox{$f=1$}) and an open Universe without dark energy (\mbox{$f=0$}).
We substitute $f$ by a new parameter $f' \equiv -0.5(3w+1)$. Thus, we obtain $f'=1$ for $\Lambda$CDM and $f'=0$ for $w$CDM with $w=-1/3$, mimicking an OCDM cosmology for which the original parameter $f$ vanished as well.

To test this simplistic approximation, we compare the computed 
corrections for \mbox{$w=(-0.5,-1.5)$} to the fitting formulae from
\citet{mtc06} in
Fig.\thinspace\ref{fi:wcorr}.
Note that we use our fiducial cosmological parameters to obtain these curves, except for \mbox{$\sigma_8=0.9$},
to match \mbox{$\sigma_8=0.897 \pm 0.097$} from \citet{mtc06}.
For most of the scales probed by our measurement the two descriptions agree
reasonably well. The modification of the halo fit follows the fits to the
simulations more accurately on large scales and at higher redshift, while it
does not reproduce the tendency of the fits by \citet{mtc06} to drop off for
large wave vectors. The precision of the modification outlined above is
sufficient for our aim to provide a proof of concept for weak lensing dark
energy measurements. However, future measurements with larger data sets will
require accurate fitting formulae for general $w$ cosmologies.

\begin{table*}[ht]
\begin{center}
\caption{Constraints on $\sigma_8\left(\Omega_\mathrm{m}/0.3\right)^{\alpha}$, $\Omega_\mathrm{m}$, $\Omega_\mathrm{DE}$, and $w$ from the COSMOS data for different 
cosmological models and analysis schemes, using our default priors. 
We  quote the marginalized mean and 68.3\% confidence limits (16th and 84th percentiles)
assuming non-linear power spectrum corrections according to \citet{spj03} and the description given in Sect.\thinspace\ref{se:constraints:modelcalc}.
Our analysis of the Millennium Simulation
(Sect.\thinspace\ref{se:cosmo:constraints:mr:wmap}) suggests that the
$\sigma_8$-estimates should be reduced by a factor $\times 0.95$ due to
biased model predictions for the non-linear power spectrum and reduced shear
corrections.
The power-law slopes $\alpha$ have typical fit uncertainties  of \mbox{$\sigma_\alpha\simeq 0.02$}.
\label{ta:results:sigma8}
}
\vspace{0.2cm}
\begin{tabular} {lllllll}
\hline
Cosmology  & Analysis & $\alpha$ & $\sigma_8\left(\Omega_\mathrm{m}/0.3\right)^{\alpha}$ & $\Omega_\mathrm{m}$ & $\Omega_\mathrm{DE}$ & $w$\\
\hline
Flat $\Lambda$CDM & 3D & $0.51$ & $0.79\pm 0.09$ & $0.32^{+0.34}_{-0.11}$ & $0.68^{+0.11}_{-0.34}$ & $-1$\\
Flat $\Lambda$CDM & 2D & $0.62$ & $0.68\pm 0.11$ & $0.30^{+0.44}_{-0.15}$ & $0.70^{+0.15}_{-0.44}$ & $-1$\\
General $\Lambda$CDM & 3D & $0.77$ & $0.74\pm 0.12$ & $0.43^{+0.40}_{-0.19}$ & $0.97^{+0.39}_{-0.60}$ & $-1$\\
Flat $w$CDM & 3D & $0.47$ & $0.79\pm 0.09$ & $0.30^{+0.39}_{-0.11}$ & $0.70^{+0.11}_{-0.39}$ & $-1.23^{+0.79}_{-0.50}$\\
\hline
\end{tabular}
\end{center}
\end{table*}

  \begin{figure}[t]
   \centering
  \includegraphics[width=8cm]{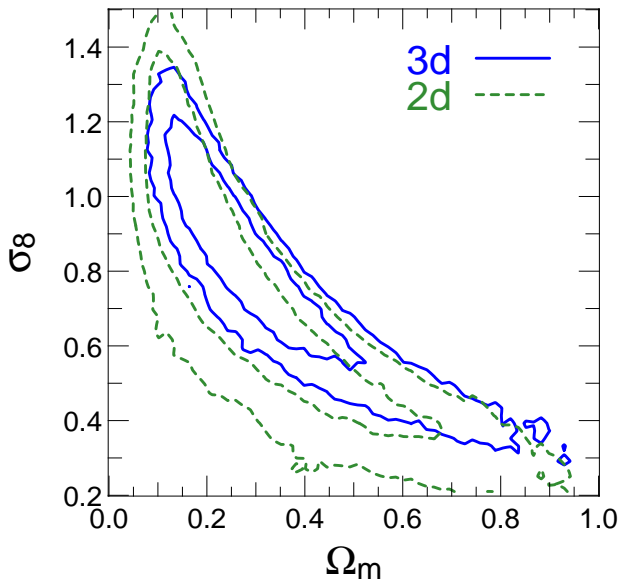}
   \caption{Comparison of our constraints on $\Omega_\mathrm{m}$ and $\sigma_8$ for a flat $\Lambda$CDM cosmology using a  3D (blue solid contours) versus a 2D weak lensing analysis (green dashed contours).
The contours show the 68.3\% and 95.4\% credibility regions, where we have marginalized over the parameters which are not shown.
The 2D analysis favours slightly lower $\sigma_8$ resulting from the lack of massive structures in the field at  low redshifts.
Nonetheless, the constraints are fully consistent as our ray-tracing covariance properly accounts for sampling variance.
}
   \label{fi:cons:3Dvs2D}
    \end{figure}

\subsection{Cosmological constraints from COSMOS}

\subsubsection{Flat $\Lambda$CDM cosmology}
\label{sec:constraints:flatlcdm}

We plot our constraints on $\Omega_\mathrm{m}$ and $\sigma_8$ for a flat
$\Lambda$CDM cosmology and our default 3D lensing analysis scheme in Fig.\thinspace\ref{fi:cons:3Dvs2D} (solid contours), 
showing the typical 'banana-shaped' degeneracy, from which we
compute\footnote{{\mt Here,
we fit a power-law
with slope $\alpha$ minimizing the separation to all posterior-weighted points in the \mbox{$\Omega_\mathrm{m}-\sigma_8$} plane, and
compute the 1D marginalized mean of
\mbox{$\sigma_8\left(\Omega_\mathrm{m}/0.3\right)^{\alpha}$} within
\mbox{$\Omega_\mathrm{m} \in [0.275,0.325]$}.}}
\begin{eqnarray}
\label{eq:results:s8om_lcdm}
\sigma_8\left(\Omega_\mathrm{m}/0.3\right)^{0.51}&=&0.79\pm 0.09\quad (68.3\%\,\mathrm{conf.}).
\nonumber
\end{eqnarray}

Here we  marginalize over 
the uncertainties in 
$h$ and the parameter $f_z$ encapsulating the uncertainty in the redshift
calibration for bin 6, where we find that $f_z$ is nearly uncorrelated with $\Omega_\mathrm{m}$, and only weakly correlated with $\sigma_8$.
The data allow us to weakly constrain \mbox{$f_z=1.03^{+0.06}_{-0.04}$}, with a maximum posterior point at \mbox{$f_z=1.05$}.
This constraint is nearly unchanged for the other cosmological models considered below.

For comparison we also conduct a classic 2D lensing analysis (dashed contours in Fig.\thinspace\ref{fi:cons:3Dvs2D}), 
where we use only the total redshift distribution and do not split galaxies into redshift bins.
We find that the 2D and 3D analyses yield consistent results with substantially overlapping $1\sigma$ regions, as expected.
Yet, the constraints from the 2D analysis shift towards lower 
\mbox{$\sigma_8\left(\Omega_\mathrm{m}/0.3\right)^{0.62}=0.68\pm 0.11$}.
The difference 
is not surprising given 
that the strongest contribution to the lensing signal in COSMOS comes from massive structures near \mbox{$z\sim0.7$} \citep{sab07,mre07}, boosting the signal for high redshift sources, but leading to a lower signal for galaxies at low and intermediate redshifts (see right panel of Fig.\thinspace\ref{fi:zscale}).
The 3D lensing analysis can properly combine these measurements, also accounting
for the larger impact of sampling variance at low redshifts.
In contrast, the 2D lensing analysis leads to a rather low (but still 
consistent) estimate for $\sigma_8$, due to the large number of low and
intermediate redshift galaxies with low shear signal.

  \begin{figure*}[t]
  \includegraphics[width=5.9cm]{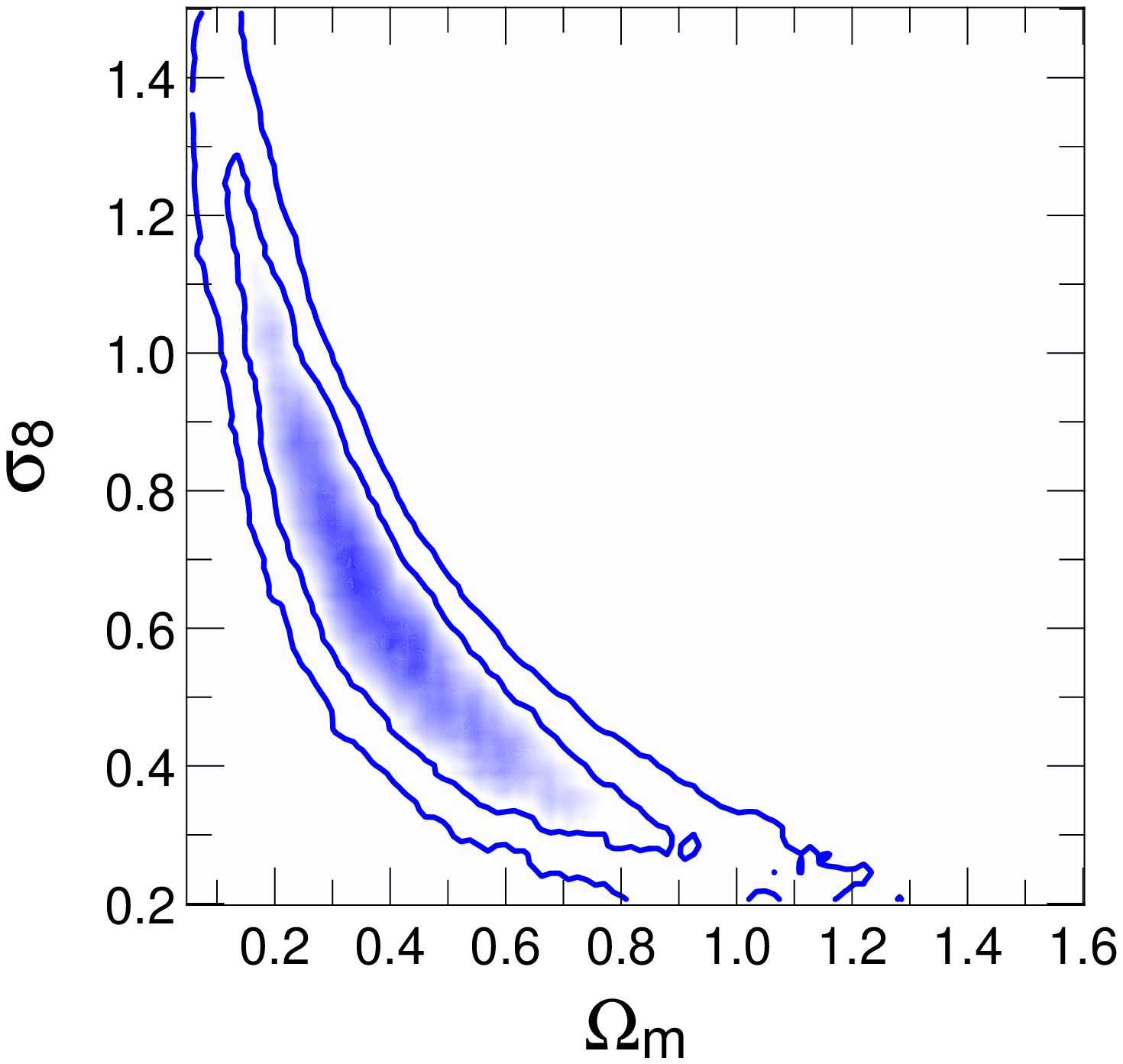}
  \includegraphics[width=5.9cm]{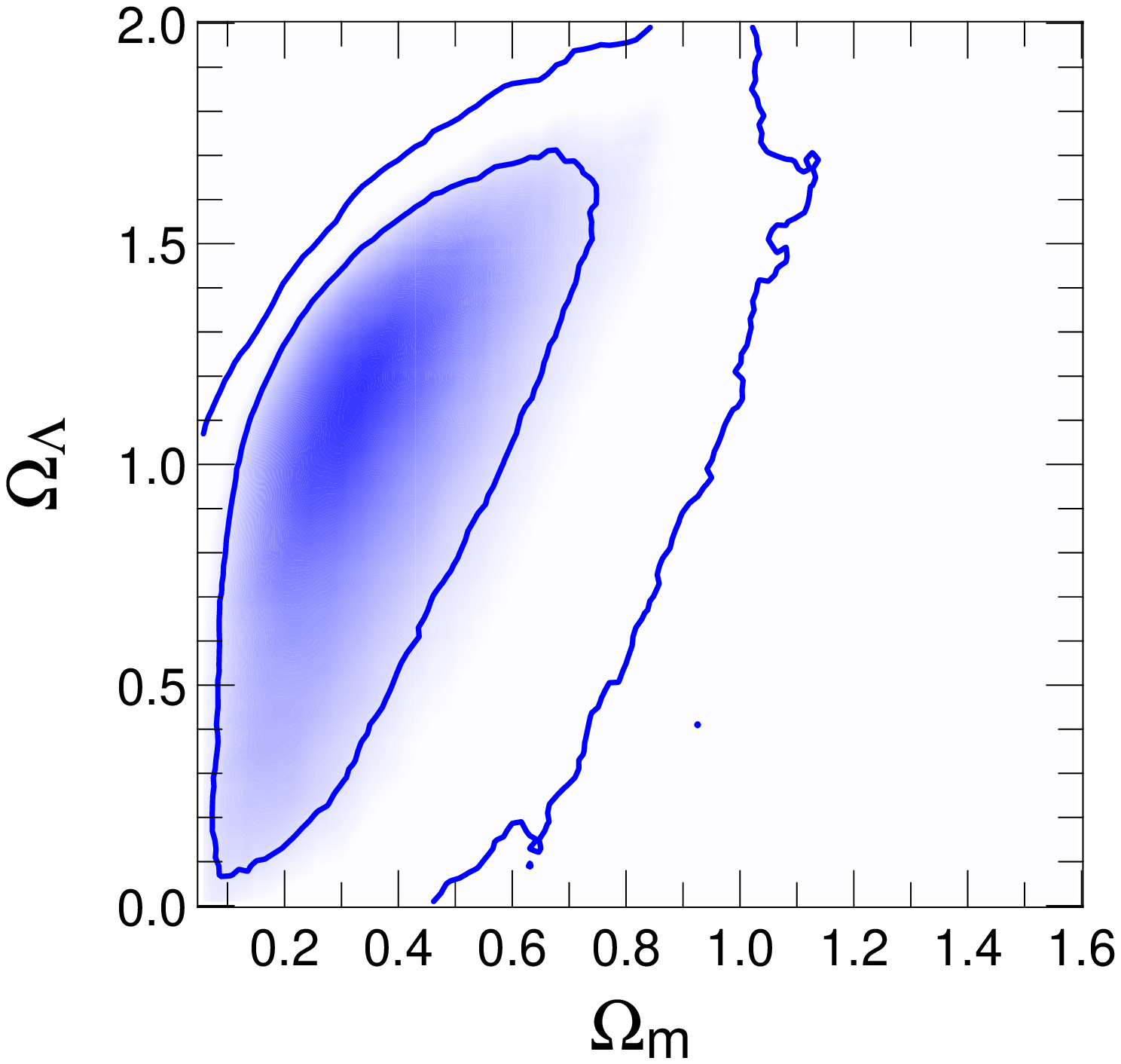}
  \includegraphics[width=5.9cm]{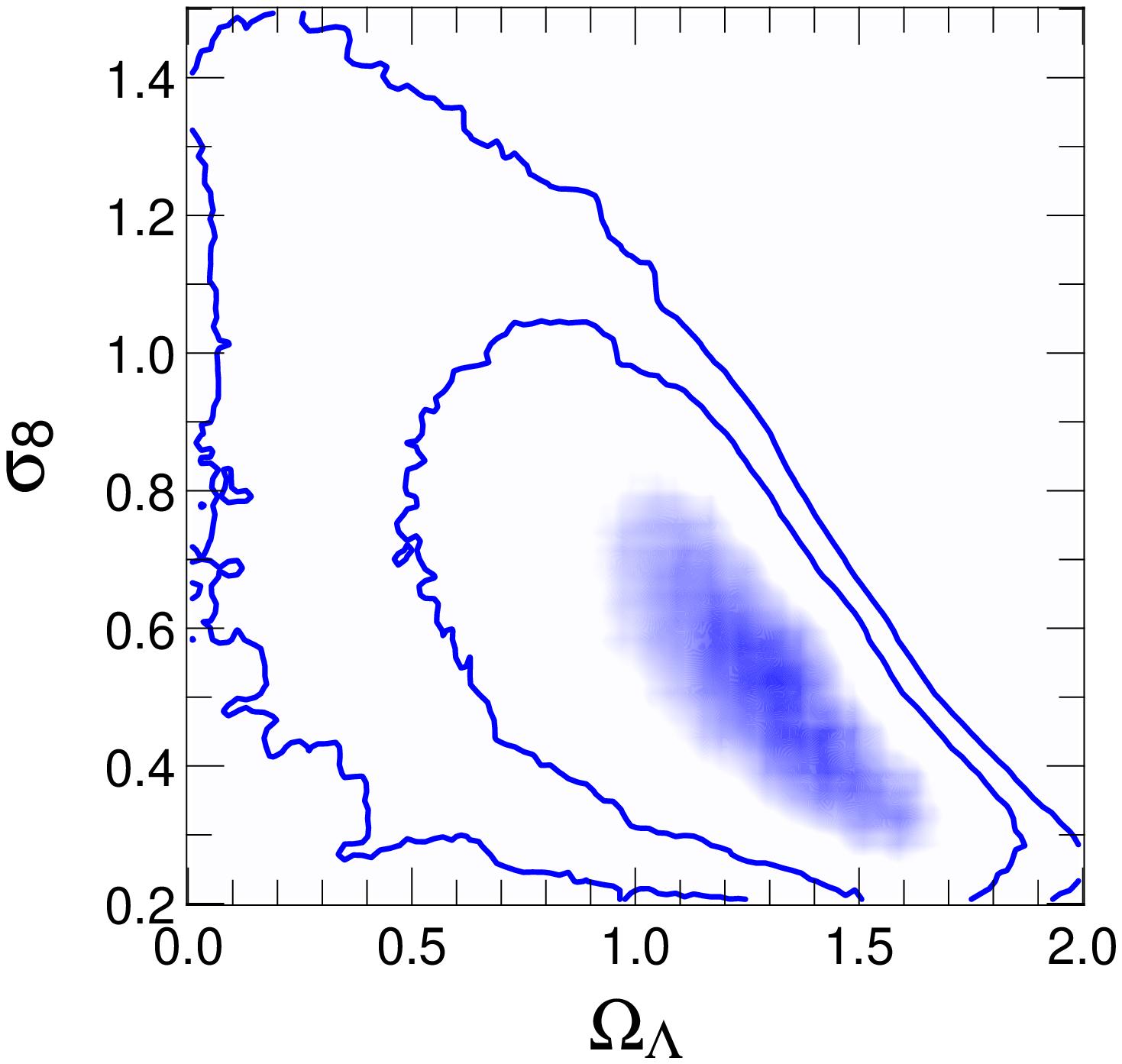}
   \caption{Constraints on $\Omega_\mathrm{m}$, $\Omega_\Lambda$, and $\sigma_8$ from our 3D weak lensing analysis of COSMOS for a general (non-flat) $\Lambda$CDM cosmology using our default priors.
The contours indicate the 68.3\% and 95.4\% credibility regions, where we have marginalized over the parameters which are not shown.
The non-linear blue-scale indicates the highest density region of the posterior.
}
   \label{fi:cons:COSMOS:nonflatlcdm}
    \end{figure*}

The tomographic analysis also reduces the degeneracy between $\Omega_\mathrm{m}$ and $\sigma_8$
by probing the redshift-dependent growth of structure and distance-redshift relation,
which differ substantially for a concordance $\Lambda$CDM cosmology and
e.g. an Einstein-de Sitter cosmology (\mbox{$\Omega_\mathrm{m}=1$}).
{ \mt
We summarize our parameter estimates 
in Table \ref{ta:results:sigma8}, also for the other cosmological models considered in
the following subsections.
}

We also test our selection criteria for the optimized data vector (Sect.\thinspace\ref{se:angularbinning}) by analysing several deviations from it for a flat $\Lambda$CDM cosmology. 
We find negligible influence if the 
 smallest angular scales \mbox{$\theta < 0\farcm5$} or LRGs 
are included, suggesting that the measurement is robust regarding the influence 
of small-scale modelling uncertainties and intrinsic alignments between galaxy shapes and their surrounding density field.
Performing the analysis using \textit{only} the usually excluded
auto-correlations of the relatively narrow redshift bins 1 to 5, we measure
a slightly lower $\sigma_8\left(\Omega_\mathrm{m}/0.3\right)^{0.52}=0.70\pm 0.13$, which 
is still consistent given the substantially degraded statistical accuracy.
If intrinsic alignments between 
physically associated galaxies contaminate the lensing measurement, we expect 
these auto-correlations to be most strongly affected. 
However, models predict an excess signal \citep[e.g.][]{hwh06}, whereas we measure a slight decrease within the statistical errors. Hence, we detect no significant indication for contamination by intrinsic galaxy alignments.

\subsubsection{General (non-flat) $\Lambda$CDM cosmology}
\label{sec:constraints:nonflatlcdm}

\begin{figure}
 \begin{center}
 \includegraphics[width=6cm]{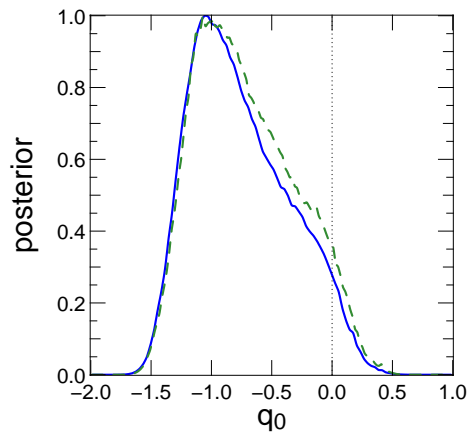}
  \caption{Posterior PDF for the deceleration parameter $q_0$ as computed
    from our constraints on $\Omega_\mathrm{m}$ and $\Omega_\Lambda$ for a general (non-flat) $\Lambda$CDM cosmology, using our
    default priors (solid curve), and using weaker priors
from the HST Key Project and Big-Bang nucleosynthesis (dashed curve). The line at \mbox{$q_0=0$} separates accelerating
    (\mbox{$q_0<0$}) and decelerating  (\mbox{$q_0>0$}) cosmologies.
    We find
\mbox{$q_0<0$} at 96.0\% confidence using our default priors, or 94.3\% confidence for the weaker priors.
   }
   \label{fi:cons:COSMOS:nonflatlcdm:q0}
\end{center}
    \end{figure}

We plot our constraints for a general $\Lambda$CDM cosmology without the assumption of flatness in Fig.\thinspace\ref{fi:cons:COSMOS:nonflatlcdm}.
From the lensing data we find 
\begin{eqnarray}
\Omega_\Lambda&>&0.32 \quad (90\%\,\mathrm{conf.}),\nonumber
\end{eqnarray}
where our prior excludes negative densities $\Omega_\Lambda<0$.
 Based on our \mbox{$\Omega_\mathrm{m}-\Omega_\Lambda$} constraints, we 
compute the posterior PDF for the
deceleration parameter
\begin{equation}
q_0=-\ddot{a}a/\dot{a}^2=\Omega_\mathrm{m}/2-\Omega_\Lambda 
\end{equation}
as shown in
Fig.\thinspace\ref{fi:cons:COSMOS:nonflatlcdm:q0}, 
which yields 
\begin{eqnarray}
q_0&<&0\quad (96.0\%\,\mathrm{conf.}).\nonumber
\end{eqnarray}
Relaxing our  priors to \mbox{$h = 0.72 \pm 0.08$}
\citep[HST Key Project,][]{fmg01}, \mbox{$\Omega_\mathrm{b}h^2 = 0.021\pm 0.001$}
\citep[Big-Bang nucleosynthesis,][]{imm09}, and \mbox{$n_\mathrm{s}\in [0.7,1.2]$},
weakens this constraint only slightly to 
\begin{eqnarray}
q_0&<&0\quad (94.3\%\,\mathrm{conf.},\,\mathrm{weak}\,\mathrm{priors}).\nonumber
\end{eqnarray}
{\mt
Employing the recent  distance ladder estimate
\mbox{$h=0.742\pm0.036$} \citep{rmc09} instead of the HST Key Project constraint, we obtain \mbox{$q_0<0$} at 94.8\% confidence.
}

Our analysis  provides evidence for the  accelerated expansion
of the Universe (\mbox{$q_0<0$}) from weak gravitational lensing.
While the statistical accuracy is still relatively weak due to the limited
size of the COSMOS field,  this evidence is  independent of external constraints on
\mbox{$\Omega_\mathrm{m}$} and \mbox{$\Omega_\Lambda$}.

{\ro
We note that the lensing data alone cannot formally exclude a non-flat OCDM cosmology. 
However, the cosmological parameters inferred for such a model would be inconsistent with various other 
cosmological probes\footnote{\ro For a lensing-only OCDM analysis the posterior peaks at \mbox{$\Omega_\mathrm{m}\simeq 0.1$, $\sigma_8\simeq 1.4$} (close to the prior boundaries). 
In the comparison with a $\Lambda$CDM analysis,
the additional parameter $\Omega_\Lambda$ causes a penalty in the Bayesian
model comparison (computed as in \citealt{kwr09}).
This leads to an only slightly larger evidence for the non-flat $\Lambda$CDM
model compared to the  OCDM model, with an inconclusive evidence ratio of 65:35.
The evidence ratio becomes a ``weak preference'' (77:23) if we employ a (still conservative) prior \mbox{$\sigma_8<1$}. Hence, with this prior the $\Lambda$CDM model makes the data more than 3 times more probable than the OCDM model.
}.
We therefore perform our analysis in the context of the well-established $\Lambda$CDM model, 
 where the lensing data provide additional evidence for cosmic acceleration.
}

\subsubsection{Flat $w$CDM cosmology}
\label{sec:constraints:flatwcdm}

  \begin{figure}
   \centering
  \includegraphics[width=6cm]{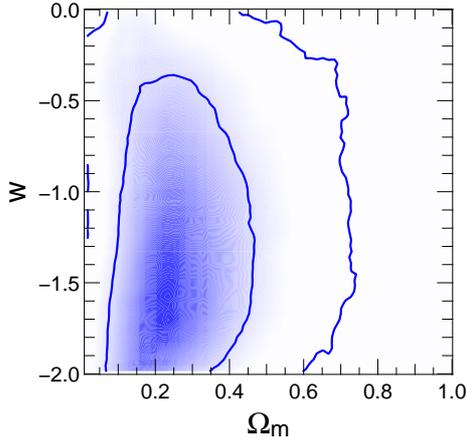}
  \caption{Constraints on $\Omega_\mathrm{m}$ and  $w$ 
from our 3D weak lensing analysis of COSMOS for a flat $w$CDM cosmology,
assuming a prior \mbox{$w\in[-2,0]$}.
The contours indicate the 68.3\% and 95.4\% credibility regions, where we have marginalized over the parameters which are not shown. The non-linear blue-scale indicates the highest density region of the posterior.
   }
  \label{fi:cons:COSMOS:flatwcdm:wOm}
    \end{figure}

For a flat $w$CDM cosmology  we 
plot our constraints 
on the (constant) dark energy equation of state parameter $w$
in Fig.\thinspace\ref{fi:cons:COSMOS:flatwcdm:wOm},
showing that the measurement is consistent with $\Lambda$CDM (\mbox{$w=-1$}).
From the posterior PDF 
we compute
\begin{eqnarray}
\label{eq:w:limit}
w&<&-0.41  \quad (90\%\,\mathrm{conf.})\nonumber
\end{eqnarray}
for the chosen prior
\mbox{$w\in[-2,0]$}.
{\ro The exact value of this upper limit depends on the lower bound of
the prior PDF given the non-closed credibility regions.}
We have chosen this prior as more negative \mbox{$w$} would require a worrisome extrapolation for the non-linear power spectrum corrections (Sect.\thinspace\ref{se:constraints:modelcalc}). 
For comparison, we repeat the analysis with a much wider prior \mbox{$w\in[-3.5,0.5]$} leading to a stronger upper limit \mbox{$w<-0.78$} ($90\%\,\mathrm{conf.}$). 
{\ro While the COSMOS data are capable to exclude very large values \mbox{$w\gg -1$},
larger lensing data-sets will be required to obtain really competitive 
constraints on $w$.}

To test the consistency of the data with $\Lambda$CDM, we compare the
Bayesian evidence for  the flat $\Lambda$CDM and $w$CDM models, which we compute
in the PMC analysis as detailed in \citet{kwr09}.
Here we find completely inconclusive  probability ratios for $w$CDM versus
$\Lambda$CDM of \mbox{$52:48$} (\mbox{$w\in[-2,0]$})
and \mbox{$45:55$} (\mbox{$w\in[-3.5,0.5]$}), 
 confirming that the data are fully consistent with $\Lambda$CDM.

 \subsection{Model recalibration with the Millennium Simulation and joint constraints with WMAP-5}
\label{se:cosmo:constraints:mr:wmap}

\citet{hww08} and \citet{hhw09} found that the \citet{spj03}  fitting functions 
slightly underestimate non-linear corrections to the power spectrum.
To test whether this has a significant influence 
on our results, we performed a 3D cosmological parameter estimation using the mean data vector of the 288
COSMOS-like ray-tracing realisations from the Millennium Simulation.
Here we modify the strong priors given in Sect.\thinspace\ref{se:cosmo:paraestimation}
to match the input values of the simulation (\mbox{$\Omega_\mathrm{m}=0.25$}, \mbox{$\sigma_8=0.9$}, \mbox{$n_\mathrm{s}=1$}, \mbox{$h=0.73$}, \mbox{$\Omega_\mathrm{b}=0.045$}),
and find 
\mbox{$\sigma_8=0.947\pm 0.006$}\footnote{Here we have scaled the uncertainty
  for the mean ray-tracing data vector from the uncertainty for a single
  COSMOS-like field assuming that all realizations are completely independent.
This is slightly optimistic given the large but finite volume of the
simulation, and fact that the realizations were cut from larger fields.}
for \mbox{$\Omega_\mathrm{m}=0.25$}.
This  confirms the result of \citet{hww08} and \citet{hhw09}, indicating that models based on \citet{spj03} slightly underestimate 
the shear signal, hence a larger $\sigma_8$ is required to fit the data.
Here we use actual  \emph{reduced shear} estimates from the
simulation,
but employ \emph{shear} predictions, as done for the real data (see Sect.\thinspace\ref{se:wl:tests}).
Using \emph{shear} estimates from the simulation yields
\mbox{$\sigma_8=0.936\pm 0.006$}.
Hence, a minor contribution to the overestimation of $\sigma_8$
is caused by the negligence of reduced shear corrections \citep[see also][]{dsw06,sha09,krh09}.  

To compensate for this underestimation of the model predictions and reduced
shear effects, we scale our derived constraints on $\sigma_8$ for a flat $\Lambda$CDM cosmology by a factor \mbox{$0.9/0.947\simeq 0.950$}\footnote{We expect that this correction factor depends on cosmological parameters.
Yet, considering the weak lensing degeneracy for $\Omega_\mathrm{m}$ and $\sigma_8$, the input values of the Millennium Simulation are quasi equivalent to \mbox{$\sigma_8\simeq 0.82$} for \mbox{$\Omega_\mathrm{m}=0.3$}, which is sufficiently close to our constraints to justify the application.}, yielding
\begin{eqnarray}
\label{eq:results:s8om_lcdm':rescaled}
\sigma_8\left(\Omega_\mathrm{m}/0.3\right)^{0.51}=0.75\pm 0.08\quad (68.3\%\,\mathrm{conf.},\,\mathrm{MS\mbox{-}calib.}).\nonumber
\end{eqnarray}
Note that we did not apply this correction for the values given in the
previous section and listed in Table \ref{ta:results:sigma8}, as we can only
test it for the case of a flat $\Lambda$CDM cosmology.
Additionally, we want to keep the results comparable to previous weak lensing studies, which we expect to be similarly affected.

\begin{figure}[h]
   \centering
 \includegraphics[width=8cm]{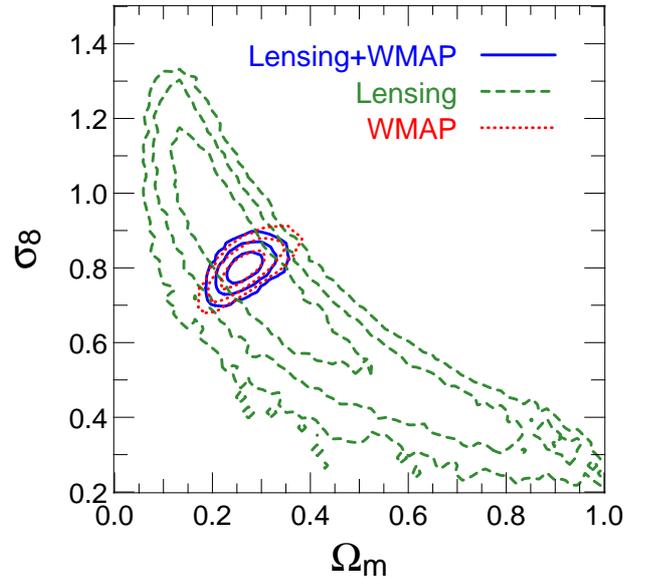}
   \caption{Comparison of the constraints on $\Omega_\mathrm{m}$ and
     $\sigma_8$ for a flat $\Lambda$CDM cosmology obtained with our COSMOS
     analysis (dashed), \mbox{WMAP-5} CMB data (dotted), and joint
     constraints (solid).  The contours indicate the 68.3\%, 95.4\%, and
     99.7\% credibility regions. Note that the weak lensing alone analysis
     uses stronger priors. The weak lensing constraints on $\sigma_8$ have
     been rescaled to account for modelling bias of the non-linear power
     spectrum and reduced shear corrections according to the ray-tracing constraints from the Millennium Simulation. }
   \label{fi:cons:withWMAP}
    \end{figure}

Having eliminated this last source of systematic uncertainty, we now 
estimate joint constraints with \mbox{WMAP-5} CMB-only data 
\citep{dkn09}, conducted similarly to the analysis by \citet{kbg09}.
Here we assume a flat $\Lambda$CDM cosmology, completely relax our priors
to \mbox{$\Omega_\mathrm{b}\in [0.01,0.1]$}, \mbox{$n_s\in [0.7,1.2]$},
 \mbox{$h\in [0.2,1.4]$}, and
scale $\sigma_8$ for the lensing model calculation according to
the Millennium Simulation results.
Here we also marginalize over an additional
2\% uncertainty in the lensing $\sigma_8$ calibration
to account for the dropped remaining mean shear calibration bias
(0.8\%, Sect.\thinspace\ref{se:wl}) and limited accuracy of the employed
residual shear correction (Sect.\thinspace\ref{se:wl:tests}), which we
estimate to be 
\mbox{$1\%$} in $\sigma_8$.
From the joint analysis with WMAP-5 we find 
\begin{eqnarray}
\label{eq:joint:lensing:wmap}
\Omega_\mathrm{m}&=&   0.266^{+0.025+0.057}_{-0.023-0.042}\,\nonumber\\
\sigma_8&=&0.802^{+0.028+0.055}_{-0.029-0.060}\,\quad (68.3\%/95.4\%\,\mathrm{conf.},\,\mathrm{MS\mbox{-}calib.}),\nonumber
\end{eqnarray}
which  reduces the size of WMAP-only $1\sigma$ ($2\sigma$) error-bars on
average by
$21\%$ ($27\%$).
We plot the joint and individual constraints  in Fig.\thinspace\ref{fi:cons:withWMAP}, illustrating the perfect agreement of
the two independent cosmological probes.

\section{Summary, discussion, and conclusions}
\label{se:dis}

We have measured weak lensing galaxy shear estimates from the HST/COSMOS
data by applying a new model  for the spatially and temporally varying ACS
PSF,
which is  based on
a principal component analysis of PSF variations in dense stellar fields.
We find that most of the PSF changes can be described with a single
parameter related to the HST focus position. 
Yet, we also correct for additional PSF variations,
which 
are coherent for neighbouring COSMOS tiles taken closely in time.
We employ updated parametric corrections for charge-transfer inefficiency,
 for both galaxies and stars, removing earlier modelling
uncertainties due to confused PSF- and CTI-induced stellar ellipticity.
Finally, we employ a simple correction for signal-to-noise dependent shear
calibration bias,
which we derive from the STEP2 simulations of
ground-based
weak lensing data.
Tests on simulated space-based data
confirm a relative shear calibration uncertainty \mbox{$|m|\le 2\%$} over the entire 
used magnitude range if this correction is applied. We  decompose the measured shear
signal into curl-free E-modes and curl-component B-modes.
As expected from pure lensing, the B-mode signal is consistent with zero for
all second-order shear statistics, providing an important confirmation for
the success of our  correction schemes for instrumental systematics.

We combine our shear catalogue with excellent ground-based photometric
redshifts from \citet{ics09} and carefully estimate the redshift distribution
for faint ACS galaxies without individual photo-$z$s.
This allows us to study weak
lensing 
cross-correlations in detail between six redshift bins, demonstrating 
that the signal indeed scales as expected from General Relativity for a concordance $\Lambda$CDM cosmology.

We employ a robust covariance matrix from 288 simulated COSMOS-like fields
obtained from ray-tracing through the Millennium Simulation \citep{hhw09}.
Using our  3D weak lensing analysis of COSMOS, we derive 
constraints 
\mbox{$\sigma_8(\Omega_m/0.3)^{0.51}=0.79\pm 0.09$}
  for a flat
$\Lambda$CDM cosmology, using non-linear power spectrum corrections from \citet{spj03}.
A recalibration of these predictions based on the ray-tracing analysis changes our constraints to
\mbox{$\sigma_8\left(\Omega_\mathrm{m}/0.3\right)^{0.51}=0.75\pm0.08$} (all 68.3\% conf.).
Our results are perfectly   consistent with WMAP-5, yielding joint constraints
\mbox{$\Omega_\mathrm{m}= 0.266^{+0.025+0.057}_{-0.023-0.042}  $}, 
\mbox{$\sigma_8=0.802^{+0.028+0.055}_{-0.029-0.060}$} (68.3\% and 95.4\%
confidence).
{ \mt
They also agree with weak lensing results from the CFHTLS-Wide \citep{fsh08} and recent galaxy cluster constraints from \citet[][]{mar09} within \mbox{$ 1\sigma$}.
}
Our errors include the full statistical uncertainty including the non-Gaussian sampling variance, 
Gaussian
photo-$z$ scatter, and marginalization over remaining parameter uncertainties, including the redshift calibration for
the faint \mbox{$i^+>25$} galaxies.

Our results are consistent with the 3D lensing constraints 
\mbox{$\sigma_8(\Omega_m/0.3)^{0.44}=0.866\pm 0.033\, (\mathrm{stat.})^{+0.052}_{-0.035}\,(\mathrm{sys.})$} 
from \citet{mrl07} assuming non-linear power spectrum corrections according to \citet{spj03}, at the \mbox{$\sim 1 \sigma$}
 level.
The analyses differ systematically in the
treatment of PSF- and CTI-effects, where the success of our methods is
confirmed by the vanishing B-mode.
Furthermore,  \citet{mrl07} employ 
earlier photo-$z$s based on fewer bands \citep{mcs07}. 
Note that the analysis of \citet{mrl07} yields 
tighter statistical errors, which may be a result of their covariance
estimate from the variation between the four COSMOS quadrants.
This potentially 
introduces a bias in the covariance inversion due to too few independent
realisations \citep{hss07b}.
While the absolute calibration accuracy  of the shear measurement method
was estimated to be the dominant source of uncertainty in their error budget,
we were able to reduce it well below the statistical error level. 
As a further difference, our analysis employs photometric redshift information to reduce potential contamination
by intrinsic galaxy alignments, where we exclude the shear-shear auto-correlations for the relatively
narrow redshift bins 1 to 5 to minimize the impact of physically associated galaxies.
In addition, we exclude  luminous red galaxies, which were found to
carry the strongest intrinsic alignment with the density field of their large-scale structure
environment causing the shear \citep[][]{hmi07}.
Finally,
we  do not include  angular scales \mbox{$\theta<0\farcm5$} due to 
{\ro increased } 
modelling uncertainties for the non-linear power spectrum.
 
{\mt
Similarly to \citet{mrl07}, we obtain a lower estimate \mbox{$\sigma_8(\Omega_m/0.3)^{0.62}=0.68\pm0.11$} for a non-tomographic (2D) analysis, assuming \citet{spj03} power spectrum corrections.
The lower signal compared to the 3D lensing
analysis is expected, given that the most massive structures in COSMOS are
located at \mbox{$0.7\lesssim z \lesssim 0.9$} \citep{sab07}, creating a strong shear signal for high redshift sources only, which is detected by the 3D analysis.
In contrast, the bulk of the galaxies in the 2D lensing analysis are located
at too low redshifts to be substantially lensed by these structures,
yielding a relatively low estimate for $\sigma_8$. Nonetheless, as sampling variance  is properly accounted for in our error analysis, the constraints are still consistent.

}

For a general (non-flat) $\Lambda$CDM cosmology, 
 we find a negative
deceleration parameter 
\mbox{$q_0<0$} at 96.0\% confidence using our default priors, and 
at
94.3\% confidence if only priors from the HST Key Project and BBN are
applied.
Hence, our
tomographic weak lensing measurement provides independent  evidence for the
accelerated expansion of the Universe.
For a flat $w$CDM cosmology we constrain the
(constant) dark energy equation of state parameter to \mbox{$w<-0.41
  \,(90\%\,\mathrm{conf.})$} for a prior \mbox{$w\in[-2,0]$},
fully consistent with $\Lambda$CDM.
Our dark energy constraints are still weak compared to recent results from
independent probes \citep[e.g.][]{kra08,hwb09,ars08,mae08,mar09,vkb09,kdn09}.
This is solely due to the limited area of COSMOS, 
leading to a dominant contribution 
to the error budget from sampling variance. 

{\ro
While the area covered by COSMOS is still small \mbox{(1.64\,$\mathrm{deg}^2$)}, the high resolution
and depth of the HST data allowed us to obtain cosmological constraints which
are comparable to results from substantially larger ground-based surveys.
However, note that HST was by no means designed for cosmic shear measurements.
In contrast, future space-based lensing mission such as Euclid\footnote{\url{http://sci.esa.int/euclid}} 
or JDEM\footnote{\url{http://jdem.gsfc.nasa.gov/}} will be highly optimised for weak lensing measurements.
High PSF stability, a much larger field-of-view providing thousands of stars for PSF measurements,
 carefully designed CCDs which minimize charge-transfer inefficiency, 
and improved algorithms will remove the need for some of the empirical calibrations employed in this paper.
}

{\mt
In order to fully exploit the information encoded in the weak lensing shear
field, second-order shear statistics, as used here, can be complemented with
higher-order shear statistics to probe the non-Gaussianity of the matter distribution \citep[e.g.][]{bar09,vlw10}.
Based on  our 
COSMOS shear catalogue,
\citet{ssw10}
present such a cosmological analysis using combined second and third-order shear statistics.
}


Finally, we  stress that weak lensing can only provide precision constraints on cosmological
parameters if sufficiently accurate models exist to compare the measurements
to.
Our analysis of the relatively small COSMOS Survey is still limited by the statistical measurement uncertainty,
for which our approximate model recalibration using the Millennium  Simulation is sufficient. 
{\ro
Most of the cosmological sensitivity in COSMOS comes from  quasi-linear
and non-linear scales.
We cut our analysis only at highly non-linear scales \mbox{$\theta<0\farcm5$},
corresponding to a comoving separation of \mbox{$\sim 360$ kpc} at \mbox{$z=0.7$}
(roughly the redshift of the most massive structures in COSMOS).
At these scales non-linear power spectrum corrections have substantial  uncertainties,
in particular  due to the influence of baryons \citep[e.g.][]{rzk08}.
Given that  our results are basically unchanged if even smaller scales are included
(insignificant increase in $\sigma_8$ by \mbox{$<1\%$}), we expect that the
model uncertainty for the larger scales should still be sub-dominant compared to our statistical errors.
}
However, analyses of large future surveys
{\ro will urgently require improved model predictions
including corrections for baryonic effects,
also for dark
energy cosmologies with \mbox{$w\ne -1$},}
and optionally also for theories of modified gravity.
{\mt
Once these are available, careful analyses 
of large 
current and future weak lensing surveys
will deliver
precision constraints on cosmological parameters and dark energy properties.
}

\begin{acknowledgements}
This work is based on observations made with the NASA/ESA \textit{Hubble Space
Telescope}, obtained from the data archives at the Space Telescope European
Coordinating Facility and the Space Telescope Science Institute.
It is a pleasure to thank the COSMOS team for making the \citet{ics09} 
photometric redshift catalogue publicly available.
We appreciate help from Richard Massey and Jason Rhodes in the creation of the
simulated space-based images.
We thank them, Maaike Damen, Catherine Heymans, Karianne Holhjem, Mike Jarvis, James Jee, Alexie Leauthaud, Mike
Lerchster, and Mischa Schirmer for useful discussions, and Steve Allen, Thomas Kitching, and Richard Massey for  helpful comments on the manuscript.
{\ro We thank the anonymous referee for his/her comments, which helped to improve this paper significantly.}
We thank the Planck-HFI and \textsc{Terapix} groups at IAP for support and computational facilities.
TS acknowledges financial support from the Netherlands Organization for Scientific Research (NWO) and the Deutsche
Forschungsgemeinschaft through SFB/Transregio 33 ``The Dark Universe''.
JH acknowledges support by the Deutsche Forschungsgemeinschaft within the Priority Programme 1177 under the project SCHN 342/6 and by the Bonn-Cologne
Graduate School of Physics and Astronomy.
BJ acknowledges support by the Deutsche Telekom Stiftung and the
Bonn-Cologne Graduate School of Physics and Astronomy.
MK is
supported by the CNRS ANR ``ECOSSTAT'', contract number
ANR-05-BLAN-0283-04, and by the Chinese National Science Foundation
Nos. 10878003 \& 10778725, 973 Program No. 2007CB 815402, Shanghai
Science Foundations and Leading Academic Discipline Project of
Shanghai Normal University (DZL805).
PSi, HHi, and MV acknowledge support by the European DUEL Research-Training
 Network (MRTN-CT-2006-036133).
MB, CDF, and PJM acknowledge support from  programs  \#HST-AR-10938 and \#HST-AR-10676, 
provided by NASA  
through  grants from the Space Telescope Science Institute (STScI), which is  
operated by the Association of Universities for Research in  
Astronomy, Incorporated, under NASA contract NAS5-26555 and
  NNX08AD79G.
HHo and SH acknowledge support from a NWO Vidi grant.
SH acknowledges support by the Deutsche Forschungsgemeinschaft within
the Priority Programme 1177 under the project SCHN 342/6.
ES acknowledges financial support from the Alexander von Humboldt
Foundation.
LVW thanks CIfAR and         
NSERC for financial support.

\end{acknowledgements}
\begin{appendix}

\section{Additional image calibrations}
\label{app:data}

In this appendix we describe  additional calibrations which we apply to the
flat-fielded \textit{\_flt} images before running
\texttt{MultiDrizzle}.

\paragraph{Background subtraction.}
We perform a
quadrant-based background subtraction due to an anomalous bias level variation between the four ACS read-out amplifiers.
Here we detect and mask objects with
\texttt{SExtractor} \citep{bea96}, combine this mask with the static bad
pixel mask, and estimate the background as the median of all non-masked
pixels in the quadrant.
We modulate the offset from the mean background level  with the
normalised inverse flat-field to correct for the fact that the improperly
bias-subtracted image has already been flat-fielded\footnote{This procedure performs well
  for relatively empty fields such as the large majority of the COSMOS
  tiles. For fields dominated by a very bright star or galaxy, 
it can, however,
  lead to erroneous jumps in the background level. Hence, we generally
 adopt a maximal accepted difference in the
background estimates of $4\,\mathrm{e}^{-}$, which, if exceeded, leads to a
subtraction of
the minimum background estimate for all quadrants.}.

\paragraph{Bad pixel masking.}
\label{app:data:masking}
We manually mask satellite trails and scattered stellar light if its
apparent sky position changes between different dither positions,
allowing us to recover otherwise unusable sky area.
In addition, we update the static bad pixel mask rejecting pixels if: 
\begin{itemize}
\item their dark current exceeds   $0.04\,\mathrm{e}^-/\mathrm{sec}$ in the
  associated dark reference file (default $0.08\,\mathrm{e}^-/\mathrm{sec}$), or
\item they are affected by variable bias structures, which we
identify in a variance image 
of five subsequent bias
  reference frames taken temporally close to the science frame considered, or
\item they show significantly positive or negative values in a median image computed from
  50 background-subtracted and object-masked COSMOS frames taken closely in time,
  indicating any other semi-persistent blemish.
\end{itemize}
The latter two masks mainly aim at the rejection of variable bias
structures which show up as positive or negative bad column
segments in the stacked image if not properly masked.
For the mask creation we utilize the \texttt{IRAF} task 
 \texttt{noao.imred.ccdmask}.
It computes the local median signal and rms variation in moving
rectangles. A pixel is then masked if its values is either \texttt{lsigma}
below or \texttt{hsigma} above the local median value.
This is done for individual pixels and sums of pixels in column sections,
where in the latter case the background dispersion 
is scaled by the square root of the number of pixels in the 
section.
Finally each column is scanned for short segments of un-flagged pixels
in between masked pixels. We additionally mask these segments if their length is
less than 15 pixels.
We summarise the values applied for the thresholds \texttt{lsigma} and
\texttt{hsigma} in Table \ref{ta:data:ccdmask}.
Due to variations in image noise properties they do not perform
optimally in all cases, so that we iteratively increase \texttt{hsigma} by
$+5$ if otherwise more than $2.5\%$ of the pixels in the bias variance image would be masked.

\begin{table}[tb]
\begin{center}
\caption{Lower and upper $\sigma$ thresholds for pixel masking with
  \texttt{ccdmask} in the bias variance and the sky-subtracted and
  object-masked median images.
}

\vspace{0.2cm}
\begin{tabular} {lll}
\hline
Image type & \texttt{lsigma} & \texttt{hsigma} \\
\hline
Bias variance & 100 & 25, +5 if more than 2.5\% masked\\
Median, gain=1 & 13 & 11\\
Median, gain=2$^*$ & 15 & 15\\
\hline
\end{tabular}
\label{ta:data:ccdmask}
\end{center}
$^*$: The COSMOS images were taken with
  \mbox{$\mathrm{gain}=1$}, whereas the \mbox{$\mathrm{gain}=2$} setting has been applied
  for some of the  HAGGLeS fields (Marshall et al.\thinspace2010,
  in prep.). While we do not include these fields in the current analysis,
  they have been processed with the same pipeline upgrades described here. Hence, we list
  these values for completeness.
\end{table}

\paragraph{Noise model.}

We compute a rms noise
model for each pixel as
\begin{equation}
\label{se:data:markII:noisemodel}
\mathit{ERR}=F^{-1} \sqrt{s F+t_\mathrm{exp} D+\sigma_\mathrm{r}^2+\gamma^2V}\quad[\mathrm{e}^-]\,,
\end{equation}
with the normalised flat-field $F$, the sky background $s$ 
$[\mathrm{e}^-]$,
the dark reference frame $D$ $[\mathrm{e}^-/\mathrm{s}]$,
the exposure time $t_\mathrm{exp} \,[\mathrm{s}]$, the  read-noise
$\sigma_\mathrm{r}\simeq5\,\mathrm{e}^-$, and the bias  variance image $V$
$[\mathrm{counts}^2]$ described in 
the previous paragraph,
which 
requires scaling with the gain $\gamma$ [$\mathrm{e}^-/\mathrm{count}$].
Containing all noise sources except object photon noise, this rms model is
used for optimal pixel weighting in \texttt{MultiDrizzle}.

\section{Correction for PSF and CTI effects}
\label{app:psf_cti}

\subsection{Summary of our KSB+ implementation}
\label{app:ksbimp}

We measure galaxy shapes 
 using the \citet{ewb01} implementation of the KSB+
formalism \citep{ksb95,luk97,hfk98},
as done in the earlier ACS weak lensing analysis of \citet{ses07}.
Object ellipticities\footnote{{\ro We adopt the widely used term ``ellipticity'' here, but note that,  strictly speaking, (\ref{eq:elli_e})  corresponds to the definition of the polarisation.
}}
\begin{equation}
\label{eq:elli_e} 
   e = e_1 + \textrm{i} e_2 = \frac{Q_{11} - Q_{22} + 2 \textrm{i} Q_{12}}{Q_{11} + Q_{22}} 
\end{equation}
are measured from weighted second-order
brightness moments
\begin{equation}
  \label{eq:qij_w}
  Q_{ij} = \int \mathrm{d}^2 \theta \, W_{r_\mathrm{g}}(|\boldsymbol{\theta}|) \, \theta_i \, \theta_j I(\boldsymbol{\theta}) \, , \quad i,j \in \{1,2\}  \, ,
\end{equation}
where $W_{r_\mathrm{g}}$
is a 
2D Gaussian with 
dispersion
$r_\mathrm{g}$.
The response of a galaxy ellipticity to reduced gravitational shear $g$ and PSF effects is given by
\begin{equation}
  \label{eq:elli_shear_psf}
  e_\alpha - e_\alpha^\mathrm{s} = P_{\alpha \beta}^g g_\beta + P_{\alpha \beta}^\mathrm{sm} q_\beta^*  \, ,
\end{equation}
with the (seeing convolved) intrinsic source ellipticity $e^\mathrm{s}$ and the ``pre-seeing'' shear polarisability
\begin{equation}
  \label{eq:pg_def}
  P^g_{\alpha \beta} = P^{\mathrm{sh}}_{\alpha \beta} - P^{\mathrm{sm}}_{\alpha \gamma} \left[ \left( P^{\mathrm{sm}*} \right)^{-1}_{\gamma \delta} P^{\mathrm{sh}*}_{\delta \beta} \right]  \,,
\end{equation}
where the shear and smear polarisability tensors $P^\mathrm{sh}$ and $P^\mathrm{sm}$ are calculated from higher-order brightness moments as detailed in \citet{hfk98}.
The PSF anisotropy kernel 
\mbox{$  q^*_\alpha = (P^{\mathrm{sm}*})^{-1}_{\alpha \beta} e_\beta^{*}$}
and ratio  of
 $P^{\mathrm{sh}*}$ and $P^{\mathrm{sm}*}$
 must be measured from stars and interpolated 
for each galaxy position, where we approximate the latter as 
\mbox{$ T^*=\mathrm{Tr}\left[P^{\mathrm{sh}*}\right]/\mathrm{Tr}\left[P^{\mathrm{sm}*}\right]$}.

In the application of the KSB+ formalism several choices 
lead to subtle differences between different KSB implementations, see
\citet{hwb06}
for a detailed comparison.
In short, we 
use sub-pixel interpolation for integral evaluations, measure galaxy
shapes with 
\mbox{$r_\mathrm{g}=r_\mathrm{f}$}, 
the \texttt{SExtractor} flux-radius,
and apply PSF measurements computed with
the same filter scale as used for the corresponding galaxy (interpolated between 24 values with \mbox{$1\le r_\mathrm{g} \le15$} pixels).
We invert the $P^g$ tensor as measured from individual galaxies using the
approximation $(P^g)^{-1}=2/\mathrm{Tr}[P^g]$ commonly applied to reduce noise
\citep{ewb01}.
In contrast to \citet{ses07} we do not apply a constant calibration correction, but employ the signal-to-noise dependent correction (\ref{eq:sndep}).

\subsection{Tests with simulated space-based data}
\label{app:sims}

We test our KSB+ shape measurement pipeline on  simulated space-based weak lensing
data with ACS-like properties, which were provided for testing in the
framework of the Shear Testing
Programme\footnote{\url{http://www.physics.ubc.ca/~heymans/step.html}}. 
The images were created with the \citet{mrc04} image simulation pipeline,
which uses shapelets \citep{reb03,mar05} to model galaxy and PSF
shapes, as already employed for the STEP2 simulations \citep{mhb07}.
All images have $4\mathrm{k}\times 4\mathrm{k}$ pixels of size 0\farcs04,
HST-like resolution, and a depth equivalent to 2ks of ACS imaging.
The data are subdivided into eight sets with different PSFs (\mbox{$|e^*|\lesssim 7\%$}), seven of which
utilize \texttt{TinyTim}\footnote{\url{http://www.stsci.edu/software/tinytim/}} ACS PSF models, and one was created by stacking stars
of similar ellipticity in an ACS stellar field (M).
One of the sets uses simplified exponential profiles 
for galaxy modelling (F), while the others include complex galaxy morphologies modelled with shapelets.
Four sets comprise 100 images, while the others include 200 frames. Within
each set, the images are split into ``rotated pairs'', 
where the intrinsic galaxy ellipticities in one frame resemble 90
degree-rotated versions from the other frame, an approach used in
\citet{mhb07} to reduce the
analysis uncertainty due to shape-noise.
Galaxies are sheared with \mbox{$|g|<0.06$} and convolved with the
PSF, both effects being constant within one frame, but with varying $g$
within one set.
Realistic image noise was added similarly to the STEP2
analysis, except that no noise correlations were introduced.

\begin{figure}
   \centering
  \includegraphics[width=8.5cm]{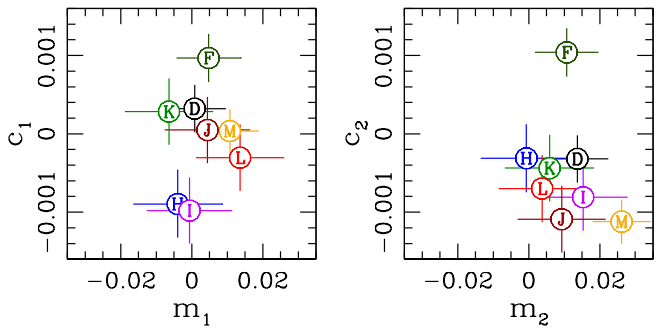}
   \caption{Shear calibration bias $m$ and PSF anisotropy residuals $c$ as measured in
the simulated ACS-like lensing data. The \textit{left} and \textit{right} panels show the results for the $\gamma_1$ and 
$\gamma_2$ shear components respectively. Each letter corresponds to a different PSF model.
Although some data-sets deviate from the optimal \mbox{$m=c=0$}, the residuals are at a level which is
negligible compared to the statistical errors for COSMOS. 
   }
   \label{fi:sim_mc}
    \end{figure}

We analyse the images with the same pipeline and cuts as the real COSMOS
data, with the only difference that the PSF is assumed to be constant across
the field, but still measured from the simulated stars.
Fig.\thinspace\ref{fi:sim_mc} shows the mean calibration bias $m$ and PSF anisotropy residuals $c$ defined in (\ref{eq:calibbias}), separately for each image simulation set, estimated from matched galaxy pairs \citep[for details on this fit see][]{mhb07}.
While some data-sets deviate from the optimal \mbox{$m=c=0$}, the residuals are at a level which is
negligible compared to the statistical uncertainty of COSMOS. Combining all sets and both shear components,
we estimate the mean calibration bias \mbox{$m=+0.008\pm0.002$}, and the scatter of the PSF anisotropy residuals
\mbox{$\sigma_c=0.0006$}.

\begin{figure}
   \centering
  \includegraphics[width=7.cm]{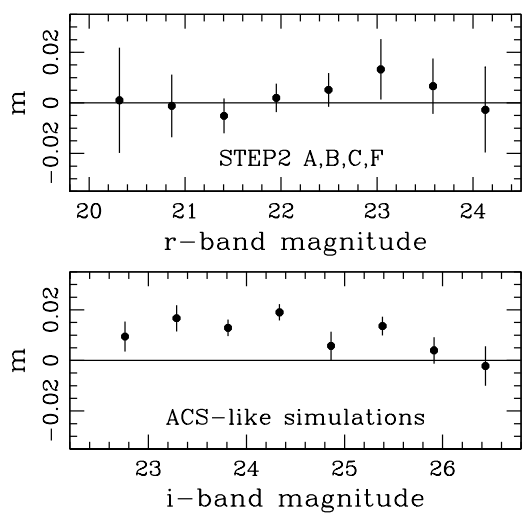}
   \caption{Magnitude-dependence of the shear calibration bias
 for our KSB implementation
after correction for $S/N$-dependent bias according to (\ref{eq:sndep}).
The \textit{top} panel shows results for the STEP2 simulations of ground-based lensing data \citep{mhb07}, which have been used to derive (\ref{eq:sndep}), where we have excluded the untypically elliptical PSFs D and E. The \textit{bottom} panel shows the remaining calibration bias for the ACS-like simulations of space-based lensing data. 
In both panels we plot the average computed from all PSF models and the two shear components, with error-bars indicating the uncertainty of the mean.
Despite the very different characteristics of the two sets of simulations, (\ref{eq:sndep}) performs also very well for the ACS-like data, with a bias \mbox{$|m|\le0.02$} over the entire magnitude range.
The remaining calibration uncertainty is negligible compared to the statistical errors for COSMOS.
   }
   \label{fi:sim_magdependence}
    \end{figure}

As discussed in Sect.\thinspace\ref{se:wl}, a possible
magnitude-dependence of the shear calibration bias $m$ is particularly
problematic for 3D weak lensing studies.
We therefore study $m$ as a function of magnitude in Fig.\thinspace\ref{fi:sim_magdependence},
both for the simulated ground-based STEP2 and the simulated space-based ACS-like data.
Although the applied correction  (\ref{eq:sndep}) was determined from the simulated ground-based data, it also performs very well for the ACS-like simulations, showing its robustness.
Over the entire magnitude range the remaining calibration bias is \mbox{$|m|\le0.02$}, which is negligible compared to our statistical errors.

\subsection{Stellar fields}
\label{sec:wl:stellarfields}

We have 
analysed 700 $i_{814}$ exposures of dense stellar fields,
which were taken between 2002 Apr 18 and 2006 Jun 03 and contain at least 300
non-saturated stars with \mbox{$S/N>50$} (for \mbox{$r_\mathrm{g}=1.5$} pixels).
This large set 
enables us to study in detail the impact of CTI on stars, as well as the temporal and
positional ACS PSF variation, which cannot be achieved from the
COSMOS exposures due to their low stellar density.

We determine both CTI and PSF models for the cosmic ray-cleansed \textit{COR} images before resampling,
and their resampled (but not stacked) counterparts (\textit{DRZ}).
The reason is that resampling unavoidably adds extra noise, 
hence it is best to fit the available stars in a galaxy field exposure before resampling.
Yet, the combined PSF model for a stack has to be determined from resampled 
image models according to the relative dithering. For the \textit{COR}-image analysis we employ a fixed Gaussian filter scale
\mbox{$r_\mathrm{g}=1.5$} pixels, in order to maximize the fitting
signal-to-noise \citep[see][]{ses07}, and characterize the PSF by the ellipticity $e^*_\alpha$ and stellar half-light radius $r_\mathrm{h}^*$ as suggested by \citet{jbs07}.
For the  \textit{DRZ} images we
require CTI-corrected PSF models 
for all 24 values of $r_\mathrm{g}$ used for the galaxy correction.

\subsection{Stellar CTI correction}
\label{sec:wl:starcte}
CTI charge trails stretch objects in the readout
$y$-direction, leading to an additional negative $e_1$ ellipticity component.
Internal calibrations \citep{mus05}, photometric studies \citep[e.g.][]{clk09},
as well as the analysis of warm pixels \citep{msl10} and cosmic rays \citep{jrf09}
 demonstrate that the influence of CTI increases linearly with time and the number of $y$-transfers,
where the latter has also been shown for the influence on galaxy
ellipticities by \citet{rma07}.
In addition, the limited depth of
charge traps leads to a stronger influence of 
CTI
 for
faint sources, which lose a larger fraction of their charge than bright sources.
Likewise, the effect is reduced for higher sky background values leading to
a fraction of continuously filled traps.
Here we only study the effect of CTI on stars, whereas galaxies
will be considered in App.\thinspace\ref{se:wl:galcor}.

Following \citet{clk09}, we assume a power-law dependence on sky background and integrated flux 
as measured in apertures
of 
\mbox{$4.5 \thinspace r_\mathrm{f}\simeq 5.8$} 
pixels,
leading to the parametric CTI model
\begin{eqnarray}
\label{eq:cte_stars}
e_1^{\mathrm{cti},*}(r_\mathrm{g})&=&-e_1^0(r_\mathrm{g})\left(\frac{\mathrm{FLUX}}{10^4\mathrm{e}^-}\right)^{-F(r_\mathrm{g})}\left(\frac{\mathrm{SKY}}{30\mathrm{e}^-}\right)^{-S(r_\mathrm{g})}\left(\frac{t}{1000\mathrm{d}}\right)\nonumber\\
&& \times\left(\frac{y_\mathrm{trans}}{2048}\right)\,,
\end{eqnarray}
with the time \mbox{$t=\mathrm{MJD}-52340$}
 since the installation of ACS on 
2002 Mar 08, and the
number of $y$-transfers $y_\mathrm{trans}$.
We expect that the normalisation \mbox{$e_1^0(r_\mathrm{g})$} and power law exponents
\mbox{$F(r_\mathrm{g})$}
 and \mbox{$S(r_\mathrm{g})$} depend on the
Gaussian filter scale $r_\mathrm{g}$ of the KSB ellipticity measurement.
E.g., for a measurement of the PSF core with small $r_\mathrm{g}$, charge traps may 
 already be filled by electrons from the outer
stellar profile, leading to an expected strong flux-dependence.
On the contrary, the PSF wings measured at large  $r_\mathrm{g}$ will be
more susceptible to trap filling by background electrons, leading to a
stronger sky-dependence.

   \begin{figure}
   \centering
   \includegraphics[height=4.35cm]{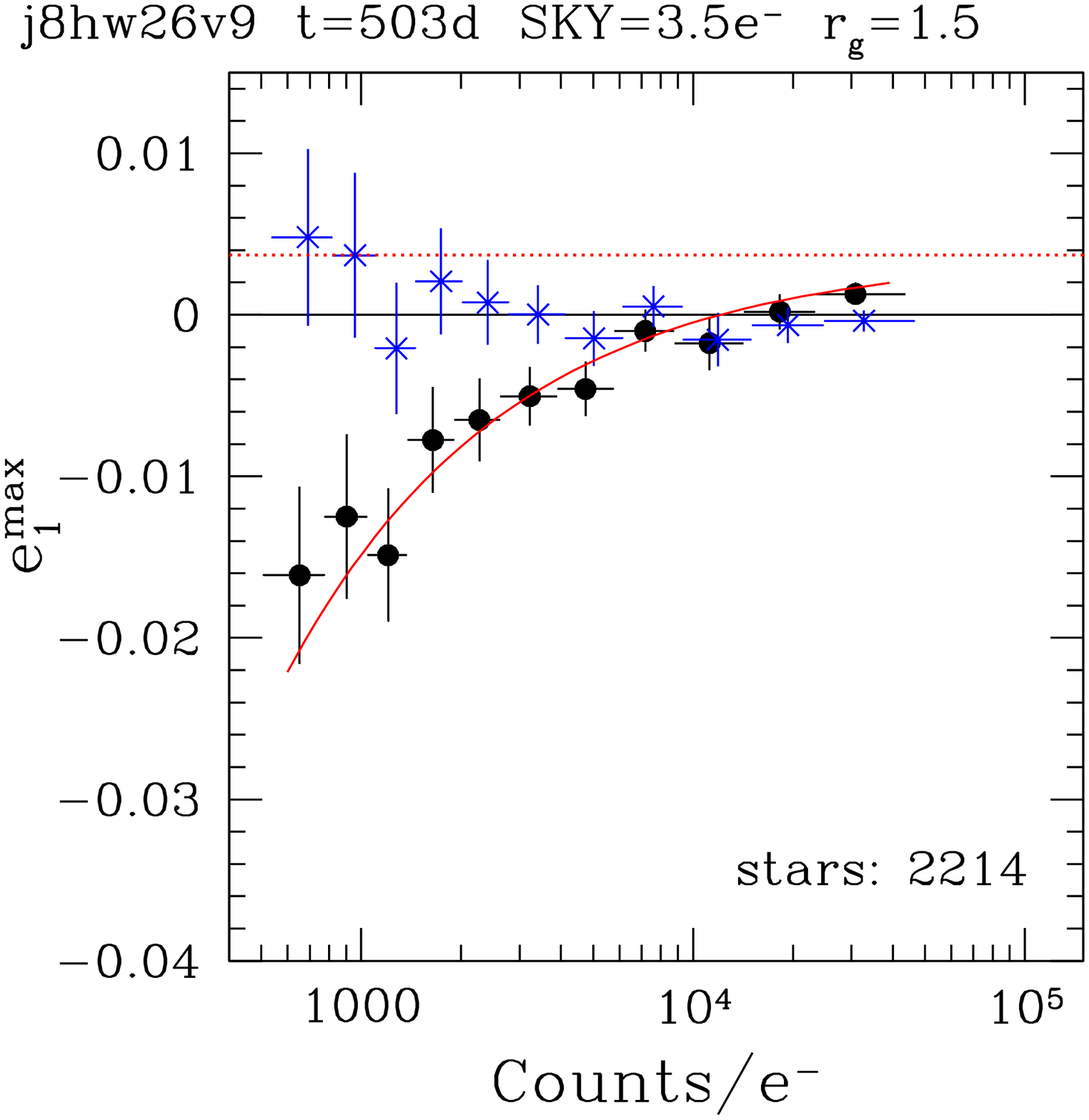}
     \includegraphics[height=4.35cm]{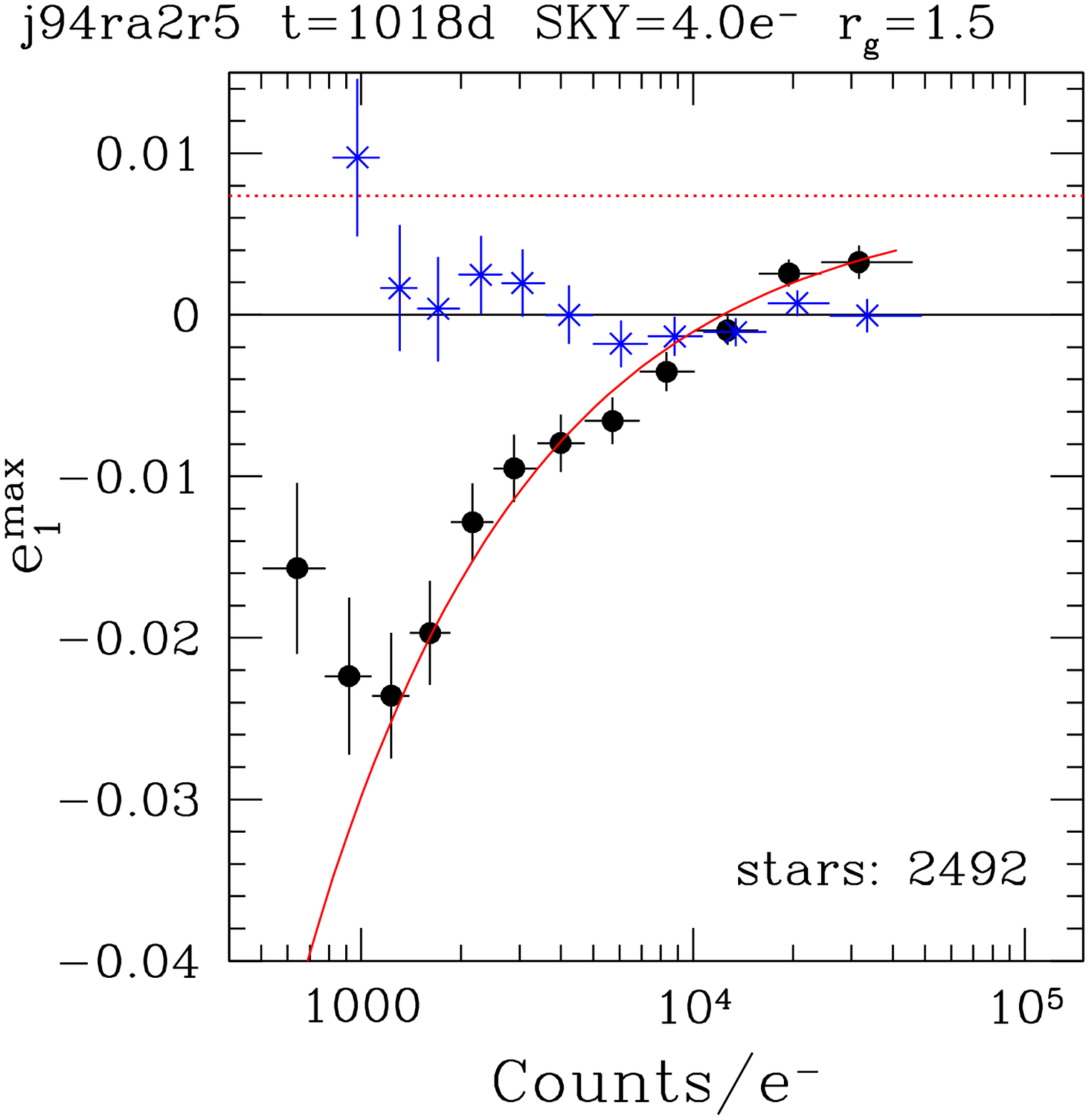}
   \includegraphics[height=4.35cm]{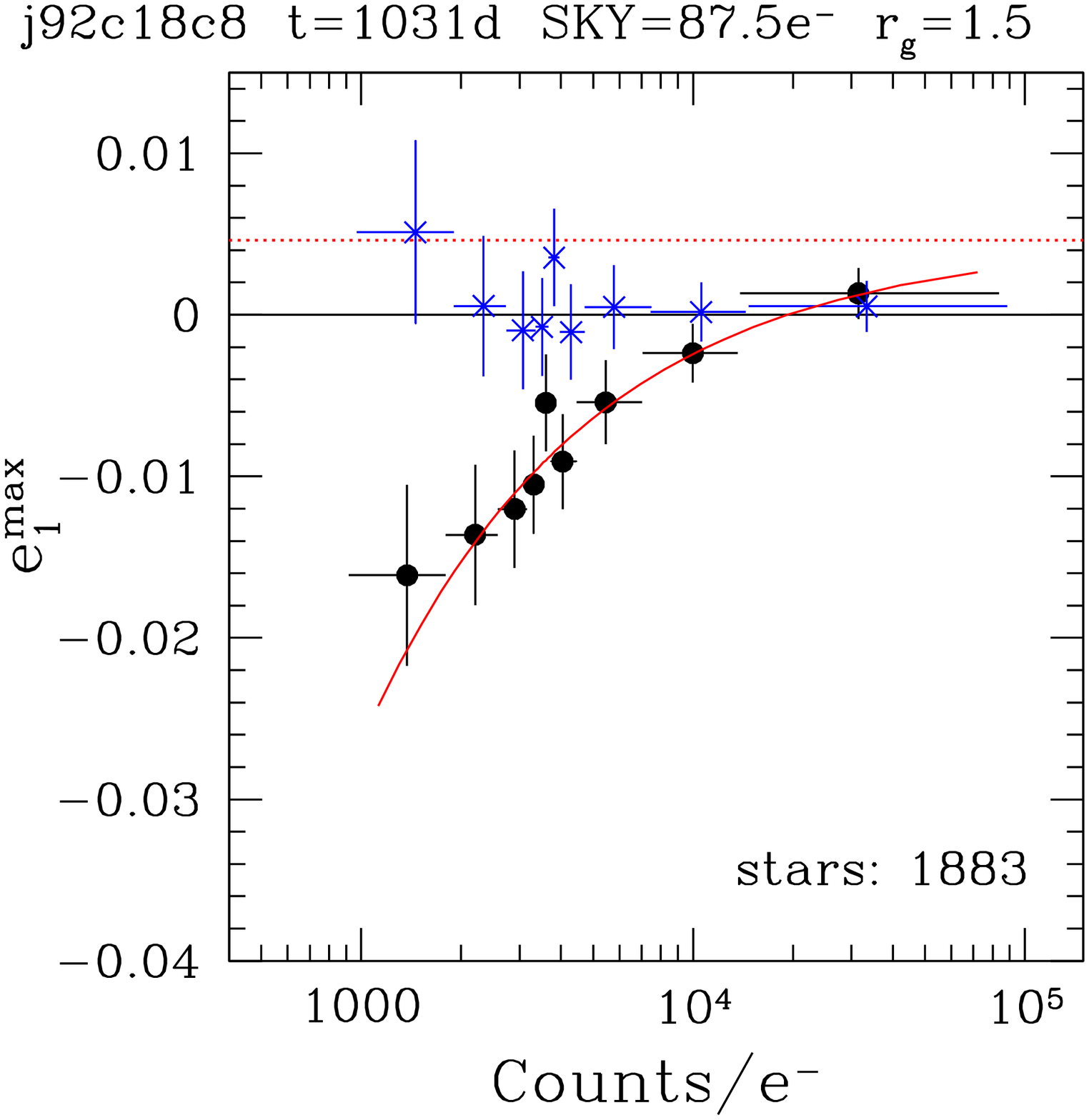}
   \includegraphics[height=4.35cm]{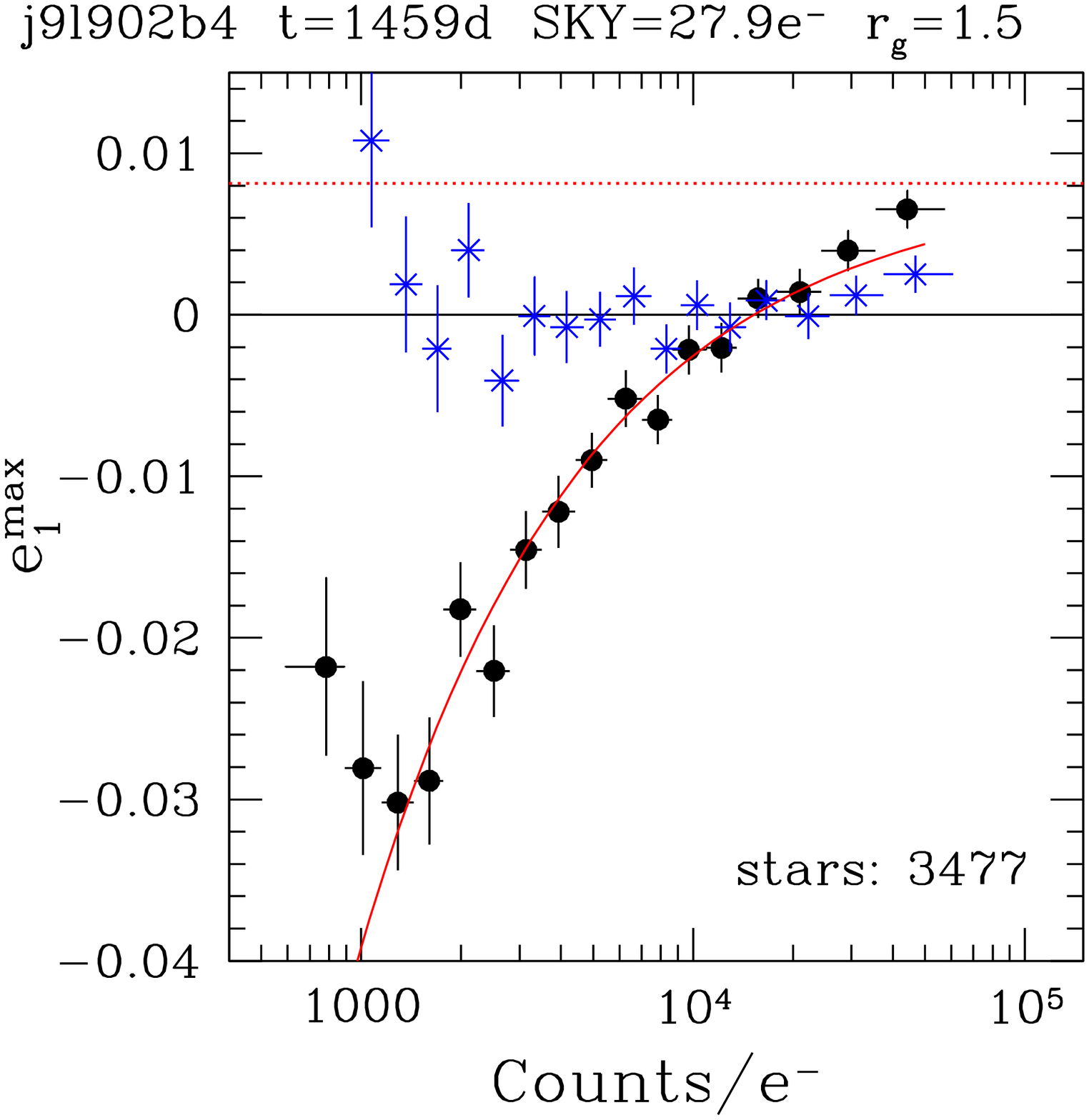}
   \caption{CTI-induced stellar ellipticity for four example stellar field exposures:
The bold points show the mean stellar $e_1$ ellipticity-component as a function of stellar
flux after subtraction of a spatial third-order polynomial model derived
from bright stars 
(\mbox{$S/N>50$}) 
to separate PSF and CTI effects. Each stellar ellipticity has
   been scaled to a reference number of $y_\mathrm{trans}=2048$ parallel
   readout transfers.
The curves show the CTI model 
(\ref{eq:cte_stars}), 
where the fit parameters
have been jointly determined from \mbox{$S/N>20$} stars in all 700 exposures, and an offset shown by the horizontal line has been
applied, corresponding to the mean CTI
model ellipticity of the bright stars used for the polynomial interpolation.
The crosses indicate  the corrected ellipticities after subtraction of the CTI
model. 
Note the strong increase of the CTI-induced ellipticity with time
(\textit{top left} to \textit{bottom right}) and moderate dependence on sky
background (\textit{top right} versus \textit{bottom left} at similar times).
Also note the turnaround occurring for faint stars at  \mbox{$\sim 1000\mathrm{e}^-$} (corresponding to \mbox{$S/N\sim5-10$}) in the \textit{right} panels, see \citet{jrf09} for a further investigation of this effect.
The plots shown correspond to the non-resampled \textit{COR}-images with
ellipticities measured using a Gaussian filter scale of \mbox{$r_\mathrm{g}=1.5$} pixels.
}
   \label{fi:cte_flux_dep}
    \end{figure}

In order to separate CTI and PSF effects we make use of the fact that
CTI-induced ellipticity is expected to depend on flux, while PSF
ellipticity is flux-independent.
In our analysis of stellar field exposures we
first fit the spatial ellipticity variation 
of bright non-saturated stars with 
\mbox{$S/N>50$} 
using a third-order
polynomial in each chip, and apply this model to all stars with
\mbox{$S/N>5$}.
For the high \mbox{$S/N$} 
 stars used in the fit, the strongest ellipticity
contribution comes from the spatially varying PSF.
Yet, for these stars the polynomial fit also corrects for the position-dependent but
\emph{flux-averaged} CTI effect, leading to a net negative  $e_1$
ellipticity for fainter than average stars (CTI under-corrected), and net
positive $e_1$ for brighter stars (CTI over-corrected).
For even fainter stars with 
\mbox{$S/N<50$}   we expect an increasingly more
negative $e_1$ ellipticity component.
Hence, the CTI influence can be measured from the flux-dependence
of the polynomial-corrected \emph{residual} ellipticity,
as illustrated in Fig.\thinspace\ref{fi:cte_flux_dep} for four example
exposures.
We note a turnaround in the CTI flux-dependence for some exposures at low 
\mbox{$S/N\sim5-10$} (right panels in Fig.\thinspace\ref{fi:cte_flux_dep}),
which was also reported for CTI measurements from cosmic rays and further investigated by \citet{jrf09}.
This does not affect our stellar models, given that we only use \mbox{$S/N>20$} stars both for PSF measurement and to constrain (\ref{eq:cte_stars}).
Yet, it suggests that CTI models may not be valid over very wide ranges in signal-to-noise, motivating the use of a separate model
for the typically much fainter galaxies in App.\thinspace\ref{se:wl:galcor}.

We determine the three fit parameters in 
(\ref{eq:cte_stars})
jointly from the polynomial-corrected residual ellipticities in all 
stellar exposures. 
For each exposure it is necessary to add an
offset, 
which 
has been 
linearly scaled with $y_\mathrm{trans}$ for each star, in order to compensate for the
flux-averaged correction included in the polynomial fit.
We compute this offset within the non-linear fitting routine\footnote{For the non-linear CTI fits we
  utilize the CERN Program Library \texttt{MINUIT}:
  \url{http://wwwasdoc.web.cern.ch/wwwasdoc/minuit/}} 
for a given set of fit parameters from the
positions and fluxes of the bright stars used in the polynomial fit, and
apply it to all stars.

   \begin{figure}
   \centering
   \includegraphics[height=6.5cm]{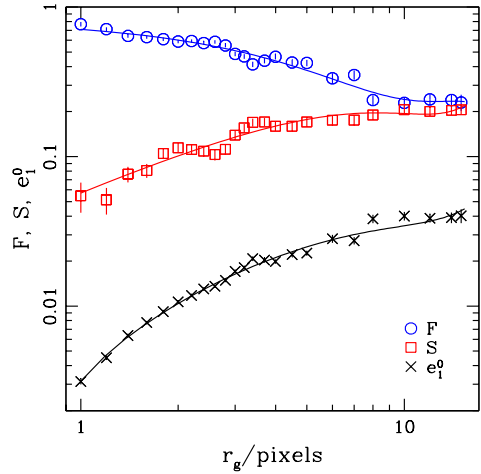}
 \caption{Dependence of the best fitting parameters of the stellar CTI model 
(\ref{eq:cte_stars}) 
on the Gaussian filter scale $r_\mathrm{g}$ used for shape measurements in
the resampled \textit{DRZ}-images. The curves correspond to the fitting
functions 
(\ref{eq:starcti_rgfitted})
.}
   \label{fi:cte_rg_dep}
    \end{figure}

We conduct this fit both for  the 
\textit{COR}-image ellipticities \mbox{($r_\mathrm{g}=1.5$)}
 yielding best fitting values
\mbox{$(e_1^0,F,S)=(0.0073\pm0.0002,0.65\pm0.02,0.06\pm0.01)$}, 
 and for the resampled \textit{DRZ}-images for all values of
$r_\mathrm{g}$.
For the latter we adjust the \mbox{$S/N$} cuts in order to keep enough stars for large $r_\mathrm{g}$.
The best fit values are shown in Fig.\thinspace\ref{fi:cte_rg_dep} as function of $r_\mathrm{g}$, indeed confirming the expected trends. 
We provide the fitting functions 
\begin{equation}
\label{eq:starcti_rgfitted}
e_1^0 (r_\mathrm{g})  =   \sum_{j=0}^{j=3} a_j r_\mathrm{g}^j\,,\,\,\, F (r_\mathrm{g})  =   \sum_{j=0}^{j=3} b_j r_\mathrm{g}^j\,,\,\,\, S (r_\mathrm{g})  =   \sum_{j=0}^{j=3} c_j r_\mathrm{g}^j\,,
\end{equation}
where the coefficients are listed in Table\thinspace\ref{tab:drz_cti_coef},
being valid for \mbox{$1\le r_\mathrm{g}\le 15$} pixels.
We correct the ellipticities of all stars both in the stellar and galaxy fields with the derived models, as implicitly assumed in the following sections.

\begin{table}[tb]
\begin{center}
\caption{Fitted coefficients for the $r_\mathrm{g}$-dependent CTI-ellipticity model (\ref{eq:starcti_rgfitted}) in the resampled \textit{DRZ} frames.
}
\vspace{0.2cm}
\begin{tabular} {crrr}
\hline
$j$ & \multicolumn{1}{c}{$a_j$} & \multicolumn{1}{c}{$b_j$} & \multicolumn{1}{c}{$c_j$} \\
\hline
0 & $- 5.623 \times10^{-3}$ & $8.371\times10^{-1}$ & $1.417\times10^{-3}$\\
1 & $9.573\times10^{-3}$ & $- 1.372\times10^{-1}$ & $6.182\times10^{-2}$\\
2 & $- 8.307\times10^{-4}$ & $1.037\times10^{-2}$ & $- 6.410\times10^{-3}$\\
3 & $2.739\times10^{-5}$ & $- 2.597\times10^{-4}$ & $2.155\times10^{-4}$\\
\hline
\end{tabular}
\label{tab:drz_cti_coef}
\end{center}
\end{table}

\subsection{Principal component correction for the time-dependent ACS PSF}
\label{su:starpca}

As discussed in Sect.\thinspace\ref{se:wl}, ACS PSF variations are expected to be mostly caused by changes in telescope focus \citep[e.g.][]{kri03,lmc06,ank06}.
If  the temporal variations indeed depend  on one physical parameter only, 
it should be possible to construct a one-parametric PSF model, which can be well constrained with the \mbox{$\sim 10-20$} stars available in an ACS field at high galactic latitudes.
Such an approach was implemented by \citet{rma07}, who
measure the mean focus offset for a COSMOS stack
from 
simulated
focus-dependent 
\texttt{TinyTim} PSF models.
They then interpolate the ACS PSF between all stars in COSMOS using polynomial functions dependent on both position and focus offset \citep{lmk07,mrl07}.
However, as suggested by the residual 
aperture mass B-mode signal 
found by \citet{mrl07}, 
this approach appears to be insufficient 
for a complete removal of systematics.
In an  alternative approach, \citet{ses07} 
fit the stars present in each galaxy field exposure using
a large library of stellar field
PSF models. 
While this approach led to no significant residual systematics within the
statistical accuracy of GEMS, it is also not sufficient for the analysis
of the much larger COSMOS data-set.
We therefore implement a new PSF interpolation scheme based on principal component analysis.
It effectively combines the idea of 
exposure-based 
empirical models, which optimally account for time variations and relative dithering
\citep{ses07}, with the aim to describe the PSF variation with a single parameter \citep{rma07}.

\citet{jaj04} introduced the application of principal component analysis 
(PCA) for ground-based PSF interpolation, which we adapt here to obtain
well-constrained PSF models for our ACS weak lensing fields.
Note that \citet{jbs07} and \citet{nbf09} employed PCA to efficiently describe the 
two-dimensional ACS PSF shape, which they then spatially interpolated with normal polynomial functions.
This is conceptually very different to the approach suggested by \citet{jaj04} and used here, 
which employs PCA for the spatial and temporal interpolation of certain quantities needed for PSF correction, such as the stellar ellipticity $e^*$. 

We represent all quantities which we want to interpolate as \mbox{$p_\alpha$}. This includes 
\mbox{$e_1^*, e_2^*, r_\mathrm{h}^*$}  measured in the \textit{COR} images for \mbox{$r_\mathrm{g}=1.5$ pixels}, but also
\mbox{$e_1^*,e_2^*,q_1^*,q_2^*,T^*$} as measured in the \textit{DRZ} images for varying $r_\mathrm{g}$.
The only exception is when we specifically allude to \textit{COR} quantities, which only includes the first group.

  \begin{figure}[t]
   \centering
   \includegraphics[width=6.5cm]{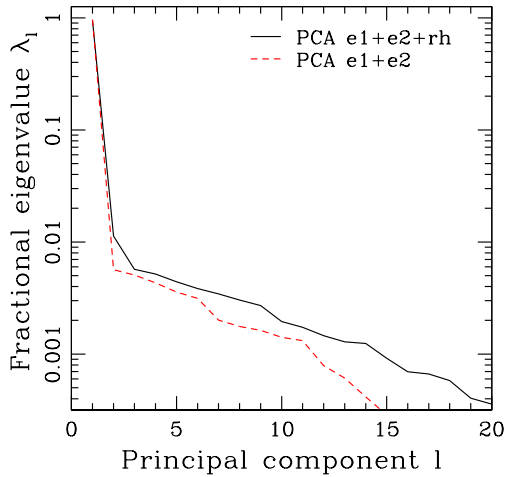}
   \caption{Fractional PCA eigenvalues for the PSF variation in 700 $i_{814}$ ACS
     stellar field exposures. The dashed (solid) curve has been computed considering
     the variation of  $e_1^*$ and $e_2^*$ ($e_1^*$, $e_2^*$, and $r_\mathrm{h}^*$).
   The dominant first principal component contains 97\% (95\%) of the variation and is caused by focus variations.
   }
   \label{fi:pca:eigenvalues}
    \end{figure}

The first step of the PCA analysis is to fit the positional variation of the three
\textit{COR} PSF quantities 
in all \mbox{$j\le N=700$} stellar exposures jointly for both
chips with 
3rd-order
polynomials 
\begin{equation}
\label{eq:polyfit}
P_{\alpha,j}^{(3)}(\hat{x},\hat{y})=\sum_{i=1}^{m}d_{ij}
\hat{x}^{\mu_i}\hat{y}^{\nu_i}\,,
\end{equation}
yielding
\mbox{$m=10$} coefficients each,
where we generally denote polynomials using a capital $P$ with the order indicated by the superscript.
Here we account for the gap
between the chips and rescale the 
pixel range to the interval \mbox{$\hat{x},\hat{y}\in [0,1]$}. 
While this fit is unable to account for some small-scale features such as a small discontinuity between the two
chips, it captures all major large-scale PSF variations and is very well
constrained by the required $\ge 300$ stars.
For each exposure we 
arrange 
the \mbox{$M=3\times m=30$} polynomial coefficients in a 
data
vector $\boldsymbol{d_j}$, 
with components $d_{ij}$
(now \mbox{$i\le M$}).
We then subtract the mean vector and divide each component with an adequately chosen normalization $n_i$,
yielding the modified data vector
$\boldsymbol{\hat{d}_j}$ with
\begin{equation}
\label{eq:pca_data_normalisation}
  \hat{d}_{ij}=\frac{d_{ij}-\overline{d_i}}{n_i} \,.
\end{equation}
We then arrange all modified data vectors
into a $M\times N$ dimensional data matrix
  \mbox{$\boldsymbol{D}=\{\boldsymbol{\hat{d}_1}, ..., \boldsymbol{\hat{d}_N}\}$}.
The central step of the PCA is a singular value decomposition 
\mbox{$\boldsymbol{D}=\boldsymbol{W\Sigma V}^T$},
where the 
orthonormal
 matrix $\boldsymbol{W}$ consists of the singular vectors of $\boldsymbol{D}$, and
the diagonal matrix 
$\boldsymbol{\Sigma}$
 contains the ordered 
singular values $s_{ll}$ of $\boldsymbol{D}$ as
diagonal elements. Here the $l$th largest singular value
corresponds to the
$l$th 
singular vector, which is also named the $l$th principal component.
In the coordinate system spanned by the singular vectors,
the matrix \mbox{$\boldsymbol{C}=\boldsymbol{D D}^T=\boldsymbol{W \Sigma \Sigma}^T \boldsymbol{W}^T=\boldsymbol{W \Lambda
    W}^T$}
corresponding to the covariance matrix for \mbox{$n_i=1$},
becomes diagonal, where the sorted eigenvalues
$\lambda_{l}=s_{ll}^2$ 
are equal
to the variance of the vectors $\boldsymbol{\hat{d}_j}$ along the direction of the
$l$th 
principal component.


 \begin{figure*}[t!]
   \centering
   \includegraphics[width=5.9cm]{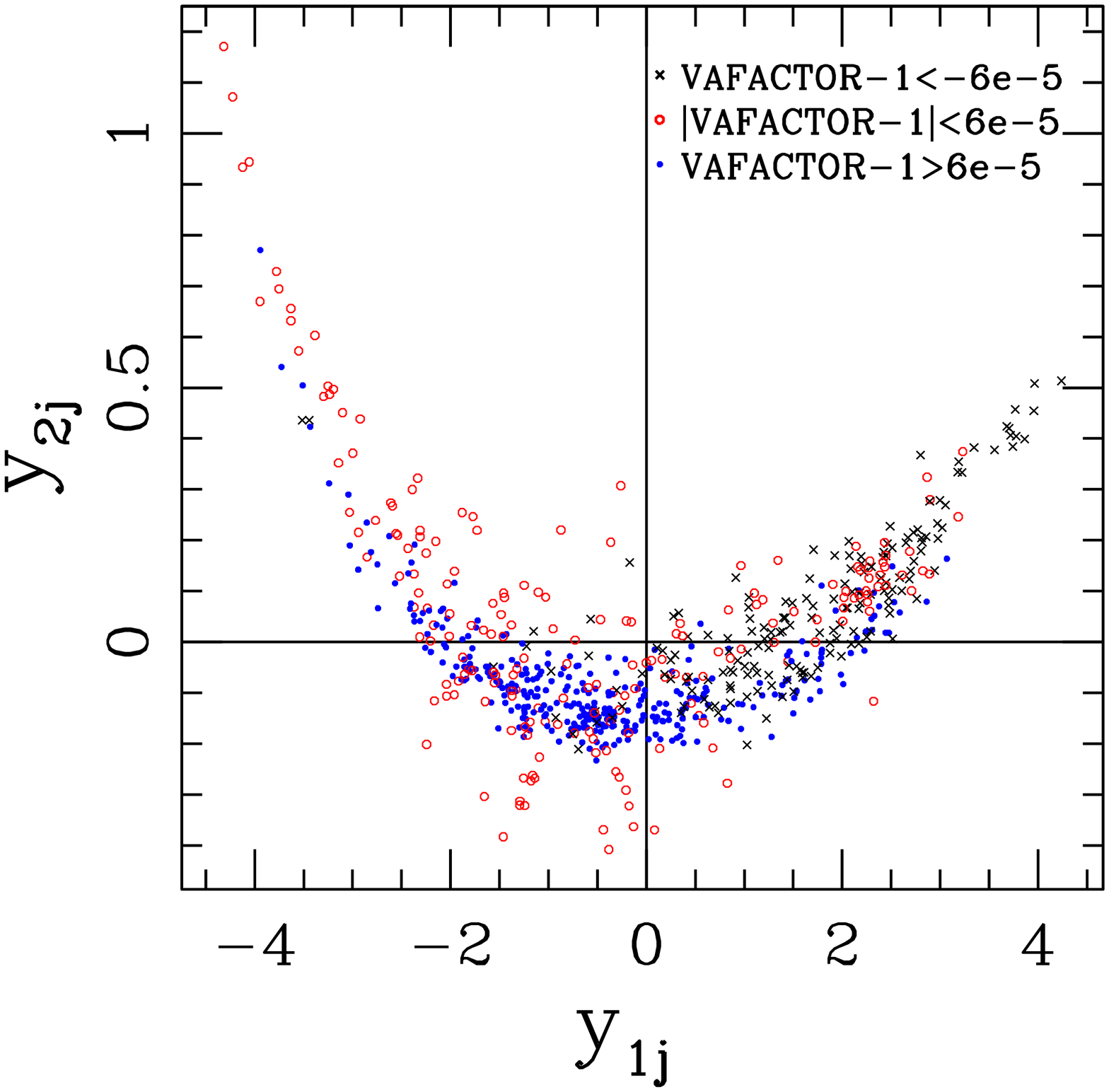}
   \includegraphics[width=5.9cm]{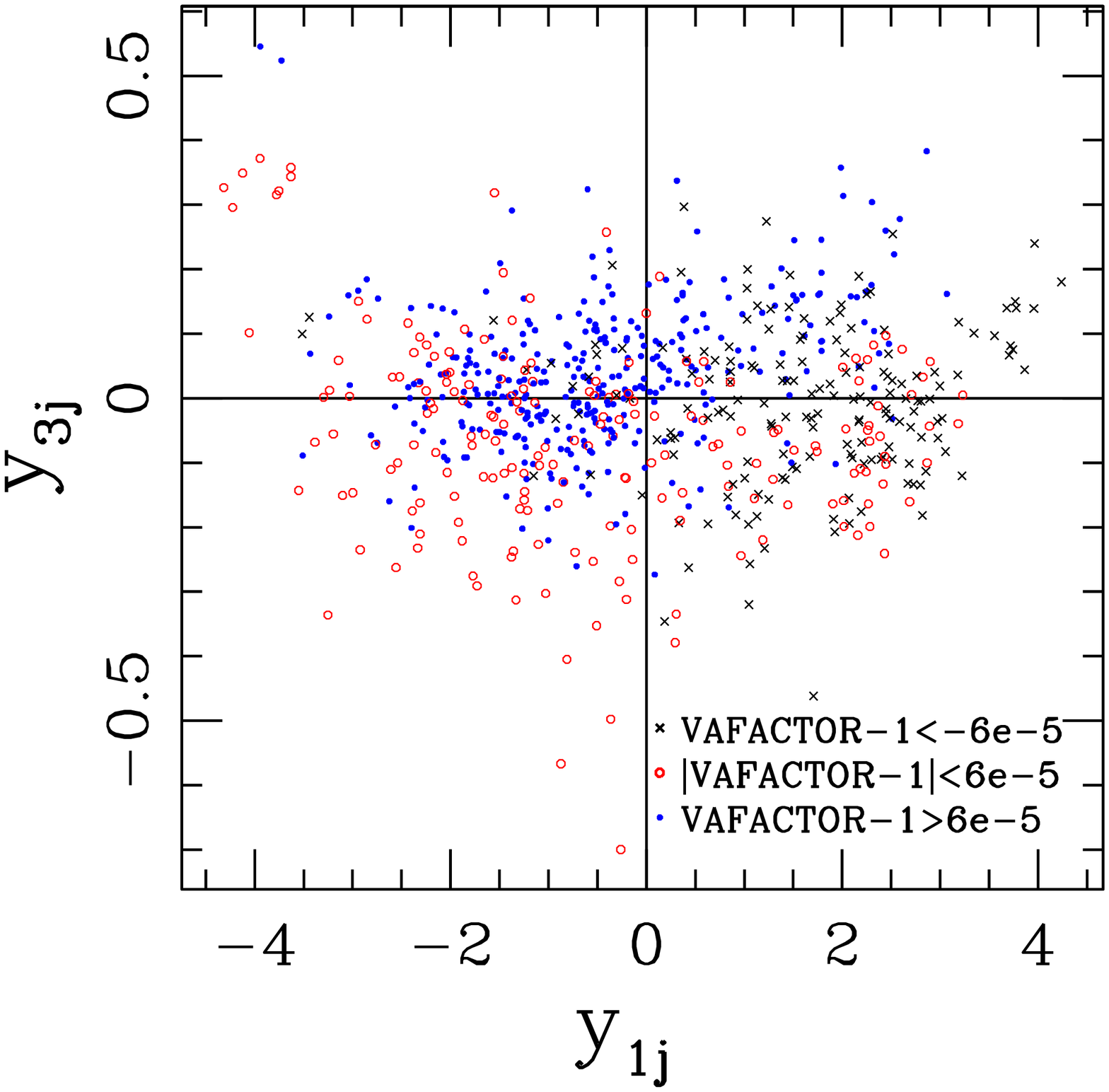}
   \includegraphics[width=5.9cm]{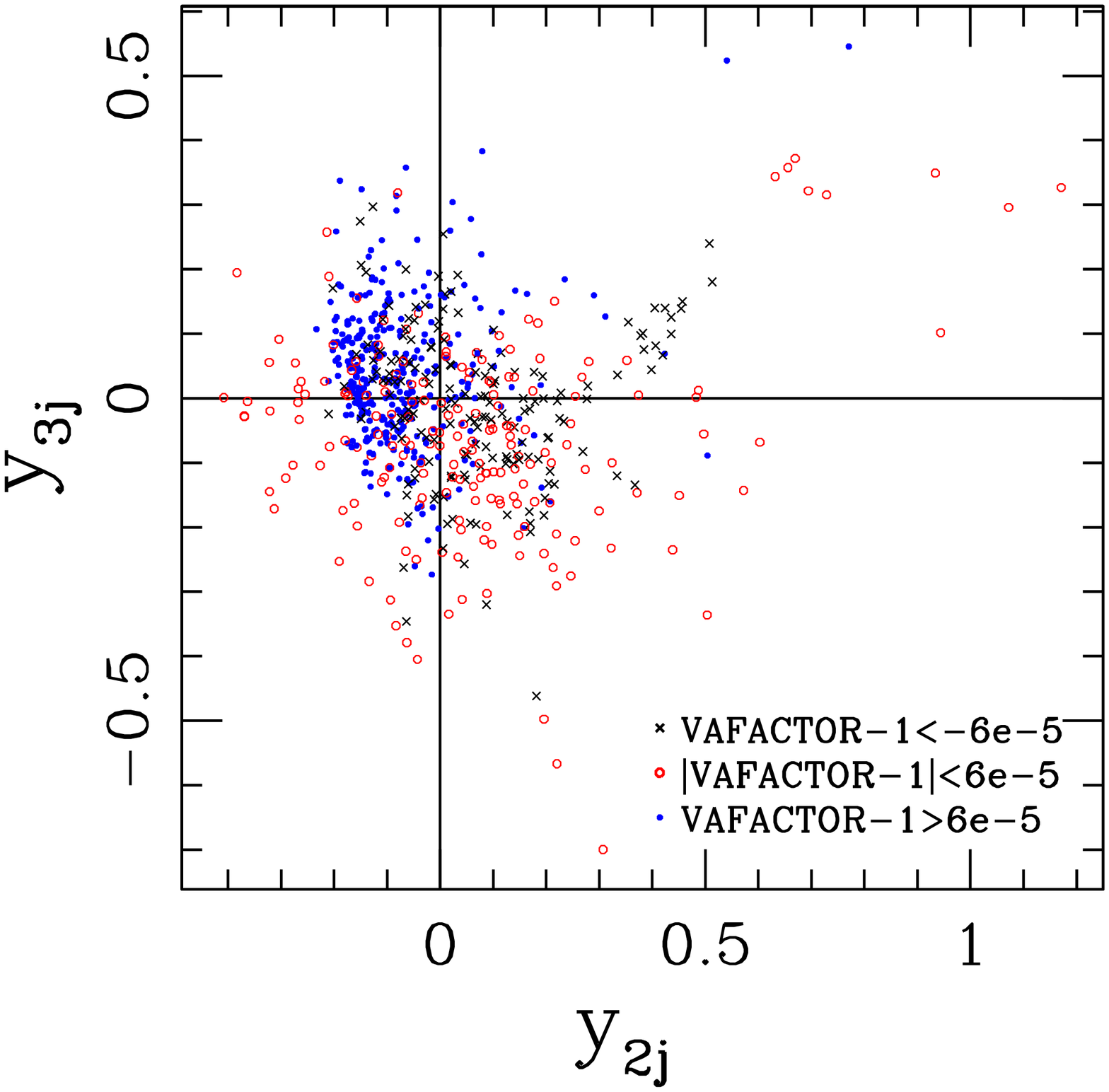}
   \caption{Variation of 700 $i_{814}$ stellar field exposures in the space
     spanned by the first three principal components, which have been computed using the polynomial coefficients of $e_1^*$, $e_2^*$, and $r_\mathrm{h}^*$. Note the different axis scales. The non-linear
     dependence in the left panel is caused by the different response of PSF ellipticity and size on defocus, and leads to the increased eigenvalue $\lambda_2$
in Fig.\thinspace\ref{fi:pca:eigenvalues} if the $r_\mathrm{h}^*$ variation is included in the PCA. The data points have been
     split according to the velocity aberration plate scale factor
     \texttt{VAFACTOR}. The fact that the three subsets scatter differently for fixed 
$y_{1j}$
 shows
     that deviations from pure focus
     variations are not completely random, but depend on orbital
     parameters and may hence be coherent for surveys such as COSMOS.
   }
   \label{fi:pca:pc1_pc2}
    \end{figure*}

Note that the relative values and absolute scale of the eigenvalues
$\lambda_{l}$ depend on the normalisations $n_i$.
Uniform \mbox{$n_i=1$} would not be adequate given that we combine PSF
quantities with different units (dimensionless $e_\alpha^*$ versus $r_\mathrm{h}^*$ in pixels). 
A correlation analysis with
\mbox{$n_i=\sigma_i=\sqrt{\sum_{j=1}^{N}(d_{ij}-\overline{d_i})^2/N}$}
could be used, 
but here
relatively stable polynomial
coefficients with small $\sigma_i$ would unnecessarily add noise,
effectively increasing the relative eigenvalues of higher principal
components.
Aiming at a compact description of most of the actual PSF variation in the
field with a small number of important principal components, we employ the
normalisation
\begin{equation}
\label{eq:new_ni}
n_i=(\mu_i+1)(\nu_i+1)\sigma_\alpha\,,
\end{equation}
where we use the mean variance of all coefficients belonging to the corresponding PSF
quantity $p_\alpha$:
\begin{equation}
\sigma_\alpha^2=\frac{1}{m}\sum_{i=i_{\mathrm{min},\alpha}}^{i_{\mathrm{max},\alpha}}\frac{1}{N}\sum_{j=1}^{N}(d_{ij}-\overline{d_i})^2\,.
\end{equation}
In this way the $\hat{d}_{ij}$ in (\ref{eq:pca_data_normalisation}) become dimensionless, all three PSF quantities contribute similarly 
to the total variation, and the undesired noise from relatively stable polynomial
coefficients is avoided, as their variation is averaged with that from the less stable coefficients.
The pre-factor 
in (\ref{eq:new_ni}) equals the inverse of the integral \mbox{$\int_0^1\mathrm{d}\hat{x}\thinspace\mathrm{d}\hat{y}\thinspace\hat{x}^{\mu_i}\hat{y}^{\nu_i}$} of
the corresponding polynomial term in (\ref{eq:polyfit}).
It accounts for our aim to scale according to the actual PSF variation,
where e.g. a $0$th-order term affects the whole field while a $3$rd-order
term with similar amplitude gets lower weight as it contributes substantially in a smaller area only.

\begin{figure}[t]
   \centering
   \includegraphics[width=6.5cm]{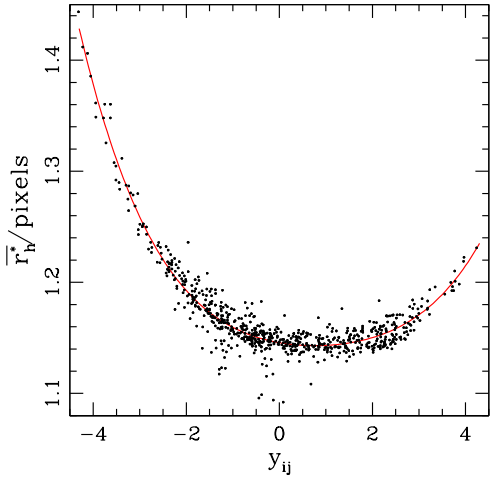}
   \caption{Mean stellar half-light radius  $\overline{r_\mathrm{h}^*}$ 
as a function
of the first principal component coefficient $y_{1j}$ for the 700 $i_{814}$ stellar
field exposures. At \mbox{$y_{1j}\simeq 1$} the telescope is optimally focused. The curve shows the best fitting fourth-order polynomial fit. The outliers are caused by crowded fields with very broad stellar locus.
   }
   \label{fi:pc1_rh}
    \end{figure}

We plot the fractional eigenvalues in Fig.\thinspace\ref{fi:pca:eigenvalues},
once using the analysis as described above (solid curve) and once
considering only the two ellipticity components without $r_\mathrm{h}^*$
(dashed curve). 
In both cases the first principal component is clearly dominant, contributing 
with $95\%$ ($97\%$) of the total variance.
We identify this variation as the influence of focus changes, which are
expected to dominate the actual PSF variation.
The reason why the second
principal component has a 
larger eigenvalue if $r_\mathrm{h}^*$ is
included in the analysis (fractional $\lambda_2=1.1\%$ versus
$\lambda_2=0.6\%$) can be seen if we project the data variation onto the
space spanned by the singular vectors $\boldsymbol{Y}=\boldsymbol{W}^T
\boldsymbol{D}$, with components $y_{lj}$.
Looking at the $y_{1j}-y_{2j}$ variation in the left panel of
Fig.\thinspace\ref{fi:pca:pc1_pc2}, where $r_\mathrm{h}^*$ has been included, 
we see that the data points roughly follow a quadratic curve in the plane 
defined by the first two singular vectors. The reason for this is the linear response of PSF ellipticity on defocus caused by
astigmatism, while PSF width responds  to
leading order quadratically \citep[see e.g.][]{jsj08}.
Given that PCA is a purely linear coordinate transformation, it is not
capable to directly capture this one-parametric variation (separation between
primary and secondary mirror) with a single principal component.
This is only possible if PSF quantities with the same dependence on
physical parameters are included, hence the smaller  $\lambda_2$ if
only the two ellipticity components are considered.
Thus, for other applications it might be more favourable to perform a PCA analysis
for each considered PSF quantity separately, as also done by \citet{jaj04}.
Yet, here we want to include the extra information encoded in the
$r_\mathrm{h}^*$ variation to constrain the galaxy field PSF models, and will
therefore account for the non-linear dependence below. The mean stellar
half-light radius in each exposure, 
$\overline{r_\mathrm{h}^*}$ 
is plotted as a function
of the first principal component coefficient in
Fig.\thinspace\ref{fi:pc1_rh}, showing that a fourth-order polynomial fit is
capable to describe the full non-linear variation.


In order to obtain a well constrained model for all $p_\alpha$ with high spatial resolution,
we jointly fit all stars from all \mbox{$j\le N$} exposures
with a model
\begin{equation}
\label{eq:highfit}
p_{\alpha,j,\mathrm{chip}}^{\mathrm{pcafit}}
(\hat{x},\hat{y})
=\sum_{l=0}^{l_\mathrm{max}}
\sum_{c_l=1}^{c_{l,\mathrm{max}}} [y_{lj}]^{c_l} P_{\alpha,\mathrm{chip},l,c_l}^{(5)}
(\hat{x},\hat{y})\,,
\end{equation}
separately for both chips, where $P_{\alpha,\mathrm{chip},l,c_l}^{(5)}(\hat{x},\hat{y})$ indicates a
fifth-order polynomial in the corresponding rescaled $\hat{x}$, $\hat{y}$
coordinates, and \mbox{$l=0$} with \mbox{$c_{0,\mathrm{max}}=1$} and \mbox{$y_{0j}=1$} corresponds
to the subtracted mean data vector, now modelled with high spatial resolution.
We aim to fit the few stars in the galaxy fields with as few parameters as
reasonably possible.
Due to the dominant role of focus changes
we hence use only the first
principal component in our analysis \mbox{$l_\mathrm{max}=1$}, but include
up to fourth-order terms (\mbox{$c_{1,\mathrm{max}}=4$}) in $y_{1j}$.
This takes out the non-linear distortion 
 visible in Figs.\thinspace\ref{fi:pca:pc1_pc2} and \ref{fi:pc1_rh}, and hence the bulk of the variation in the
second principal component.
This combination yields a total of \mbox{$(1+4)\times 21 = 105$} coefficients
per PSF quantity and chip, which are very well constrained from a total of $5\times 10^5$ stars per chip.

\begin{figure*}[t]
   \centering
   \includegraphics[width=7.8cm]{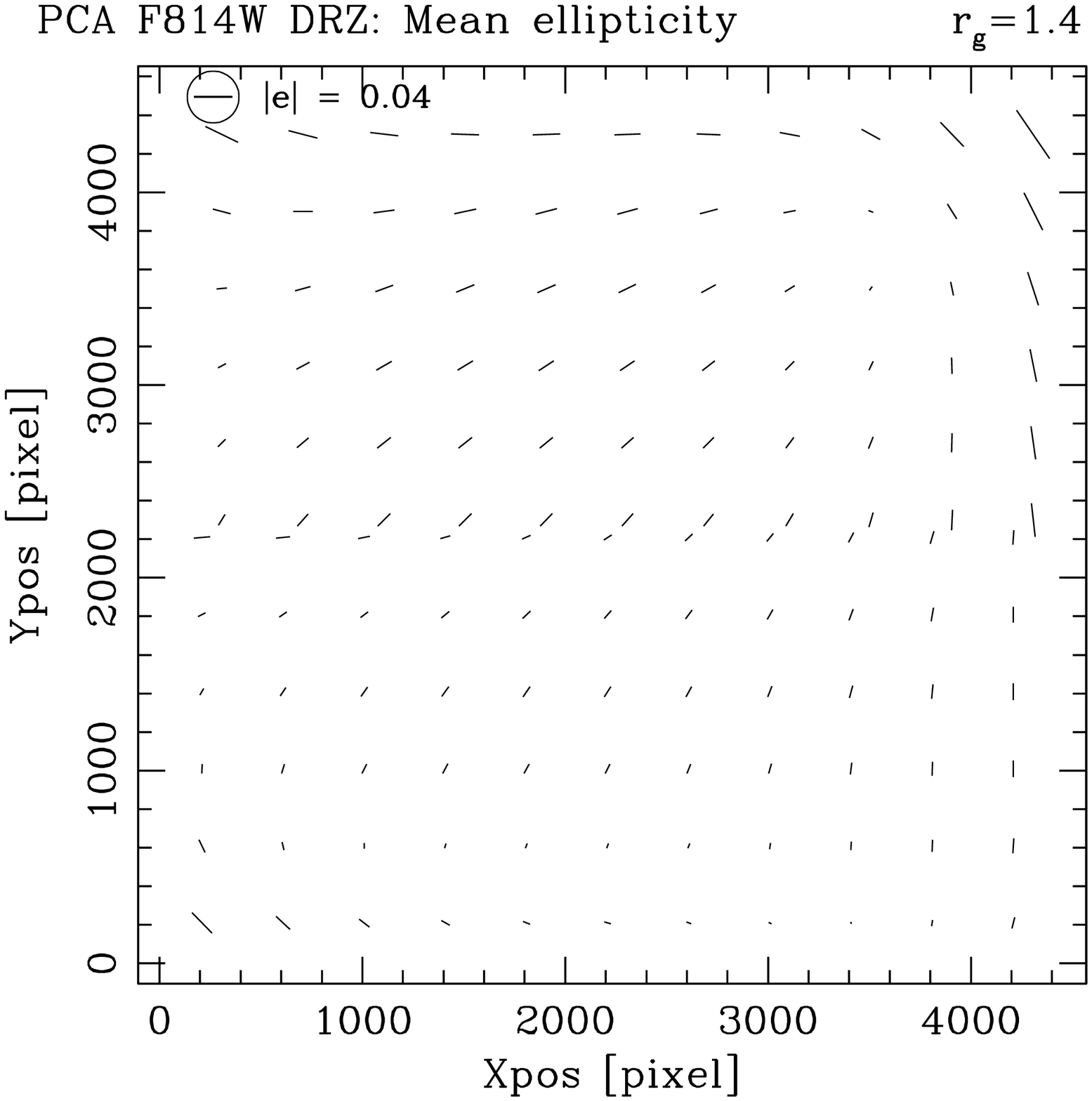}
   \includegraphics[width=7.8cm]{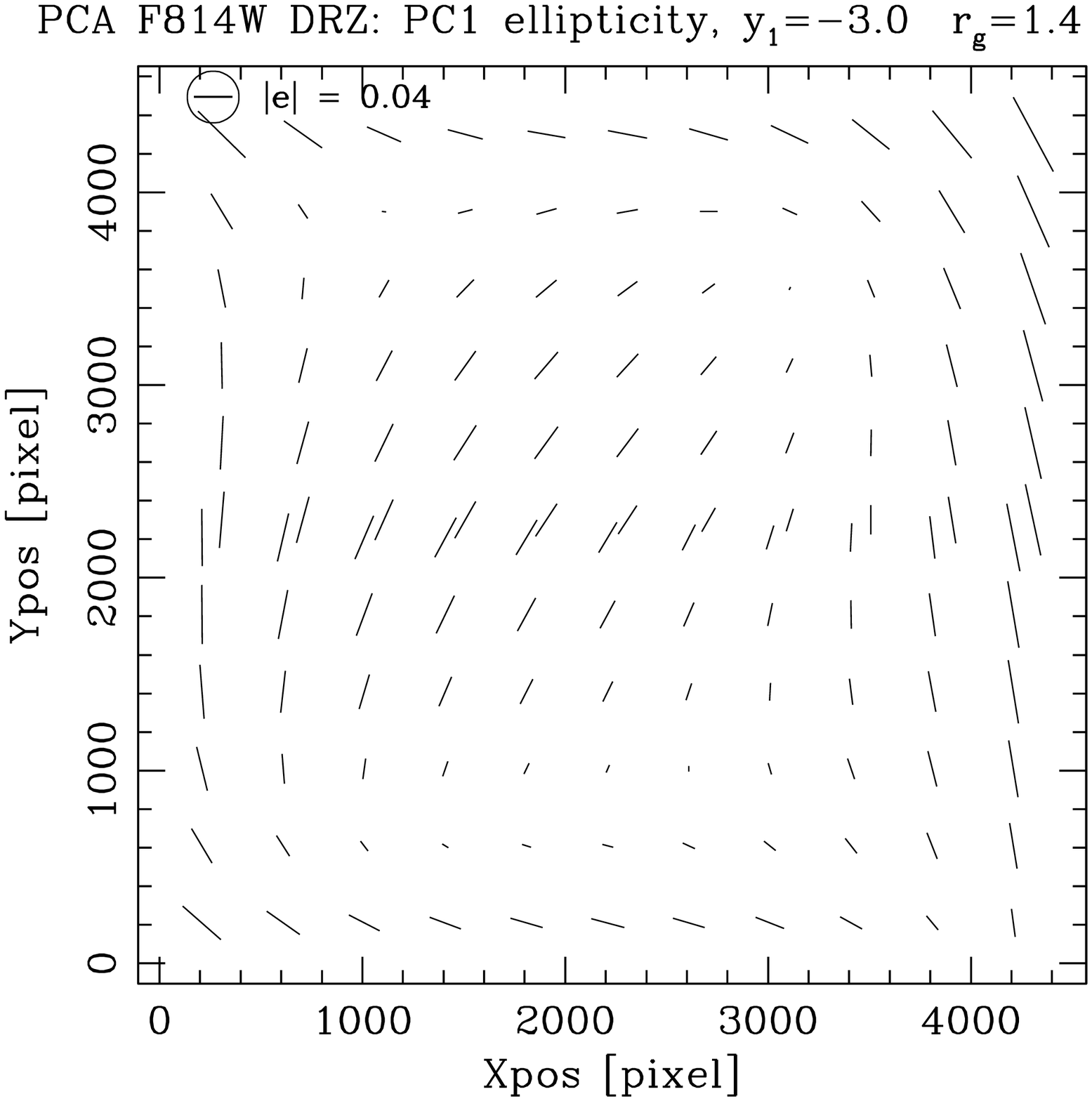}
   \caption{PCA PSF model (\ref{eq:highfit}) for the \textit{DRZ}
     field-of-view ellipticity variation measured with
     \mbox{$r_\mathrm{g}=1.4$ pixels}. The \textit{left} panel shows the
     mean ellipticity (\mbox{$l=0$}), whereas the \textit{right} panel
     depicts the first singular vector (\mbox{$l=1$, $c_l=1$})  which
     corresponds to focus changes, for an arbitrary scale
     \mbox{$y_{1}=-3.0$} (for positive $y_{1}$ the ellipticities are
     rotated by $90^\circ$).
   }
   \label{fi:mean_and_pc1}
    \end{figure*}

For illustration we plot the field-of-view dependence of the high-resolution \textit{DRZ} ellipticity model measured for $r_\mathrm{g}=1.4$ pixels in Fig.\thinspace\ref{fi:mean_and_pc1}, where the left panel shows the mean PSF ellipticity (\mbox{$l=0$}), while the right panel depicts the first singular vector (\mbox{$l=1$}).
Note the slight discontinuity of the mean PSF ellipticity between the chips, which is likely caused by small height differences between the CCDs as reported by \citet{kri03}.
See also \citet{rma07} who measure a stronger discontinuity in the \texttt{TinyTim} PSF model but not for stars in COSMOS, and \citet{jbs07} who notice it in the PSF size but not ellipticity variation.

To obtain PSF models for our COSMOS stacks, 
we fit  \mbox{$e_1^*,
  e_2^*, r_\mathrm{h}^*$} of all stars in the single COSMOS \textit{COR} exposures
with the PCA model (\ref{eq:highfit}) to determine the first principal component coefficient $y_{1j}$ for this exposure. 
We then average the corresponding  \textit{DRZ}-image PSF models of all
exposures contributing to a tile, taking their relative dither offsets and
rotations into account, as detailed in \citet{ses07}.

\begin{figure*}
   \centering
   \includegraphics[width=17cm]{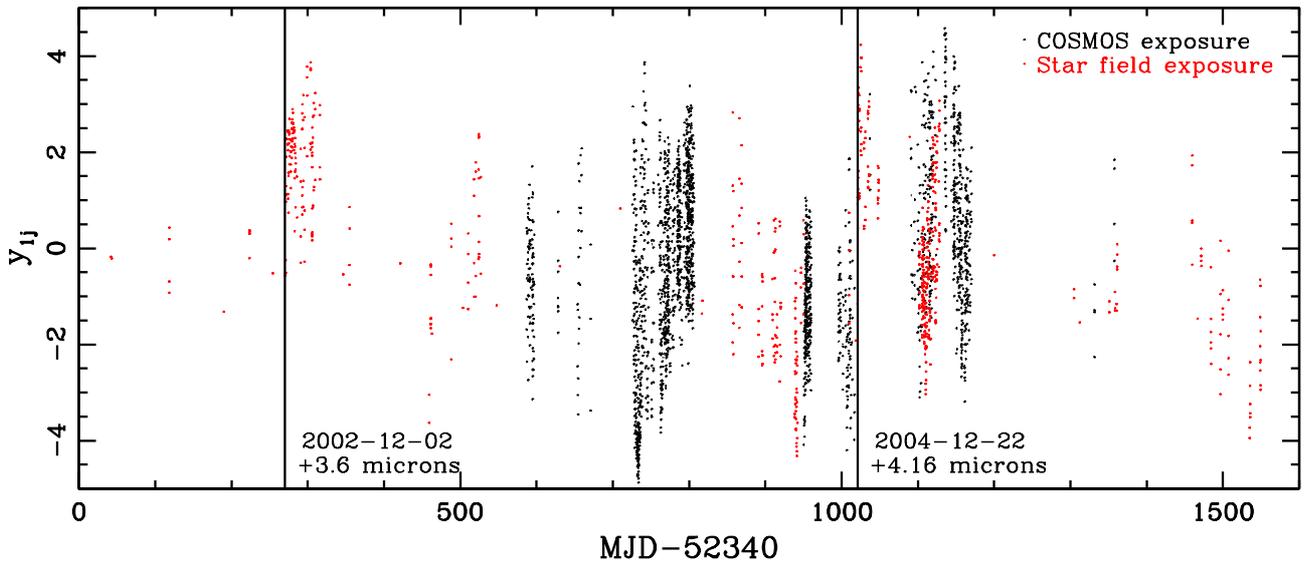}
  \caption{Temporal variation of the first principal component coefficient $y_{1j}$, which is related to the HST focus position, measured in the stellar field and COSMOS exposures. The long-term shrinkage of the OTA is  well visible as a decrease in the mean $y_{1j}$, which was compensated with the marked focus adjustments.
The large spread at a given date is not caused by measurement errors but orbital breathing leading to substantial short-term focus variations.
   }
   \label{fi:pc1_time_dep}
    \end{figure*}

We plot the time dependence of the estimated coefficient $y_{1j}$ for both the COSMOS and stellar field exposures in Fig.\thinspace\ref{fi:pc1_time_dep}.
Note that HST has been refocused at several
occasions to compensate  long-term shrinkage of the OTA, with one correction by
$+4.2$ microns being applied during the time-span of the COSMOS observations on
2004 Dec 22\footnote{\url{http://www.stsci.edu/hst/observatory/focus/} \url{mirrormoves.html}}. 
To ease the comparison, all plots shown in Figs.\thinspace\ref{fi:pca:eigenvalues} to \ref{fi:pc1_time_dep} have been created using a single PCA model determined from all star fields. Yet, to exclude any possible influence of the refocusing, we
actually use separate (but very similar) PCA models for the two epochs in our weak lensing analysis.

While the fit (\ref{eq:highfit}) captures $\sim 97\%$ of the total PSF
variation in the stellar fields and metric defined
above, it is important to realize that further PSF variations
beyond focus changes do actually occur.
These are indicated by the higher principal components and the additional
scatter beyond the curved distortion in the second principal component. 
The subdivision of
fields according to the velocity aberration plate scale factor
\texttt{VAFACTOR} in Fig.\thinspace\ref{fi:pca:pc1_pc2}, which depends on the
angle between the pointing and the telescope orbital velocity vector \citep[see
  e.g.][]{cog02}, indicates that these distortions are not random
but may be coherent for neighbouring fields observed under similar
conditions.
This is not surprising given that HST undergoes substantial temperature
changes and the relative angle towards the sun may lead to
pointing-dependent 
effects\footnote{Note that the actual impact of velocity aberration on object shapes is negligible for our analysis, as long as it is properly accounted for in the image stacking, as done by \texttt{MultiDrizzle}.}.
For a survey like COSMOS, where neighbouring fields have often been observed under
similar conditions, we 
hence
expect 
coherent residual PSF
distortions beyond the one-parameter model introduced here.

{\mt These residuals cannot be constrained reliably from the few stars present in
a single ACS galaxy field, as one would have to fit \mbox{$\gtrsim 10$}
principal components given the slow decline of the  \mbox{$l \ge 3$} eigenvalues
(Fig.\thinspace\ref{fi:pca:eigenvalues}).
However, under the  assumption that they are semi-stable for fields
observed under similar conditions, we can 
constrain these PSF residuals
by combining the stars of multiple COSMOS tiles taken closely in time.
}

\begin{figure*}
   \centering
   \includegraphics[width=7.8cm]{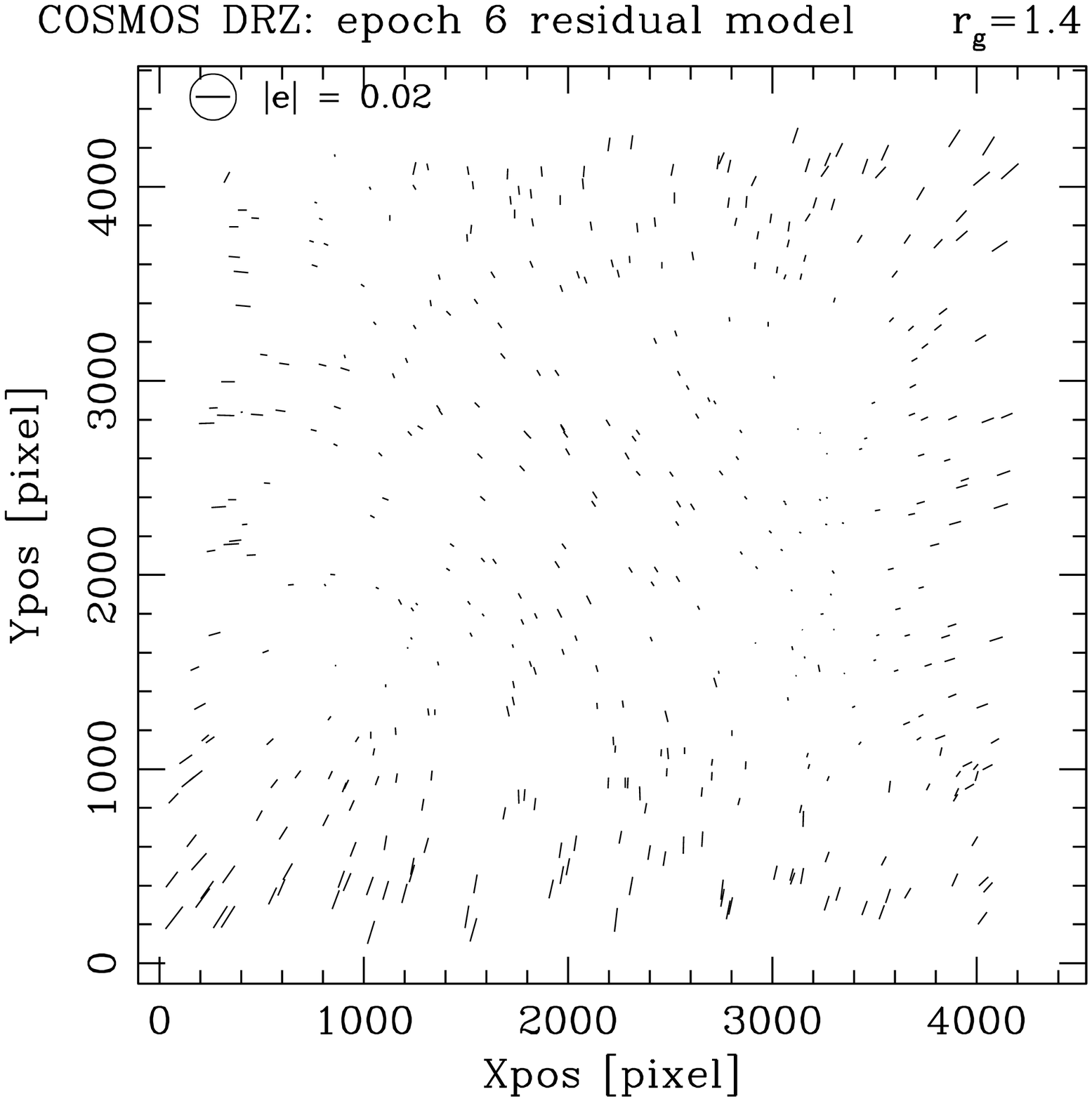}
   \includegraphics[width=7.8cm]{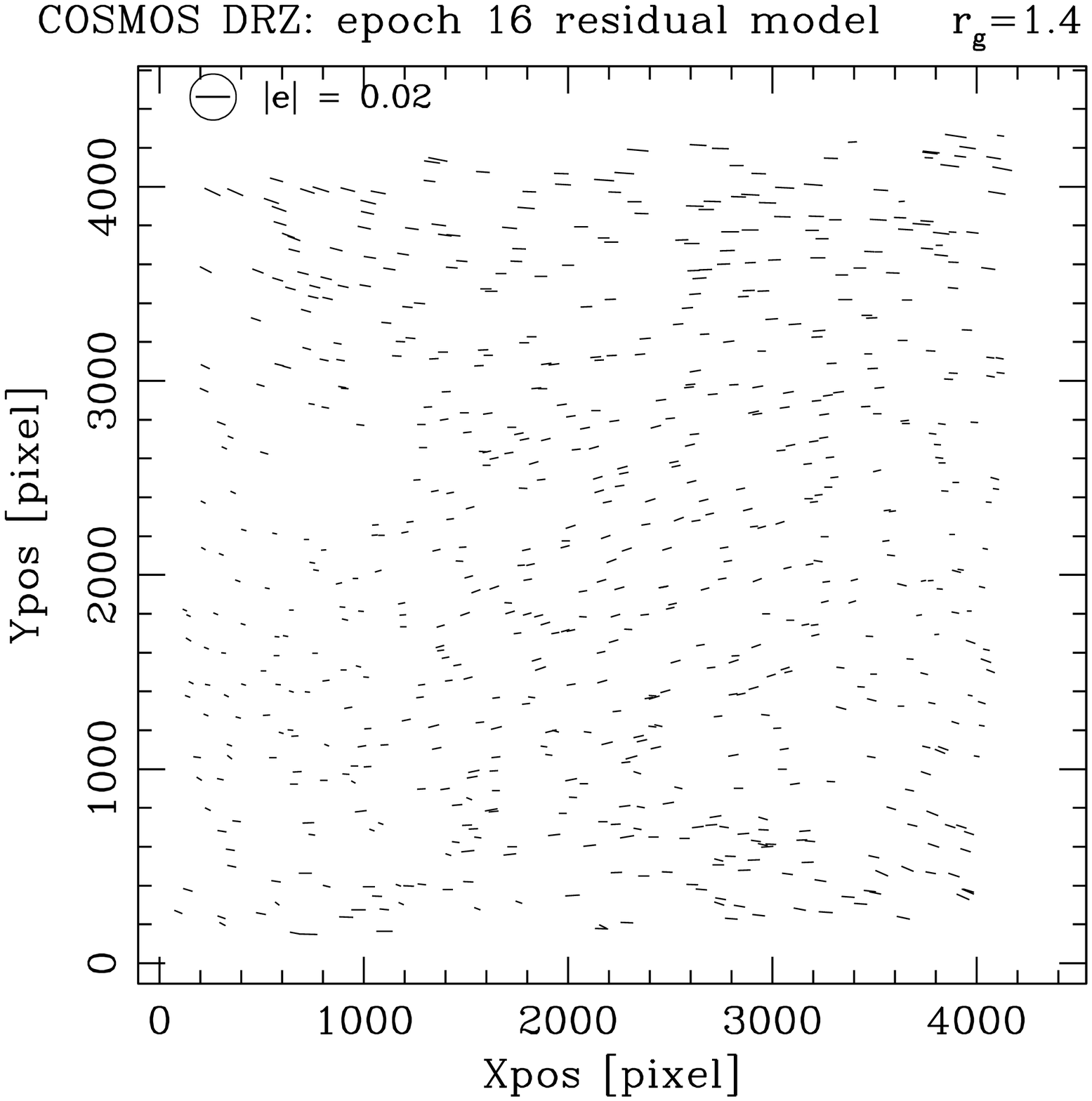}
   \caption{Examples for the residual ellipticity model (\ref{eq:resfit}) determined after subtraction of the 1-parametric PCA model (\ref{eq:highfit}) from the stellar ellipticities  measured in COSMOS stacks with \mbox{$r_\mathrm{g}=1.4$ pixels}. 
The \textit{left} (\textit{right}) plot has been determined from all COSMOS fields with \mbox{$732<t<735$} (\mbox{$950<t<954.5$}), where \mbox{$t=\mathrm{MJD}-52340$}.
Each whisker represents the residual ellipticity model for one star in the epoch. In some cases the model appears to be discontinuous due to the dependence on $\overline{y_{1j}}$ or focus. Note the different scale compared to Fig.\thinspace\ref{fi:mean_and_pc1}.
   }
   \label{fi:res_elli_model}
    \end{figure*}

For this purpose, we split the COSMOS fields into 24 epochs.
Within each epoch the data were taken closely in time, with the same orientation and similar sun angles.
The only exception are two tiles which were reobserved between  2005 Oct 28 and  2005 Nov 24 due to previous guide-star failures \citep{kac07}, which we add to epochs observed one year earlier under similar conditions.
Within each epoch we combine the stars of all tiles and compute residuals of the 
\textit{DRZ} image
PSF quantities 
\mbox{$e_1^*, e_2^*,q_1^*, q_2^*,T^*$} 
by correcting their values measured from the stacks
with
the (dithered and averaged) models (\ref{eq:highfit}).
We then fit these \emph{residuals} as
\begin{equation}
\label{eq:resfit}
p_{\alpha,j}^{\mathrm{resfit}}
(\hat{x},\hat{y})
=P_{\alpha,0}^{(2)}
(\hat{x},\hat{y})
+
\overline{y_{1j}}
P_{\alpha,1}^{(2)}
(\hat{x},\hat{y})
\end{equation}
where $\overline{y_{1j}}$ has been averaged between the four exposures contributing to the stack $j$, and $P_{\alpha,0}^{(2)}$ and $P_{\alpha,1}^{(2)}$ indicate second-order polynomials in the rescaled coordinates $\hat{x}$, $\hat{y}$ 
determined for both chips together.
Here we assume that the additional 
PSF variations
are in principle stable during each epoch, but their impact might depend on the actual focus position and hence $\overline{y_{1j}}$.
Note that we do not use higher-order polynomials in $\hat{x}$, $\hat{y}$ or non-linear powers of $\overline{y_{1j}}$,
as we would otherwise risk over-fitting for epochs with few contributing exposures.
Yet, we tested slightly higher orders for those epochs containing sufficiently many stars, yielding nearly unchanged results.
In general we found that the fitted coherent PSF residuals are small, with a mean rms model ellipticity of 0.3\% (for \mbox{$r_\mathrm{g}=1.4$ pixels}). However, some epochs showed somewhat enhanced ellipticity residuals, with two examples given in Fig.\thinspace\ref{fi:res_elli_model}, motivating us to include this extra term in the galaxy PSF correction.
In contrast we found that the residuals for $T^*$ are negligible.
Also note that we deviate from our philosophy to obtain purely exposure-based models at this point, which is justified by the small and smooth (low-order) corrections applied, which are only 	marginally affected by dithering.

\subsection{Galaxy correction and selection}
\label{se:wl:galcor}

We measure galaxy shapes and correct for PSF effects as detailed in the previous subsections. We then select galaxies with cuts 
\mbox{$r_\mathrm{h}>1.2  r_\mathrm{h}^{*,\mathrm{max}}$}, where $r_\mathrm{h}^{*,\mathrm{max}}$ is the maximum  half-light radius of the $0.25$ pixel wide, automatically determined stellar locus in the image, 
\mbox{$S/N>2.0$}, and \mbox{$\mathrm{Tr}[P^g]/2>0.1$}, identical to the cuts applied to the simulated data in App.\thinspace\ref{app:sims}.
We also reject  saturated stars and galaxies containing masked pixels (Sect.\thinspace\ref{se:data}).
In order to correct galaxy shapes for spurious CTI ellipticity, we fit the PSF anisotropy-corrected 
galaxy ellipticity component \mbox{$e_1^\mathrm{ani}=e_1-P_{1 \beta}^\mathrm{sm} q_\beta^*$} with 
the power law model 
\begin{eqnarray}
\label{eq:galaxycti}
e_1^\mathrm{cti,gal}&=&-e_1^0\left(\frac{\mathrm{FLUX}}{10^3\mathrm{e}^-}\right)^{-F}\left(\frac{r_\mathrm{f}}{3\thinspace\mathrm{pixel}}\right)^{-R}\left(\frac{\mathrm{SKY}}{30\mathrm{e}^-}\right)^{-S}\nonumber\\
&&\times \left(\frac{t}{1000\mathrm{d}}\right)\left(\frac{y_\mathrm{trans}}{2048}\right)\,,
\end{eqnarray}
with the mean sky level of the contributing exposures \texttt{SKY}, the time   \mbox{$t=\mathrm{MJD}-52340$}  since the installation of ACS,
the number of $y$-transfers $y_\mathrm{trans}$,
the \texttt{SExtractor}  flux-radius $r_\mathrm{f}$ (\texttt{FLUX\_RADIUS}), and the  mean integrated flux  per exposure \mbox{$\mathrm{FLUX}=\overline{t_\mathrm{exp}}\cdot\mathtt{FLUX\_AUTO}$} measured by  \texttt{SExtractor}. We scale the latter with the mean exposure time per exposure as the stacks are in units of $\mathrm{e}^-/\mathrm{s}$. 
This model is similar to the one employed by \citet{rma07}, but additionally accounts for the sky background-dependence of CTI effects and allows us to separate the dependence on galaxy flux and size.
Despite the similarity to the stellar model (\ref{eq:cte_stars}), we do not determine a common CTI model for the typically bright stars and faint galaxies, as a simple power law fit is not guaranteed to work well over such a wide range in $S/N$.
Considering all selected COSMOS galaxies we determine best fitting parameters \mbox{$(e_1^0,F,R,S)=(0.0230,0.134,0.638,1.46)$}.
The correction for field distortion leads to a mean rotation of the original
$y$-axis and hence readout direction in ACS stacks and \textit{DRZ} exposures by \mbox{$\phi\sim -2.5^\circ$}.
Thus, CTE degradation has also a minor effect on the $e_2$ ellipticity
component, which we account for in both the galaxy and stellar correction as
\mbox{$e_2^\mathrm{cte}=\tan{(2\phi)}\,e_1^\mathrm{cte}\simeq -0.087 \,e_1^\mathrm{cte}$}.
Note that CTI affects an image after convolution with the PSF. Hence, one would ideally wish to correct for it first. 
Yet, in order to determine the impact of CTI, we need to correct for PSF anisotropy first, which would otherwise dominate the mean $e_1$ ellipticity. We then subtract the CTI model (\ref{eq:galaxycti}) and compute the fully corrected galaxy ellipticity \mbox{$e_\alpha^\mathrm{iso}=(2/\mathrm{Tr}[P^g])(e_\alpha^\mathrm{ani}-e_\alpha^\mathrm{cti,gal})$} with (\ref{eq:elli_shear_psf}), which is an unbiased estimate for the shear $g_\alpha$ if (\ref{eq:calibbias}), (\ref{eq:sndep}) are taken into account.
As it may be easier applicable for non-KSB methods, we also quote best-fitting parameters  \mbox{$(e_1^0,F,R,S)=(0.0342,0.068,1.31,1.26)$} if the actual shear estimates are fitted instead of the PSF anisotropy-corrected ellipticities, where the difference is caused by the PSF seeing correction blowing up the CTI ellipticity.

{
\mt
As a test for residual instrumental signatures 
we create a stacked shear catalogue from all COSMOS tiles.
Doing this, we marginally detect a very weak residual shear pattern,
which
changes with cuts on $\overline{y_{1j}}$.
To quantify and model this residual pattern, we fit it from the 
PSF anisotropy and CTI-corrected \emph{galaxy} ellipticities 
with a focus-dependent, second-order model (\ref{eq:resfit}) jointly for all fields,
yielding a very low 
rms
ellipticity correction of \mbox{$\sim 0.003$}.
}
One possible explanation for these residuals could be the limited capability of KSB+ to fully correct for a complex 
space-based 
PSF,
despite the very good performance on the simulated data in App.\thinspace\ref{app:sims}. 
Alternatively  the limited number of stars per field may ultimately limit
the possible PSF modelling accuracy.
In order to assess if these residuals have any significant impact on our results, we have performed
our science analysis twice, once with and once without subtraction of this residual model. 
The resulting changes in our constraints on $\sigma_8$ are at the $2\%$
level, which is 
negligible compared to the statistical uncertainties. Also the E/B-mode decomposition is nearly unchanged (Fig.\thinspace\ref{fi:tests:eb}). We only detect a significant influence for the star-galaxy cross-correlation, which is strictly consistent with zero only if this correction is applied, but even without correction it is negligible compared to the expected  cosmological signal (Fig.\thinspace\ref{fi:xisys}).

As last step in the catalogue preparation, we create a joint mosaic shear
catalogue from all fields, carefully rejecting double detections in
neighbouring tiles, where we keep the detection with higher $S/N$ and refine
relative shifts between tiles. 
{\ro In the case of close galaxy pairs with separations \mbox{$<0\farcs5$} we exclude the fainter component.}
Our filtered shear catalogue contains 472\,991
galaxies, corresponding to 80 galaxies$/\mathrm{arcmin}^2$, with a mean
ellipticity dispersion per component $\sigma_{e,\alpha}=0.31$.
To limit the redshift extrapolation in
Sect.\thinspace\ref{se:tomo:extrapolate},
we apply an additional cut \mbox{$i_{814}<26.7$}, leaving 446\,934 galaxies,
or 76  galaxies$/\mathrm{arcmin}^2$.

We rotate all shear estimates to common coordinates, and  accordingly create a joint mosaic star catalogue for the analysis in Sect.\thinspace\ref{se:wl:tests}.

\end{appendix}
\bibliographystyle{aa}
\bibliography{h4491,paper1d}

\end{document}